\documentclass[fleqn,usenatbib]{mnras}

\usepackage{newtxtext,newtxmath}

\usepackage[T1]{fontenc}

\DeclareRobustCommand{\VAN}[3]{#2}
\let\VANthebibliography\thebibliography
\def\thebibliography{\DeclareRobustCommand{\VAN}[3]{##3}\VANthebibliography}

\usepackage{graphicx}	
\usepackage{threeparttable}
\usepackage{amsmath}    
\usepackage[left = `, right = ', leftsub = ``, rightsub = '']{dirtytalk}
\usepackage{multirow}
\usepackage{gensymb}
\usepackage{color}
\usepackage[singlelinecheck=off]{subcaption}
\usepackage[dvipsnames]{xcolor}

\usepackage{newtxtext,newtxmath}

\usepackage{anyfontsize}

\definecolor{myblue}{HTML}{1F77B4}
\definecolor{mygreen}{HTML}{2CA02C}
\definecolor{myred}{HTML}{D62728}
\definecolor{mymagenta}{HTML}{D33682}
\definecolor{codepurple}{HTML}{C42043}

\hypersetup{
unicode=true,           
colorlinks=true,        
linkcolor=myblue,        
citecolor=myblue,       
urlcolor=myblue        
}

\title[MUSE on TDE host galaxies]{MUSE IFU observations of galaxies hosting Tidal Disruption Events}

\author[M. Pursiainen et al.]{M. Pursiainen,$^{1}$\thanks{E-mail: Miika.Pursiainen@warwick.ac.uk}
G. Leloudas,$^{2}$
J. Lyman,$^{1}$
C.~M Byrne,$^{1}$
P. Charalampopoulos,$^{3}$
P. Ramsden,$^{4,5}$
S. Kim,$^{6}$
\newauthor
S. Schulze,$^{7}$
J.~P. Anderson,$^{6,8}$
F.~E. Bauer,$^{9}$
L. Dai,$^{10}$
L. Galbany,$^{11,12}$
H. Kuncarayakti,$^{3}$
M. Nicholl,$^{5}$
\newauthor
T. Pessi,$^{6}$
J.~L. Prieto,$^{13,8}$
S.~F. Sanchez$^{14,15}$
\\
$^{1}$ Department of Physics, University of Warwick, Gibbet Hill Road, Coventry, CV4 7AL, UK\\
$^{2}$ DTU Space, National Space Institute, Technical University of Denmark, Elektrovej 327, 2800 Kgs. Lyngby, Denmark\\
$^{3}$ Department of Physics and Astronomy, University of Turku, 20014 Turku, Finland\\
$^{4}$ School of Physics and Astronomy, University of Birmingham, Birmingham B15 2TT, UK\\
$^{5}$ Astrophysics Research Centre, School of Mathematics and Physics, Queens University Belfast, Belfast BT7 1NN, UK\\ 
$^{6}$ European Southern Observatory, Alonso de C\'ordova 3107, Casilla 19, Santiago, Chile\\ 
$^{7}$ Center for Interdisciplinary Exploration and Research in Astrophysics (CIERA), 1800 Sherman Ave., Evanston, IL 60201, USA\\
$^{8}$ Millennium Institute of Astrophysics MAS, Nuncio Monse\~nor Sotero Sanz 100, Off. 104, Providencia, Santiago, Chile\\ 
$^{9}$ Instituto de Alta Investigaci{\'{o}}n, Universidad de Tarapac{\'{a}}, Casilla 7D, Arica, Chile\\ 
$^{10}$ Department of Physics, The University of Hong Kong, Pokfulam Road, Hong Kong SAR\\ 
$^{11}$ Institut d’Estudis Espacials de Catalunya (IEEC), E-08034 Barcelona, Spain\\ 
$^{12}$ Institute of Space Sciences (ICE, CSIC), Campus UAB, Carrer de Can Magrans, s/n, E-08193 Barcelona, Spain\\ 
$^{13}$ Instituto de Estudios Astrof\'isicos, Facultad de Ingenier\'ia y Ciencias, Universidad Diego Portales, Avenida Ej\'ercito Libertador 441, Santiago, Chile\\ 
$^{14}$ Universidad Nacional Autónoma de México, Instituto de Astronomía, AP 106, Ensenada 22800, BC, Mexico\\
$^{15}$ Instituto de Astrofísica de Canarias, Vía Láctea s/n, 38205 La Laguna, Tenerife, Spain
}
\date{Accepted XXX. Received YYY; in original form ZZZ}

\pubyear{\the\year{}}

\begin{document}
\label{firstpage}
\pagerange{\pageref{firstpage}--\pageref{lastpage}}
\maketitle

\begin{abstract}
We present an analysis of twenty tidal disruption event (TDE) host galaxies observed with the MUSE integral-field spectrograph on ESO VLT. We investigate the presence of extended emission line regions (EELRs) and study stellar populations mostly at sub-kpc scale around the host nuclei. EELRs are detected in 5/20 hosts, including two unreported systems. All EELRs are found at $z<0.045$, suggesting a distance bias and faint EELRs may be missed at higher redshift. EELRs only appear in post-merger systems and all such hosts at $z<0.045$ show them. Thus, we conclude that TDEs and galaxy mergers have a strong relation, and $>45$\% of post-merger hosts in the sample exhibit EELRs. Furthermore, we constrained the distributions of stellar masses near the central black holes (BHs), using the spectral synthesis code \texttt{Starlight} and \texttt{BPASS} stellar evolution models. The youngest nuclear populations have typical ages of $\sim$1\,Gyr and stellar masses below $2.5M_\odot$. The populations that can produce observable TDEs around non-rotating BHs are dominated by subsolar-mass stars. 3/4 TDEs requiring larger stellar masses exhibit multi-peaked light curves, possibly implying relation to repeated partial disruptions of high-mass stars. The found distributions are in tension with the masses of the stars derived using light curve models. Mass segregation of the disrupted stars can enhance the rate of TDEs from supersolar-mass stars but our study implies that low-mass TDEs should still be abundant and even dominate the distribution, unless there is a mechanism that prohibits low-mass TDEs or their detection.

\end{abstract}

\begin{keywords}
transients: tidal disruption events -- galaxies: starburst -- galaxies: star formation
\end{keywords}



\section{Introduction}

When a star wanders too close to a supermassive black hole (SMBH) residing in the nucleus of a galaxy, it gets shred apart by the gravitational forces producing a bright flare. Such phenomena were first hypothesised already in the 1970s \citep{Hills1975}, but the first tidal disruption events (TDEs) were discovered in X-rays decades later \citep[e.g.][]{Komossa1999TheScenarios}. The discoveries were rare to begin with, but the advent of wide-field, transient surveys, such as the Sloan Digital Sky Survey \citep[SDSS;][]{Gunn2006TheSurvey}, the La Silla-QUEST \citep[LSQ;][]{Hadjiyska2012LaHemisphere}, All-Sky Automated Survey for Supernovae \citep[ASASSN;][]{Shappee2014AllAssassin}, (intermediate) Palomar Transient Factory \citep[iPTF;][]{Law2009}, the Asteroid Terrestrial impact Last Alert System \citep[ATLAS;][]{Tonry2018} and the Zwicky Transient Facility \citep[ZTF; e.g.][]{Bellm2019}, has resulted in large numbers of optically bright TDEs. Despite being discovered mostly in the optical due to the abundant surveys, TDEs are luminous in multiple wavelength regimes from radio \citep[e.g.][]{VanVelzen2016AASASSN-14li} to X-rays and $\gamma$-rays \citep[e.g. ][]{Zauderer2011BirthJ164449.3+573451}, and some are even faint in the optical with promiminent emission in infrared \citep[e.g.][]{Masterson2024}.

One of the early peculiarities of TDEs, was their host galaxy distribution. \citet{Arcavi2014} noted that their sample showed a significant preference for peculiar E+A galaxies that are otherwise passive with no significant star-formation, but show strong Balmer absorption lines implying a large population of A type stars. The combination implies a strong starburst in the recent ($\lesssim1$\,Gyr) past \citep[e.g.][]{Dressler1983SpectroscopyCluster.}, possibly triggered by a galaxy merger \citep[e.g.][]{Zabludoff1996TheGalaxies}. Later studies have quantified that the over-enhancement factor of TDEs in such post-starburst (PSB) galaxies to be $\gtrsim20$ over the generic galaxy population \citep[e.g.][]{French2016, Law-Smith2017, Graur2018, Hammerstein2021TDESystem}. The TDE host galaxies also stand out photometrically. \citet{Hammerstein2021TDESystem} notes that 63\% of their 19 analysed TDE hosts are found in the \say{green valley} between blue star-forming galaxies, and red passive ones. The green valley is relatively bare, and the study concludes that while these galaxies represent only $13$\% of SDSS galaxies, $61\%$ of their TDE sample is found in such galaxies indicating a clear causal relationship between TDEs and their peculiar host galaxies. Together these galaxy properties can be understood with galaxy mergers producing more concentrated central stellar distributions \citep[e.g.][]{Hammerstein2021TDESystem}, thus enhancing the TDE rate in these peculiar galaxies \citep[][]{Stone2016AN3156, Stone2016RatesFunction, French2020TheParsecs}. 

The TDE hosts also often exhibit extended emission line regions (EELRs) of low velocity, non-turbulent gas that do not follow the general galaxy morphology. \citet{Prieto2016} analysed an Integral Field Unit (IFU) data of host galaxy of ASASSN-14li taken with the Multi-Unit Spectroscopic Explorer \citep[MUSE; ][]{Bacon2010} spectrograph on the Very Large Telescope (VLT), and identified significant emission of ionised [\ion{O}{III}] $\lambda 5007$ and [\ion{N}{II}] $\lambda6583$ lines, extending far ($\gtrsim10$\,kpc) from the PSB host galaxy. Given the emission line strengths and widths together with the assymetric filamentry distributions of the EELRs, the study concluded that the EELRs were likely a remnant of a past merger of the galaxy hosting a weak active galactic nucleus (AGN). Similar features have since been identified in a few TDE hosts. \cite{French2023} searched for similar EELRs in a sample 93 PSB galaxies and found them in six. Coincidentally, one of the galaxies hosted a TDE: AT\,2019azh  \citep{Hinkle2021DiscoveryGalaxy}. Further, \cite{Wevers2024ExtendedEvents} recently analysed the MUSE cube of the host of iPTF16fnl and identified prominent [\ion{O}{III}] EELRs. The study notes that the three galaxies (of ASASSN-14li, AT\,2019azh and iPTF16fnl) are all PSB galaxies and exhibit low-velocity ($\lesssim50$\,km/s) EELRs, leading to a conclusion that the post-merger host environment is crucial to understand the TDE rate. In absence of a powerful enough AGN to ionise the EELRs, they further suggest that either the EELRs are powered by frequent TDEs or that the fading AGN may be associated with increasing TDE rate or increasing rate of detecting TDEs.

One of the key TDE observables to understand, is the distribution of stars that are disrupted. The stellar masses are typically derived using light curve fitters such as Modular Open Source Fitter for Transients  \citep[\texttt{MOSFiT};][]{Guillochon2018} showing a disribution from $0.1M_\odot$ to few solar masses \citep[see e.g. ][]{Nicholl2022}. Furthermore, \citet{Mockler2022EvidenceHoles} used from Hubble Space Telescope (HST) UV spectra of three TDEs to suggest that the high nitrogen-to-oxygen abundance ratio \citep[$N/C\geq10$; see also][]{Yang2017TheEvents} of the disrupted stars is compatible only with TDEs from intermediate mass stars (1\,--\,2$M_\odot$). As a consequence, they concluded that TDEs from such stars are over-enhanced by factor of $\gtrsim100$ over a simple initial mass function (IMF). However, given the central SMBHs of TDE host galaxies typically have low masses \citep[$\lesssim10^7 M_\odot$, e.g.][]{Ramsden2022TheMass, Hammerstein2023a}, disruption of low-mass stars is not prohibited by the Hills mass \citep{Hills1975}, and they should produce a large number of observable TDEs. In fact, \citet{Kochanek2016} predicted that the theoretically likeliest TDEs should be ones arising from a disruption of a $\sim0.3M_\odot$ star, and TDEs from low-mass stars should dominate the population in galaxies with low-mass SMBHs. To understand the distribution of disrupted stars, a detailed spectrosopcic analysis to investigate the stellar population and the stellar mass distribution in the near vicinity of the SMBH is needed.

IFU studies of supernovae (SNe) and other extragalactic transients can offer complementary constraints on the progenitors and/or explosion mechanism of individual objects of interest through characterisation of their local environment within a census of their wider host galaxy properties \citep[e.g.][]{Lyman2020, Singleton2025AnMUSE}. Additional power in probing progenitor channels (proxied by the stellar populations of their host galaxies) is achieved by population studies \citep[e.g.][]{Galbany2018PISCO:Compilation, Pessi2023AProperties}, which can unveil new insights not possible without fully spatially-resolved spectroscopic information \citep[e.g.][]{Pessi2023ASupernovae}. For TDEs, IFU observations allow to investigate the presence of the EELRs efficiently, but the data can also be used to extract volume-limited spectra of the nuclear regions needed to understand the local stellar populations. One IFU sample study of TDE host galaxies has so far been performed. \citet{Hammerstein2023a}  analysed IFU data of 13 TDE host galaxies obtained with the Keck Cosmic Web Imager \citep[KCWI;][]{Morrissey2018}, and used the high-resolution spectra to estimate the black hole masses of the hosts owning to the high spectral resolution of the instrument. Unfortunately, due to the narrow wavelength range and the small Field-of-View (FoV) KCWI, the study could not investigate the presence of EELRs or touch on the stellar populations present in the nuclear regions.

In this manuscript we present an analysis of a sample of 20 TDE host galaxies observed with MUSE, with a focus on two particular topics. The wide wavelength range 4650\,--\,9300\,Å combined with the suberb spatial (0.2"/px) and wavelength resolution ($R\sim3000$)\footnote{\url{https://www.eso.org/sci/facilities/paranal/instruments/muse/inst.html}}, allows a detailed investigation of the EELRs characteristics, and enables determining the stellar populations in the near-vicinity of the SMBHs. The paper is structured follows. In Section \ref{sec:sample} we present the host sample and discuss the sample demographics, in \ref{sec:EELRs} we investigate the presence and properties of the EELRs in the sample, and in \ref{sec:SMDs} we determine the stellar mass distribution in the near vicinity of the SMBHs and discuss the implications for the TDEs. Finally, in Section \ref{sec:conclusions}, we conclude our findings.

\section{Sample and Observations}
\label{sec:sample}

\begin{table*}
    \def\arraystretch{1.1}%
    \setlength\tabcolsep{5pt}
    \centering
    \fontsize{9}{11}\selectfont
    \caption{TDE host galaxies observed with MUSE. Approximate date of the optical peak of the TDE, the date of MUSE observations, coordinates (J2000) spatial scale of the MUSE observation in kiloparsecs per arcsecond (kpc/$\arcsec$) and per MUSE spaxel (kpc/px) at the redshift of each host galaxy are provided. The host galaxy types are Post-Starburst (PSB), Quiescent Balmer Strong (QBS), Quiescent (Q) and Star-forming (SF) as presented in \citet{French2016}, with the addition of Shocked PSB \citep[SPOG;][]{Alatalo2016}. The TDE spectral types are He, H+He and He following the classification scheme of \citet{VanVelzen2021SeventeenStudies}.}
    \begin{threeparttable}
    \begin{tabular}{l c c c c c c c l l}
    \hline
    \hline
        \multicolumn{1}{c}{TDE} & Peak Date & Obs. Date & RA & DEC & $z$ & kpc/$\arcsec$ & kpc/px & \multicolumn{1}{c}{Host Type} & \multicolumn{1}{c}{Spec. Type}   \\
    \hline
D23H-1        & 2007-09-29 & 2019-07-28 & 23:31:59.54  & $+$00:17:14.58 & $ 0.1855$ & $  3.11$ & $  0.62$ & SF\tnote{\bf 1 }  & --\tnote{\bf   }   \\
PTF09axc      & 2009-07-23 & 2019-05-06 & 14:53:13.07  & $+$22:14:32.2  & $ 0.1146$ & $  2.08$ & $  0.42$ & PSB\tnote{\bf 1 }  & H\tnote{\bf 7 }   \\
SDSSJ134244   & 2010-01-14 & 2019-05-06 & 13:42:44.42  & $+$05:30:56.14 & $ 0.0366$ & $  0.73$ & $  0.15$ & SF\tnote{\bf 1 }  & --\tnote{\bf   }   \\
LSQ12dyw\tnote{\bf a}      & 2012-08-04 & 2021-08-16 & 23:29:24.87  & $-$07:23:20.39 & $ 0.0900$ & $  1.68$ & $  0.34$ & SF\tnote{\bf   }  & H\tnote{\bf 8 }   \\
ASASSN-14ae   & 2014-02-01 & 2021-03-13 & 11:08:40.15  & $+$34:05:51.99 & $ 0.0436$ & $  0.86$ & $  0.17$ & QBS\tnote{\bf 1 }  & H\tnote{\bf 7 }   \\
ASASSN-14ko   & 2014-07-23 & 2015-12-02 & 05:25:18.12  & $-$46:00:20.52 & $ 0.0425$ & $  0.84$ & $  0.17$ & SF\tnote{\bf   }  & --\tnote{\bf   }   \\
ASASSN-14li   & 2014-12-04 & 2016-01-21 & 12:48:15.23  & $+$17:46:26.45 & $ 0.0206$ & $  0.42$ & $  0.08$ & PSB\tnote{\bf 1 }  & H+He\tnote{\bf 7 }   \\
iPTF15af\tnote{\bf a}      & 2015-01-15 & 2021-02-09 & 08:48:28.15  & $+$22:03:33.52 & $ 0.0790$ & $  1.49$ & $  0.30$ & QBS\tnote{\bf 1 }  & H+He\tnote{\bf 7 }   \\
ASASSN-15lh   & 2015-06-05 & 2016-08-11 & 22:02:15.42  & $-$61:39:34.99 & $ 0.2326$ & $  3.71$ & $  0.74$ & Q\tnote{\bf   }  & --   \\
ASASSN-15oi   & 2015-08-14 & 2019-05-06 & 20:39:09.14  & $-$30:45:20.60 & $ 0.0484$ & $  0.95$ & $  0.19$ & QBS\tnote{\bf 1 }  & He\tnote{\bf 7 }   \\
iPTF16fnl\tnote{\bf a}     & 2016-09-01 & 2021-08-16 & 00:29:57.04  & $+$32:53:37.19 & $ 0.0163$ & $  0.33$ & $  0.07$ & PSB\tnote{\bf 1 }  & H+He\tnote{\bf 7 }   \\
AT2017eqx\tnote{\bf a}     & 2017-06-17 & 2021-09-30 & 22:26:48.39  & $+$17:08:51.95 & $ 0.1089$ & $  1.99$ & $  0.40$ & PSB\tnote{\bf 2 }  & H+He\tnote{\bf 7 }   \\
AT2018dyb     & 2018-08-10 & 2019-04-12 & 16:10:58.87  & $-$60:55:24.3  & $ 0.0180$ & $  0.37$ & $  0.07$ & Q\tnote{\bf 3 }  & H+He\tnote{\bf 7 }   \\
AT2018fyk     & 2018-09-08 & 2019-06-10 & 22:50:16.09  & $-$44:51:53.5  & $ 0.0590$ & $  1.14$ & $  0.23$ & Q\tnote{\bf 4 }  & H+He\tnote{\bf 7 }   \\
AT2018hyz     & 2018-11-07 & 2021-02-05 & 10:06:50.87  & $+$01:41:34.05 & $ 0.0458$ & $  0.90$ & $  0.18$ & QBS/PSB\tnote{\bf 5 }  & H+He\tnote{\bf 10}   \\
AT2019ahk     & 2019-03-04 & 2019-09-02 & 07:00:11.39  & $-$66:02:24.7  & $ 0.0262$ & $  0.53$ & $  0.10$ & SPOG\tnote{\bf 6 }  & H\tnote{\bf 7 }   \\
AT2019dsg\tnote{\bf a}     & 2019-05-14 & 2021-08-05 & 20:57:02.96  & $+$14:12:16.44 & $ 0.0501$ & $  0.98$ & $  0.20$ & SF\tnote{\bf   }  & H+He\tnote{\bf 10}   \\
AT2019lwu\tnote{\bf a}     & 2019-07-30 & 2021-09-30 & 23:11:12.30  & $-$01:00:10.71 & $ 0.1170$ & $  2.12$ & $  0.42$ & QBS/PSB\tnote{\bf   }  & H\tnote{\bf 7 }   \\
AT2019qiz\tnote{\bf a}     & 2019-10-07 & 2021-02-08 & 04:46:37.87  & $-$10:13:34.71 & $ 0.0151$ & $  0.31$ & $  0.06$ & SF\tnote{\bf   }  & H+He\tnote{\bf 10}   \\
AT2023clx     & 2023-02-22 & 2021-03-05 & 11:40:09.44  & $+$15:19:39.03 & $ 0.0111$ & $  0.23$ & $  0.05$ & SF\tnote{\bf   }  & H+He\tnote{\bf 11}   \\
    \hline
    \hline
  \end{tabular}
    \begin{tablenotes}
    \item[\bf a] Obtained with adaptive optics.
    \item[] References: [1] \citet{French2020}; [2] \citet{Nicholl2019}; [3] \citet{Leloudas2019}; [4] \citet{Wevers2019}; [5] \citet{Short2020}; [6] \citet{Holoien2019a}; [7] \citet{VanVelzen2020}; [8] \citet{Charalampopoulos2022}; [9] \citet{Leloudas2017}; [10] \citet{Hammerstein2023}; [11] \citet{Zhu2023a}.
    \end{tablenotes}
    \end{threeparttable}    
\label{tab:tde_list}
\end{table*}

\begin{figure}
    \centering
    \includegraphics[width=0.49\textwidth]{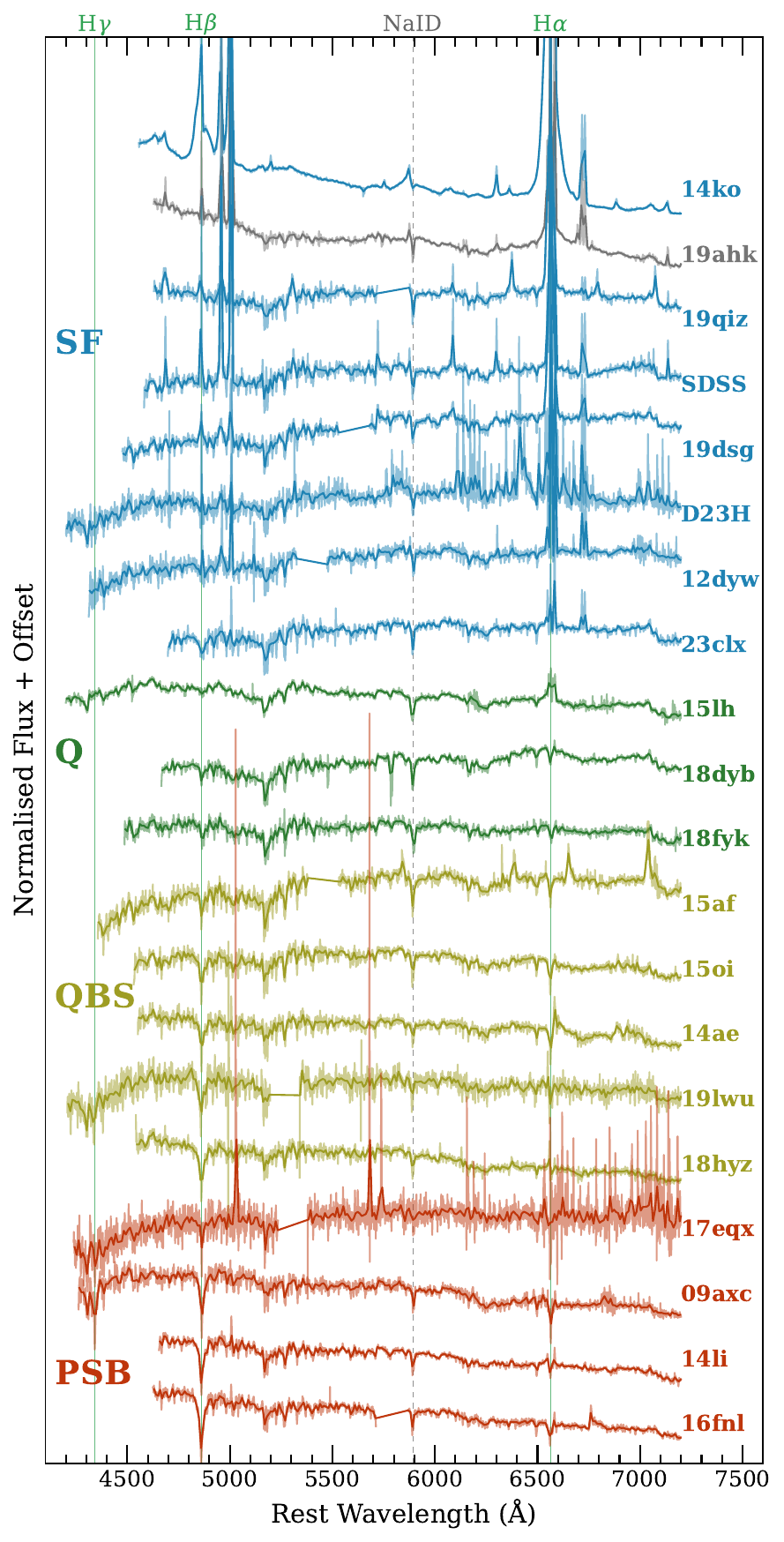}

    \caption{Nuclear spectra of the 20 TDE host galaxies in our sample (lighter shade) under spectra binned to 10\,Å (darker shade). The spectra were extracted with an aperture equal to the FWHM measured for stars in the MUSE cubes (see Table \ref{tab:fwhms}). The galaxy types are Star-Forming (SF, first from the top), Quiescent (Q, second), Quiescent, Balmer Strong (QBS, third) and Post-Starburst (PSB, fourth) following \citet{French2020}. Note that AT\,2019ahk is classfied as a shocked PSB galaxy (SPOG) despite prominent emission lines.}
    \label{fig:nuclear_spec}
\end{figure}

In this paper we analyse a sample of 20 TDE host galaxies observed with the MUSE IFU. This includes all TDE hosts that have been targeted by MUSE and are publicly available. Nine of the cubes were obtained with a dedicated programme (PI: Kim), a further nine by the all-weather MUSE supernova integral field nearby (AMUSING; \citealt{Galbany2016a}; \citealt{Lopez-Coba2020}; Galbany et al. in prep.) survey, and two by smaller programmes. Seven host galaxies in our dedicated programme were observed with the adaptive optics mode, improving both the spatial resolution and the S/N of the observations. The cubes of ASASSN-14ko \citep{Tucker2021}, ASASSN-15lh \citep{Kruhler2018}, ASASSN-14li \citep{Prieto2016}, iPTF16fnl \citep{Wevers2024ExtendedEvents}, AT\,2018fyk \citep{Wevers2022} and AT\,2019qiz \citep{Xiong2025ExtendedQPE} have been previously analysed in the literature. We used the default reductions provided by European Southern Observatory (ESO) for all data cubes. The data have been corrected for Galactic extinction using \citet{Schlafly2011} reddening maps assuming $R_V=3.1$. Throughout the paper, we assume a flat $\Lambda$CDM cosmology with $\Omega_\mathrm{M}=0.3$ and H$_\mathrm{0}$ = 70 km s$^{-1}$ Mpc$^{-1}$. The sample is summarised in Table \ref{tab:tde_list}, and colour images generated based on the MUSE cubes are shown in Figure \ref{fig:VRI_crop}.

We note that the sample includes two events that are outliers in the TDE populations in terms of their optical properties. First, ASASSN-14ko exhibits repeated flares from the nucleus of a merging galaxy system \citep[e.g.][]{Payne2021}. While atypical, the event can be understood as a repeating partial TDE, where the star would be partially stripped at roughly constant intervals as it passes through an AGN disk \citep[e.g.][]{Payne2021, Linial2023PeriodNuclei}, and thus we have included the event in our sample. On the other hand, ASASSN-15lh is one of the most remarkable optical transients in recent years, reaching a peak luminosity of $M_V=-23.5$. The event has been discussed both as a superluminous supernova \citep[SLSN;][]{Dong2016} and a TDE \citep{Leloudas2017}. However, the TDE classification is supported based on its host galaxy properties. ASASSN-15lh is spatially coincident with the nucleus of a galaxy hosting a massive SMBH \citep[$M_\mathrm{BH}\sim5\times10^8M_\odot$;][]{Kruhler2018}, and further the galaxy does not host notable young stellar populations required for a stripped massive star progenitor of a SLSN \citep[e.g.][]{Kruhler2018, Lunnan2014, Leloudas2015, Schulze2018}. As such, we consider ASASSN-15lh a TDE and include it in our sample. 

\subsection{Sample Demographics}
\label{subsec:sample_demo}
Given the MUSE data were obtained by a number of different ESO programmes, some targeting individual host galaxies and some samples of hosts, it is important to verify how representative the sample is. To investigate this, we compare the demographics of the sample with \citet[][see also \citealt{Auchettl2017NewWavelengths}]{French2020}. The work included all 41 identified TDEs with pre-TDE host spectroscopy available at the time, and offers the best comparison sample to date, in the absence of a large representative TDE host study. The hosts are labelled Post-Starburst (PSB), Quiescent Balmer Strong (QBS), Quiescent, and \say{Star-forming} (SF), based on the Balmer emission/absorption line strengths in their nuclear spectra. Galaxies that show strong H$\delta$ absorption are classified as QBS (Lick $\mathrm{H}\delta_A > 1.31$\,Å) or PSB ($\mathrm{H}\delta_A -\sigma(\mathrm{H}\delta_A) > 4$\,Å). Those that show strong H$\alpha$ emission ($\mathrm{EW}<-3$\,Å\footnote{Using traditional definition where negative $EW$ refers to emission.}) are labelled as SF galaxies, but note that as the classification relates solely to H$\alpha$ emission in a nuclear spectrum, it does not necessarily imply actual star-formation as emission could arise from another physical process \citep[such as AGN, see e.g. ][]{Sanchez2020SpatiallyGalaxies}. For instance, the host spectrum of AT\,2023clx is similar to the Quiescent galaxies (see Fig \ref{fig:nuclear_spec}), but it exhibits H$\alpha$ emission due to an low-ionisation nuclear emission-line region (LINER) listed in NASA/IPAC Extragalactic Database (NED\footnote{\url{ned.ipac.caltech.edu}}) and in SIMBAD\footnote{\url{simbad.u-strasbg.fr}}. The host also shows strong star formation in the spiral arms (see Figure \ref{fig:VRI_crop}), and we adopt the SF label. Finally, galaxies that show both weak Balmer absorption and emission are labelled Quiescent. 

\begin{table}
    \def\arraystretch{1.2}%
    \setlength\tabcolsep{6pt}
    \centering
    \fontsize{9}{11}\selectfont
    \caption{The comparison of the MUSE sample and the \citet{French2020} TDE sample in terms of their host galaxy types. Total sample size ($n$) and the numbers in each galaxy types are shown. The values in parentheses refer to the percentage in the subtype.}
    \begin{tabular}{l c c c c c}
    \hline
    \hline
\bf Sample & $n$          & Q          & QBS/PSB    & SF         & SPOG       \\
\hline
\bf French &   41 &   13 (32\%) &   13 (32\%) &   15 (37\%) &    0 (0\%)  \\
\bf MUSE   &   20 &    3 (15\%) &    9 (22\%) &    7 (35\%) &    1 (5\%)  \\

    \hline
    \hline
  \end{tabular}
\label{tab:sample_comp}
\end{table}

The galaxy types demographics of the samples are listed in Table~\ref{tab:sample_comp}. In \citet{French2020}, $\sim32$\% of galaxies were QBS or PSB, $\sim32$\% were Quiescent and $\sim37$\% SF galaxies. In our MUSE sample there are eight shared objects with \citet{French2020} and for these we adopt their classifications. In addition, further five hosts have been classified in single-object publications in the literature. For the rest, we perform a qualitative classification with visual comparison of the spectra presented in Figure \ref{fig:nuclear_spec}. As the discriminant H$\delta$ line lies beyond the wavelength range of MUSE for the whole sample, distinguishing between PSB, QBS and Q galaxies is not directly possible. We use the H$\alpha$ emission to identify the SF galaxies and H$\beta$ absorption to loosely classify the PSB, QBS and Q galaxies. Qualitative assessment has inherent uncertainties, but it is sufficient for our needs to compare sample demographics in broad terms. We find that in our sample 35\% (7/20) are SF, 45\% (9/20) are QBS/PSB and 15\% (3/20) are Quiescent as shown in Table \ref{tab:tde_list}. In addition, the host of AT\,2019ahk was classified as a \say{Shocked Post-Starburst galaxy} (SPOG) by \citet[][]{Holoien2019a}, and we retain their classification. SPOGs exhibit both strong Balmer absorption and emission with line ratios inconsistent with star formation, and instead the emission lines are likely ionised via shocks. In general, such galaxies appear to have star formation histories (SFHs) similar to PSB galaxies, but their stellar populations appear to be younger on average \citep{French2018a}, and they may represent a phase in galaxy transformation before traditionally PSB galaxies \citep{Alatalo2016}. 

In comparison to \citet{French2020}, the Quiescent host galaxies appear to be under-represented in our sample, while QBS/PSB are over-represented. To investigate this, we tested if the two samples are statistically different using the Fisher's exact test, by comparing each of the three classes (QBS/PSB, Q and SF) against the rest of the galaxies. We find that in all three cases $p>0.22$, implying that the samples are not statistically significantly different from each other. However, we note that in the absence of a large, representative TDE host galaxy sample, the exact demographics are uncertain. For instance, \citet{Hammerstein2021TDESystem} find that only 2/19 of their TDE host sample exhibit significant H$\delta$ absorption indicative of a QBS/PSB nature, and it is possible that the MUSE sample might be biased towards the QBS/PSB hosts. 

\begin{figure}
    \centering
    \includegraphics[width=0.44\textwidth]{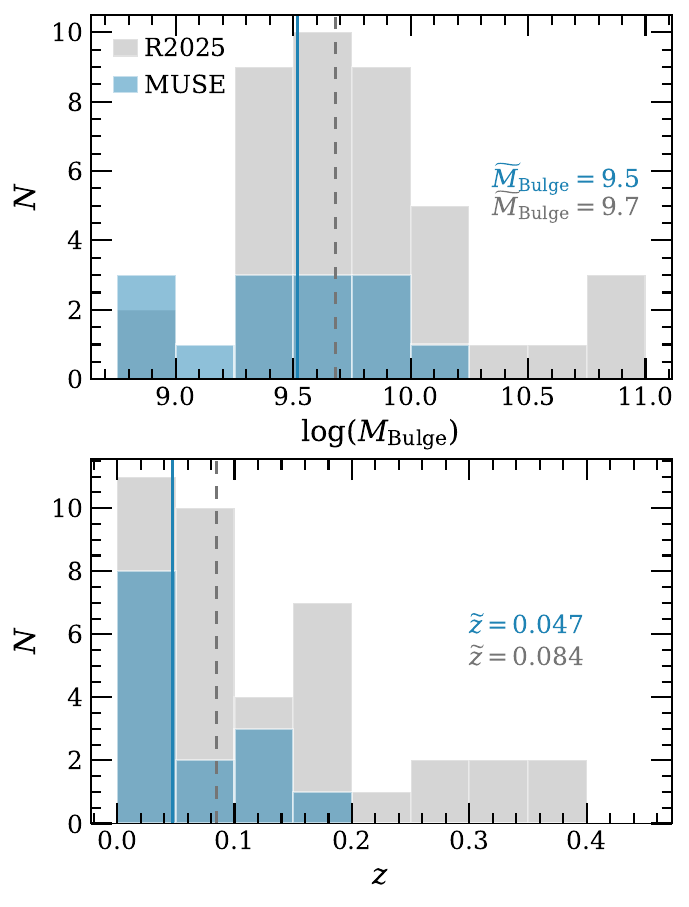}
    \caption{Bulge mass and redshift comparsion between the MUSE sample and sample of \citet{Ramsden2025} (R2025). The MUSE sample concentrates on lower bulge masses and lower redshifts than R2025. The samplesa are nominally consistent, as the MUSE sample is less biased towards more distant brighter galaxies possibly explaining the minor discrepancy. The median values ($\widetilde{M}_\mathrm{Bulge}$, $\widetilde{z}$) of are shown with a solid lines for the MUSE sample and dashed lines for R2025.}
    \label{fig:bulge_mass_hist}
\end{figure}

We also examined whether the central bulge masses of the MUSE sample are representative of the 40 hosts analysed in \citet{Ramsden2025} \citep[see also][]{Ramsden2022TheMass}. As only eight galaxies are common between the samples, we estimated the bulge masses for the MUSE sample following the methodology. Values were derived for 14 systems, as for six hosts (LSQ12dyw, ASASSN-14ko, ASASSN15lh, AT2018dyb, AT2018fyk, AT2019ahk) sufficient host photometry needed by the modelling was not found. As shown in Figure \ref{fig:bulge_mass_hist}, the bulge masses of the MUSE sample are much smaller. The two samples are nominally consistent according to the Kolmogorov-Smirnov test ($p = 0.35$), especially as  the MUSE sample is concentrated at much lower redshift in comparison to \citet{Ramsden2025} (median $z=0.047$ vs. $z=0.084$). As such. the sample is less biased towards brighter and more distant galaxies, possibly explaining the minor discrapancy.

Finally, we inspected the sample demographics in terms of TDE spectral types. The TDEs can be divided into those that show hydrogen (H), hydrogen and helium (H+He) and only helium (He) following \citet{VanVelzen2021SeventeenStudies}. Furthermore, \citet{Hammerstein2023} presented a fourth class of TDEs that exhibit featureless spectra based on a uniformly selected sample of 30 TDEs from ZTF Phase I operations. Their sample consisted of $13.3\%$ featureless, $10.0\%$ He, $56.7\%$ H+He and $20.0\%$ H TDEs. In our sample, a total of 16 TDEs have spectroscopic classifications in the literature \citep{VanVelzen2020, Hammerstein2023, Charalampopoulos2022, Zhu2023a, Hoogendam2024, Charalampopoulos2024}. Note that we have left ASASSN-15lh unclassified due to its unique spectral evolution \citep[see e.g.][]{Leloudas2017}, although parts of its evolution mostly resemble the featureless class of \citet{Hammerstein2023}. The subset consists of one He ($6.3\%$), ten H+He ($62.5\%$), five H ($31.3\%$) TDEs and no featureless TDEs. In comparison to \citet{Hammerstein2023}, the MUSE sample slightly under-represents the two rarest subtypes, but the sample has roughly the same fractions of H+He and H TDEs. Using Fisher's exact test we verified that the samples can be drawn from the same distribution and $p$-values of all spectral types are $>0.28$, indicating that the samples are not statistically significantly different from each other. As such, we consider the sample representative of TDEs in terms of their spectral types.

\begin{table}
    \def\arraystretch{1.2}%
    \setlength\tabcolsep{4pt}
    \centering
    \fontsize{9}{11}\selectfont
    \caption{Details of the identified EELR bins. $n$ states the number of bins identified per galaxy, $R$ the range of projected offset of the bins from the nucleus of the galaxy and $L$ the range of the [\ion{O}{III}] $\lambda5007$ emission line luminosities. We also show the minimum required luminosity to ionise the EELRs ($L_\mathrm{ion}$), and the estimated integrated nuclear luminosity ($L_\mathrm{nucl}$).}
    \begin{threeparttable}
    \begin{tabular}{l c c c c r}
    \hline
    \hline
TDE             &  $n$    &  $R$  &  $L$ ($10^{38}$)  &  Log($L_\mathrm{ion}$) &  Log($L_\mathrm{nucl}$)  \\
             &      & (kpc) &  (erg/s)  &  (erg/s) &  (erg/s)  \\

    \hline
14ae     & $     7$ & $   0.6$\,--\,$   2.4$ & $ 0.98$\,--\,$ 9.90$  & $  >41.8$  & $  <41.3$ \\
14li     & $    32$ & $   0.6$\,--\,$   6.9$ & $ 0.11$\,--\,$ 6.63$  & $  >42.4$  & $  41.3$ \\
14ko     & $    66$ & $   2.4$\,--\,$  22.9$ & $ 0.53$\,--\,$22.10$  & $  >44.2$  & $  43.4$ \\
16fnl       & $    53$ & $   0.7$\,--\,$   7.5$ & $ 0.09$\,--\,$ 4.60$  & $  >42.8$  & $  40.8$ \\
19ahk       & $    65$ & $   1.9$\,--\,$   6.2$ & $ 0.17$\,--\,$ 9.27$  & $  >43.1$  & $  42.9$ \\

    \hline
    \hline
  \end{tabular}
    \end{threeparttable}    
\label{tab:EELR_details}
\end{table}

\section{Extended Emission Line Regions}

We investigated the presence of extended emission line regions (EELRs) that do not follow the host galaxy morphology by using the [\ion{O}{III}] $\lambda5007$ emission line. While other emission lines, such as [\ion{N}{II}] $\lambda6583$, have also been used to identify EELRs \citep[e.g.][]{Prieto2016}, [\ion{O}{III}] was chosen as it has been the strongest EELR emission line identified in TDE host galaxies \citep[i.e.][]{Wevers2024ExtendedEvents}. After visual inspection of the MUSE cubes we identified EELRs in five of the 20 TDE host galaxies. Three of them have been analysed in the literature (ASASSN-14li: \citealt{Prieto2016}; ASASSN-14ko: \citealt{Tucker2021}; iPTF16fnl: \citealt{Wevers2024ExtendedEvents}), but we report two additional events, ASASSN-14ae and AT2019ahk, where we see evidence of extended emission. For these five MUSE cubes, we ran the \texttt{\ion{H}{II} explorer} \citep{Sanchez2012} routine, as implemented in the \texttt{IFUanal} package \citep{Lyman2018}, on [\ion{O}{III}] $\lambda5007$ emission line maps to spatially bin the EELRs to quantify the extent and amount of the emission. Line maps were generated by subtracting a linear continuum from an image extracted with $10$\,Å width centered at $\lambda5007$. While the approach does not utilise a full stellar population modelling as done by \texttt{Pipe3D} \citep[e.g.][]{Sanchez2016Pipe3DFIT3D}, the data-driven continuum subtraction should provide comparable results. False-colour images highlighting [\ion{O}{III}] and H$\alpha$ emission are shown alongside grey-scale images of the identified bins in Figures \ref{fig:EELR_figs1} and \ref{fig:EELR_figs2}. The extracted spectra corresponding to the bins are shown in Figures \ref{fig:emission_line_spectra_ASASSN-14ae_14li_14ko} and \ref{fig:emission_line_spectra_iPTF16fnl_AT2019ahk}.

\begin{table}
    \def\arraystretch{1.1}%
    \setlength\tabcolsep{4pt}
    \centering
    \fontsize{9}{11}\selectfont
    \caption{FWHMs and the aperture sizes used to estimate the [\ion{O}{III}] detection limits. If FWHM was larger than 0.75\,kpc, the aperture was set to be FWHM. $d$ denotes the used diameter and $n$ the number of spaxels in the aperture.}
    \begin{threeparttable}
    \begin{tabular}{l c c c c c}
    \hline
    \hline
        \multicolumn{1}{c}{TDE} & FWHM        & FWHM  & $d$         & $d$ & $n$  \\
                                & ($\arcsec$) & (kpc) & ($\arcsec$) & (kpc)  \\
    \hline
D23H-1        & $  1.13$ & $  3.51$ & $  1.13$ & $  3.51$ & $20    $   \\
PTF09axc      & $  0.82$ & $  1.70$ & $  0.82$ & $  1.70$ & $8     $   \\
SDSSJ134244   & $  0.82$ & $  0.60$ & $  1.03$ & $  0.75$ & $15    $   \\
LSQ12dyw      & $  0.95$ & $  1.60$ & $  0.95$ & $  1.60$ & $15    $   \\
ASASSN-14ae   & $  1.20$ & $  1.03$ & $  1.20$ & $  1.03$ & $24    $   \\
ASASSN-14ko   & $  0.71$ & $  0.59$ & $  0.89$ & $  0.75$ & $9     $   \\
ASASSN-14li   & $  0.99$ & $  0.41$ & $  1.80$ & $  0.75$ & $59    $   \\
iPTF15af      & $  0.81$ & $  1.21$ & $  0.81$ & $  1.21$ & $7     $   \\
ASASSN-15lh   & $  1.01$ & $  3.74$ & $  1.01$ & $  3.74$ & $15    $   \\
ASASSN-15oi   & $  0.75$ & $  0.71$ & $  0.79$ & $  0.75$ & $8     $   \\
iPTF16fnl     & $  0.97$ & $  0.32$ & $  2.26$ & $  0.75$ & $87    $   \\
AT2017eqx     & $  1.62$ & $  3.22$ & $  1.62$ & $  3.22$ & $44    $   \\
AT2018dyb     & $  1.83$ & $  0.67$ & $  2.05$ & $  0.75$ & $72    $   \\
AT2018fyk     & $  1.34$ & $  1.53$ & $  1.34$ & $  1.53$ & $31    $   \\
AT2018hyz     & $  0.74$ & $  0.67$ & $  0.83$ & $  0.75$ & $7     $   \\
AT2019ahk     & $  1.61$ & $  0.85$ & $  1.61$ & $  0.85$ & $44    $   \\
AT2019dsg     & $  0.98$ & $  0.96$ & $  0.98$ & $  0.96$ & $15    $   \\
AT2019lwu     & $  0.77$ & $  1.63$ & $  0.77$ & $  1.63$ & $8     $   \\
AT2019qiz     & $  0.71$ & $  0.22$ & $  2.44$ & $  0.75$ & $108   $   \\
AT2023clx     & $  1.15$ & $  0.26$ & $  3.29$ & $  0.75$ & $196   $   \\
    \hline
    \hline
  \end{tabular}
    \end{threeparttable}    
\label{tab:fwhms}
\end{table}

To characterise the line luminosity, the spectra of the bins were fit for the [\ion{O}{III}] emission line in each bin with a simple Gaussian model with $\mathrm{FWHM}=170$\,km/s as measured from the MUSE line-spread function shown in \citet{Guerou2017}. The flux limit of the line emission is set very low in the binning routine to thoroughly investigate regions with low level emission. The choice resulted in a number of bins where no real emission line was detected, which were excluded by setting a strict detection limit of $5\sigma$, measured as the ratio of the Gaussian peak luminosity and the standard deviation of the residual in the fitting interval. Note that as we were solely interested in extended regions of emission, we excluded any bins identified within the half-light radius of the host of ASASSN-14ko and AT\,2019ahk where significant nuclear line emission is present. The details of the identified bins are listed in Table \ref{tab:EELR_details} with the number of bins per galaxy and range of their projected offsets and [\ion{O}{III}] emission luminosity. 

The EELRs are exclusively identified in systems that are local ($z<0.045$; $D_\mathrm{L}<200$\,Mpc), implying a possible detection bias against distant hosts. To investigate this, we ran the same line-fitting routine on spectra extracted at a visually empty region of the 20 MUSE cubes to estimate the upper limit at which the line emission could have gone unnoticed. Given the redshift range of the sample is very wide ($0.0111$ -- $0.2326$), the MUSE spaxels probe very different physical scales ($0.05$ -- $0.74$\,kpc/px), even before considering the effect of atmoshperic smearing. As emission lines are not present in these spectra, we have adopted a strategy to estimate the limit of emission arising from a circular aperture with $d=0.75$\,kpc, chosen as it is comparable to the size of one spaxel at the highest redshift of the sample. To account for atmospheric smearing, we estimated the FWHM of the MUSE cubes using stars present in the FoV. If the FWHM was $<0.75$\,kpc, we used an aperture of the size of the physical scale, but if the FWHM was larger we assumed an aperture equal to the FWHM of the cube. As shown in Table \ref{tab:fwhms}, in approximately half of the MUSE cubes the physical size was adopted.

\label{sec:EELRs}
 
\begin{figure}
    \centering
    \includegraphics[width=0.49\textwidth]{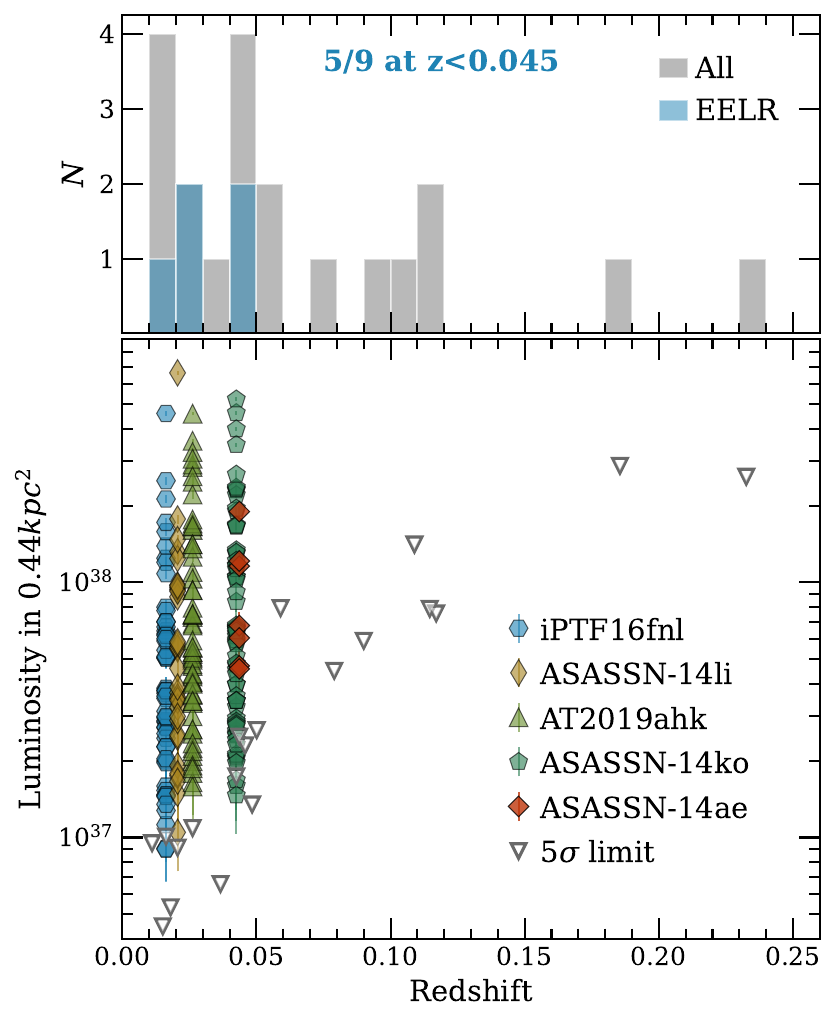}

    \caption{The [\ion{O}{III}] $\lambda5007$ luminosity of the EELR bins identified in the five TDE host galaxies (coloured markers), and the $5\sigma$ detection limits of each MUSE cube against redshift (open triangles). The brightest EELRs bins should have been seen in all the host galaxies regardless of redshift, but there may be bias against galaxies with lower luminosity EELRs.}
    \label{fig:L_OIII}
\end{figure}

Together with the measured EELR luminosities, we can investigate if the emission seen in these five galaxies would have been seen in the other 15 hosts as shown in Figure \ref{fig:L_OIII}. Limits are shown as explained above, but for the detected EELRs we have rescaled the luminosities to what would be seen of each individual bin at the aperture size used for the detection limits ($d=0.75$\,kpc). This is done as a multiplication of average luminosity per spaxel in each bin by the number of spaxels the aperture size would encompass at the redshift of the host. In case the number of spaxels is smaller than covered by the aperture, the average luminosity was multiplied by the actual number of pixels (i.e. luminosity of a bin is not higher than actually measured). The figure shows that the detection limits of the [\ion{O}{III}] $\lambda5007$ emission for all hosts are lower than the brightest identified EELRs. In fact, the brightest EELR bins ($\sim4\times10^{38}$\,erg/s) of 4/5 galaxies should have been seen even for the host of ASASSN-15lh at $z=0.23$, if they were present. However, most of the bins have a significantly lower luminosity (down to $\gtrsim10^{37}$\,erg/s) and for instance the detections in host of ASASSN-14ae are actually comparable to detection limits of any host above $z>0.1$. This suggests that low-luminosity EELRs exist in the five hosts and such EELRs can be confidently ruled out only in galaxies below $z\sim0.05$. Furthermore, it is possible that the physical scales the MUSE spaxels probe at different redshifts also play a significant role. The scale shown in Figure \ref{fig:L_OIII} ($d=0.75$\,kpc; $A=0.44$\,kpc$^2$) is covered by 101.4\,spx at the redshift of iPTF16fnl ($z=0.0163$) but by 14.9 at the redshift of ASASSN-14ae ($z=0.0436$), so the pixel size is reduced by factor of seven. At low redshifts it is possible to resolve small-scale EELRs, but as the redshift increases the regions become unresolvable and might therefore be virtually impossible to identify, even if the luminosity was nominally above the non-detection limit for the cube. 

\begin{figure*}
    \centering
     \begin{subfigure}[b]{0.33\textwidth}
         \centering
        \includegraphics[width=\textwidth]{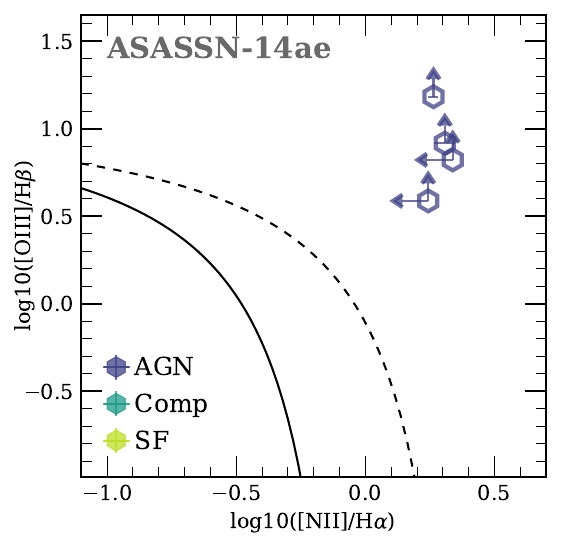}
    \end{subfigure} %
    \begin{subfigure}[b]{0.33\textwidth}
         \centering
        \includegraphics[width=\textwidth]{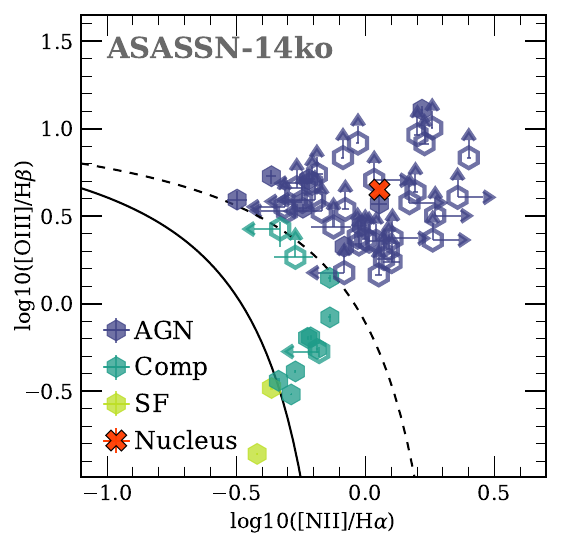}
    \end{subfigure} %
    \begin{subfigure}[b]{0.33\textwidth}
         \centering
        \includegraphics[width=\textwidth]{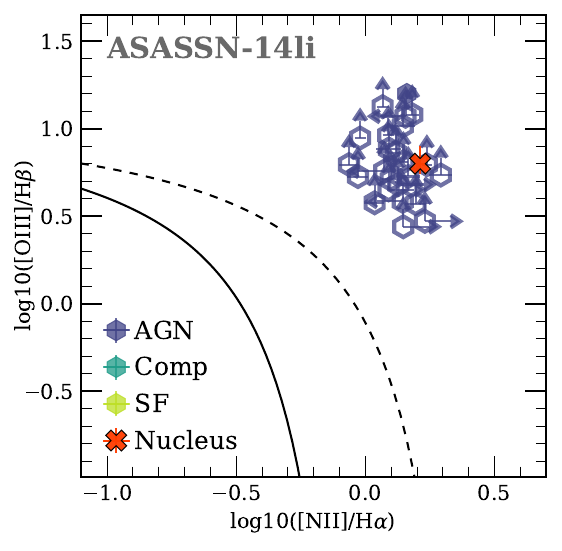}
    \end{subfigure} %
    \begin{subfigure}[b]{0.33\textwidth}
         \centering
        \includegraphics[width=\textwidth]{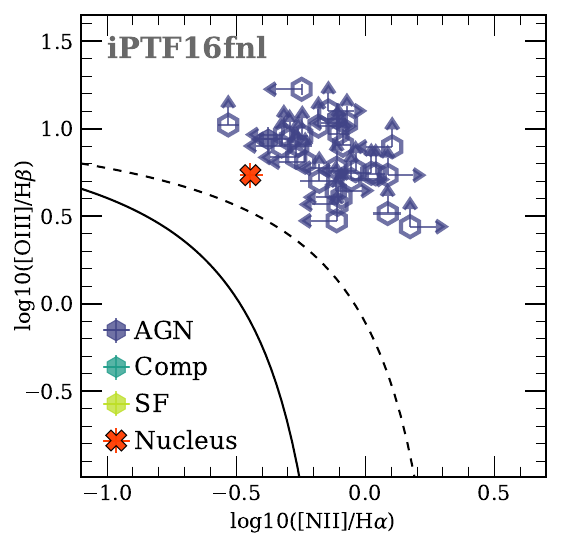}
    \end{subfigure} %
    \begin{subfigure}[b]{0.33\textwidth}
         \centering
        \includegraphics[width=\textwidth]{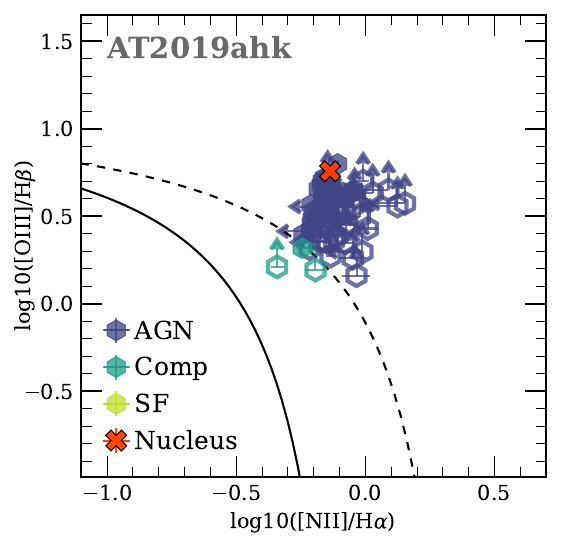}
    \end{subfigure} %
    \caption{The BPT diagrams for the EELRs and the nuclei of the five TDE host galaxies with detected EELRs. The values are derived from spectra after subtracting the best-fitting \texttt{Starlight} model (see Section \ref{sec:SMDs} for details).  Below the solid line, the dominant powering mechanism is assumed to be star-formation \citep{Kauffmann2003}, while above the dashed line the harder continuum emission (i.e. AGN) is required \citep{Kewley2001}. Between the two lines, in the composite region, both mechanisms are viable. Bins where both ratios were estimated are marked with solid markers, while limiting values are shown with open markers. In the nuclear spectrum of ASASSN-14ae H$\alpha$ and H$\beta$ were not detected and [\ion{N}{II}] falls on a strong telluric so the value is not shown on the diagram. Note that the few bins where [\ion{N}{II}]/H$\alpha$ could not be estimated are not shown. Except a few bins in ASASSN-14ko, all EELRs require hard AGN continuum. }
    \label{fig:EELR_BPTs}
\end{figure*}

\begin{figure*}
    \centering
     \begin{subfigure}[b]{0.33\textwidth}
         \centering
        \includegraphics[width=\textwidth]{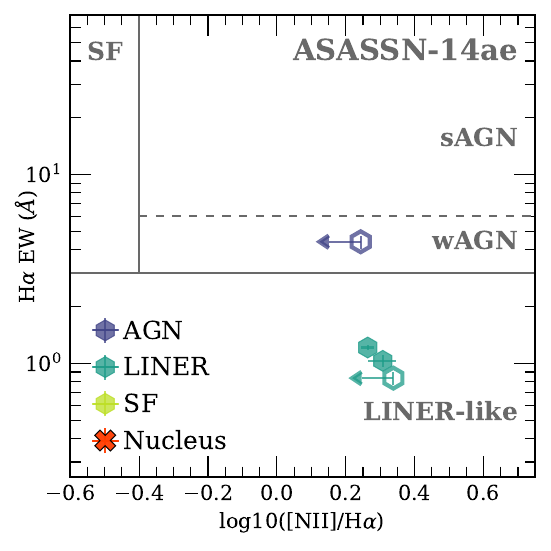}
    \end{subfigure} %
    \begin{subfigure}[b]{0.33\textwidth}
         \centering
        \includegraphics[width=\textwidth]{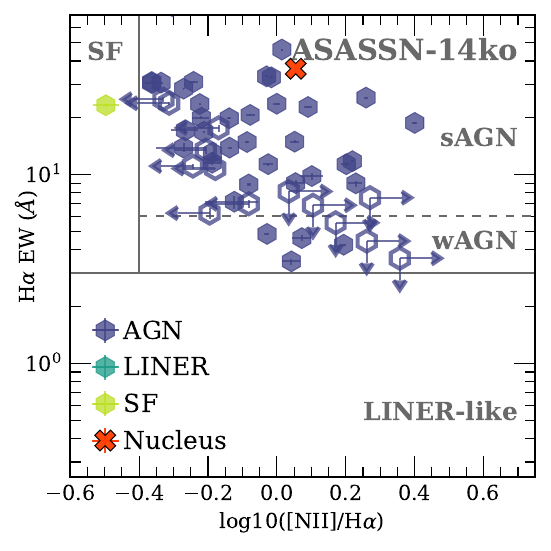}
    \end{subfigure} %
    \begin{subfigure}[b]{0.33\textwidth}
         \centering
        \includegraphics[width=\textwidth]{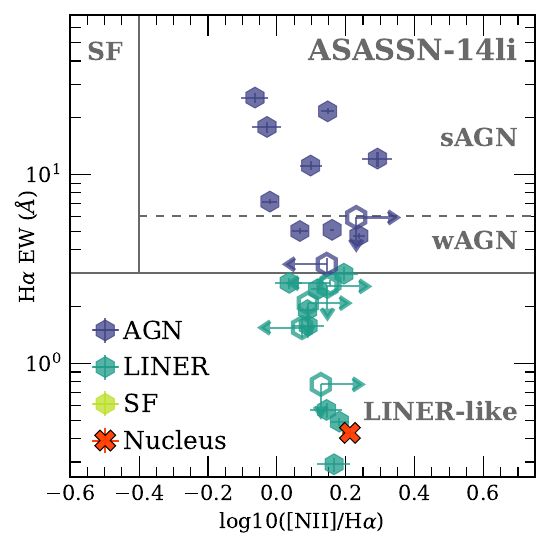}
    \end{subfigure} %
    \begin{subfigure}[b]{0.33\textwidth}
         \centering
        \includegraphics[width=\textwidth]{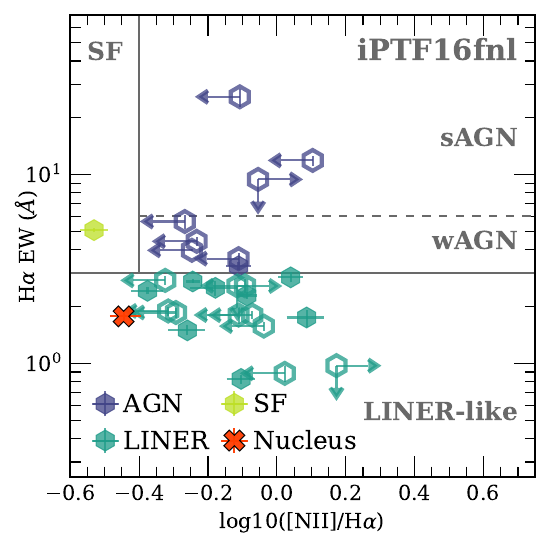}
    \end{subfigure} %
    \begin{subfigure}[b]{0.33\textwidth}
         \centering
        \includegraphics[width=\textwidth]{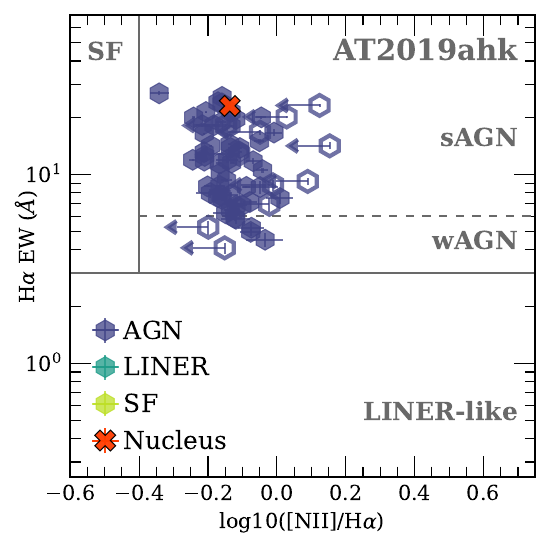}
    \end{subfigure} %
    \caption{The WHAN diagrams \citep{CidFernandes2011AAGN} of the five TDE hosts that exhibit EELRs. The required ionising source is divided into star-formation (SF), strong or weak AGN (sAGN, wAGN), or LINER-like based on their H$\alpha$ EW and [\ion{N}{II}] to H$\alpha$ ratio. All hosts exhibit some EELRs that require AGN-like ionising continuum. The nuclear spectra (red cross) show that ASASSN-14ko and AT\,2019ahk are consistent with sAGN, possibly due to the presence of TDE emission in the MUSE cube, while iPTF16fnl and ASASSN-14li are LINER-like. The $3\sigma$ upper limit of H$\alpha$ $\mathrm{EW}$ for ASASSN-14ae ($3.7$) effectively places it in LINER-like region, implying that no significant AGN activity is present, but no marker is shown as a strong telluric is located over the [\ion{N}{II}] emission.}
    \label{fig:EELR_WHANs}
\end{figure*}

\subsection{EELR demographics}
The galaxies that host EELRs are similar in the fact that they have had significant starbursts in the recent past. The hosts of ASASSN-14li \citep{Prieto2016} and iPTF16fnl \citep{Wevers2024ExtendedEvents} are PSB galaxies and ASASSN-14ae is a QBS galaxy \citep{French2016}, while the host of AT\,2019ahk is consistent with SPOGs \citep{Holoien2019a}. A past starburst has not been verified only for the host of ASASSN-14ko, but the system is a known galaxy merger and it is reasonable to assume that the merger has caused a starburst. Assuming that the QBS/PSB/SPOG galaxies in this MUSE sample are a result of galaxy mergers \citep[e.g.][]{Zabludoff1996TheGalaxies}, 5/11 of the post-merger hosts (including ASASSN-14ko) exhibit EELRs. More significantly, EELRs are identified in all such galaxies at $z<0.045$, clearly implying a distance bias as discussed above. It is thus possible that EELRs could have been missed in the remaining six QBS/PSB hosts at higher redshift. Two of these galaxies are just outside the redshift range: ASASSN-15oi ($z=0.0484$) and AT\,2018hyz ($z=0.0458$). To investigate whether EELRs are present, we ran the same procedure to quantify the EELRs in these systems. In both hosts, we identified a single bin at a distance of $\sim5$\,kpc from the nucleus that showed tentative [\ion{O}{III}] emission at $\sim3\sigma$ level, corresponding to $L\sim2\times10^{37}$\,erg/s. At such a low luminosity, the emission would be well below what was detected in ASASSN-14ae and comparable to the faintest identified bins in the other four systems with EELRs. While faint, the emission is thus on a level expected for EELRs in the TDE hosts in the sample. As there appears to be a distance bias against discovering EELRs in host at $z>0.045$, we conclude that in the MUSE sample EELRs are present in minimum $45\%$ (5/11) post-merger host galaxies of TDEs. The real fraction is likely higher also because we did not investigate the presence of EELRs deeply embedded in the host galaxies, as was done in \citet{Xiong2025ExtendedQPE} which identified EELRs near the nuclear region of AT\,2019qiz (within 3.7\,kpc). If the sample is representative in terms of the host galaxies, the result applies to all TDEs in post-merger hosts. However, as discussed in Section \ref{subsec:sample_demo}, the MUSE sample may be biased towards QBS/PSB galaxies and the true occurance rate is uncertain.

We verified that EELRs are over-represented in TDE hosts by visually inspecting the host galaxies of a MUSE sample of Type Ia SNe from \citet{Castrillo2021}. After selecting the galaxies found in the same redshift range as the TDEs, and the ones where the whole galaxy could fit the MUSE FoV, we were left with 46 systems. None of them exhibit EELRs similar to those seen in the TDE host galaxies. While the SN sample is likely not representative of the Ia SN hosts due to selection effects, Type Ia SNe are expected to occur in all types of galaxies, and there should not be a significant bias against ones with EELRs. The lack of EELRs in the Type Ia SN hosts directly implies a low occurrence rate of EELRs in the general galaxy population. As such, incidencies of $25$\% for the whole sample and $45\%$ for the post-merger TDE hosts, indicate a very significant over-representation of EELRs. The result is in agreement with the literature. \citet{Wevers2024ExtendedEvents} suggest that EELRs are over-represented in PSB galaxies that have hosted TDEs by a factor of $\sim2-5$ over a generic galaxy population \citep[e.g.][]{Lopez-Coba2020,Keel2024TheSample}, and the study determined that $60^{+40}_{-33}$\% of the PSB-TDE galaxies exhibit EELRs. Assuming that the PSB galaxies result from galaxy merger, the study concludes that the post-merger environment is a crucial factor in driving TDE rates. Our study performed on a larger dataset agrees with the conclusion.

\subsection{The source of ionising emission}
We have verified that at least some of the identified EELRs have to be ionised by nuclear activity using the Baldwin-Philips-Terlevich \citep[BPT;][]{Baldwin1981} and WHAN diagrams \citep{CidFernandes2011AAGN}. Apart from a few EELR bins for ASASSN-14ko, all of these regions require the ionising emission to be AGN-like based on the BPT diagram shown in Figure \ref{fig:EELR_BPTs}, and star-formation can be safely excluded. We also note that as no significant H$\alpha$ emission is present in the false-colour images in Figures \ref{fig:EELR_figs1} and \ref{fig:EELR_figs2}, except for ASASSN-14ko where some star-formation appears to be present along the spiral arm, our focus on the [\ion{O}{III}] $\lambda5007$ emission line to identify EELRs did not bias us against star-forming regions where Balmer emission lines would dominate. Furthermore, the WHAN diagrams shown in Figure \ref{fig:EELR_WHANs}, imply that AGN-like hard emission is required to explain the high EW of H$\alpha$ emission ($>3$\,Å) with the high [\ion{N}{II}] $\lambda6583$ to H$\alpha$ ratio. Even in ASASSN-14ae, ASASSN-14li and iPTF16fnl where some EELRs are found in LINER-like region of the diagram, individual EELRs are located in the AGN regime, where alternative ionising sources such as post-asymptotic giant branch (post-AGB) stars and low-velocity shocks have difficulties producing the observed line strengths \citep[see e.g.][]{Sanchez2020SpatiallyGalaxies, Sanchez2021FromGalaxies}. While excitation by fast shocks ($v>200$\,km/s) could result in similar diagnostic line strengths, both \citet{Prieto2016} and \citet{Wevers2024ExtendedEvents} note that the intrinsically narrow emission line ($v\sim40$\,km/s) of the EELRs in ASASSN-14li and iPTF16fnl are inconsistent with shocks, favouring the AGN interpretation.  Due to the strong Balmer absorption lines present in some of the spectra, the best-fitting simple stellar population \texttt{Starlight} models were subtracted from them before estimating the line fluxes. The details for the models are presented in Section \ref{sec:SMDs}. 

The BPT and WHAN diagrams imply that a significant level of nuclear activity must have occurred in the recent past for the ionising emission to reach these EELRs. To investigate whether the current nuclear luminosity is sufficient to light up the EELRs, we estimated the integrated nuclear luminosities using the H$\beta$ and [\ion{O}{III}] $\lambda5007$ emission lines seen the in spectra presented in Figure \ref{fig:nuclear_spec} using the bolometric correction from \citet{Netzer2009AccretionNuclei}, as well as the minimum ionising luminosity ($L_\mathrm{ion}$) required by the EELRs following \citet{French2023}. The resulting values are shown in Table \ref{tab:EELR_details}. The nuclear luminosity of ASASSN-14ae is quoted as an upper limit, due to non-detection of H$\beta$. Furthermore, we note that the estimates are approximate, and for instance, a more conservative approach using NIR images has resulted in $\sim1$\,dex higher upper limits for the nuclear luminosities \citep[e.g.][]{Wevers2024ExtendedEvents}. 

For ASASSN-14ko and AT\,2019ahk the nuclear luminosities appear to be high, but as the MUSE cubes were obtained only $\sim6$ months after the TDEs (see Table \ref{tab:tde_list}), it is difficult to estimate the actual pre-TDE luminosities. For instance, the nuclear spectrum of AT\,2019ahk is visibly blue and exhibits broad emission line wings (see Figure \ref{fig:nuclear_spec}), but the archival host spectrum presented by \citet{Holoien2019a} is red with only narrow lines, implying that a significant level of TDE-related activity remains present in the MUSE cube. The two nuclei are also found on the strong-AGN (sAGN) regime of the WHAN diagrams (Figure \ref{fig:EELR_WHANs}), implying hard AGN-like continuum emission. As such, it is plausible that the nuclear luminosity was significantly lower than  $L_\mathrm{ion}$. For the remaining three hosts, the level of nuclear emission is clearly below the luminosity required to ionise the EELRs and the nuclei are located in the LINER-like region of the WHAN diagrams. As noted by \citet{Wevers2024ExtendedEvents}, this possibly implies higher AGN activity that has since turned off, but past TDEs cannot be ruled out as the cause, even if they would have to be frequent in order to keep the EELRs ionised. As shown by modelling of \citet{Mummery2025GalaxyEchoes}, galaxies with higher intrinsic rate of TDEs are more likely to show EELRs. With the range of projected offsets of the EELR bins shown in Table \ref{tab:EELR_details}, we can constrain the time when the nuclei had to be active. Assuming that the projected offsets are the actual light-travel distances, the hard ionising emission had to be emitted $\sim2000$\,yr to $\sim75000$\,yr before the TDEs. The real values could be higher, but given all five hosts exhibit EELRs at projected offsets $\lesssim2$kpc it is likely that there are EELRs at such distances. Given bright AGN phases are predicted to last $\sim10^4$\,--\,$10^5$\,yr \citep[e.g.][]{Schawinski2015ActiveYr, French2023}, the timescale in which the AGN would have had to turn off is remarkably short to still power the seen EELRs. Further, it would also be peculiar to see an over-representation of TDEs during such a short-lived phase in the galactic evolution, unless there was some causal relation. Following \citet{Wevers2024ExtendedEvents}, the results imply that either the AGN turn-off has to be somehow related to the TDEs (or simply the detection efficiency of TDEs), or alternatively the TDE rate in these post-merger galaxies is significant enough for TDEs to provide the required hard emission to power the EELRs.

\section{Stellar Populations near to the SMBHs}
\label{sec:SMDs}
\begin{table}
    \def\arraystretch{1.2}%
    \setlength\tabcolsep{12pt}
    \centering
    \fontsize{9}{11}\selectfont
    \caption{The adopted black hole masses. The literature values are estimated using M-$\sigma$ relation based on spectra from high-resolution spectrographs. Where high-quality values were not available, we used the velocity dispersion $\sigma$ from the \texttt{Starlight} fits to the nuclear MUSE spectra to determine M-$\sigma$ estimates (AT\,2019ahk, LSQ12dyw). Note that for AT\,2017eqx and AT\,2019lwu the low S/N spectra did not allow estimating $\sigma$ and in lack of literature M-$\sigma$ we used black hole masses based on TDE light curve fits. For ASASSN-14ko we adopted the value estimated in \citet{Payne2021}, which is based on several methods.}
    \begin{threeparttable}
    \begin{tabular}{l l l }
    \hline
    \hline
        \multicolumn{1}{c}{TDE} & log$M_\mathrm{BH, lit}$        & log$M_\mathrm{BH, MUSE}$  \\
    \hline
      iPTF16fnl & $  5.50_{ -0.42}^{+  0.42}$~[1]  & $  6.24_{ -0.20}^{+  0.20}$ \\
      AT2019qiz & $  6.52_{ -0.34}^{+  0.34}$~[2]  & $  6.50_{ -0.18}^{+  0.18}$ \\
    ASASSN-14ko & $  7.86_{ -0.41}^{+  0.31}$~[3]  & $  9.05_{ -0.09}^{+  0.09}$ \\
      AT2019ahk &    -                             & $  7.78_{ -0.10}^{+  0.10}$ \\
       iPTF15af & $  6.88_{ -0.38}^{+  0.38}$~[1]  & $  6.88_{ -0.16}^{+  0.16}$ \\
      AT2018hyz & $  5.25_{ -0.39}^{+  0.39}$~[4]  & $  5.95_{ -0.22}^{+  0.22}$ \\
    ASASSN-14ae & $  5.42_{ -0.46}^{+  0.56}$~[1]  & - \\
      AT2023clx & $  5.71_{ -0.40}^{+  0.40}$~[5]  & $  6.12_{ -0.21}^{+  0.21}$ \\
    ASASSN-14li & $  6.23_{ -0.40}^{+  0.39}$~[1]  & $  6.51_{ -0.18}^{+  0.18}$ \\
       PTF09axc & $  5.68_{ -0.49}^{+  0.48}$~[1]  & $  5.11_{ -0.27}^{+  0.27}$ \\
      AT2018dyb & $  6.51_{ -0.40}^{+  0.40}$~[6]  & $  7.24_{ -0.13}^{+  0.13}$ \\
    ASASSN-15oi & $  5.93_{ -0.60}^{+  0.60}$~[7]  & $  5.51_{ -0.25}^{+  0.25}$ \\
      AT2019dsg & $  6.90_{ -0.32}^{+  0.32}$~[8]  & $  6.75_{ -0.16}^{+  0.16}$ \\
    ASASSN-15lh & $  8.70_{ -0.40}^{+  0.41}$~[9]  & $  8.99_{ -0.08}^{+  0.08}$ \\
      AT2017eqx & $  6.30$~[10]                    & - \\     
      AT2018fyk & $  7.69_{ -0.39}^{+  0.39}$~[7]  & $  7.72_{ -0.10}^{+  0.10}$ \\
      AT2019lwu & $  6.37_{ -0.20}^{+  0.21}$~[11] & - \\
       LSQ12dyw &    -                             & $  5.67_{ -0.24}^{+  0.24}$ \\
         D23H-1 & $  6.39_{ -0.44}^{+  0.44}$~[12] & $  6.11_{ -0.21}^{+  0.21}$ \\
    SDSSJ134244 & $  6.06_{ -0.52}^{+  0.51}$~[12] & $  6.28_{ -0.20}^{+  0.20}$ \\

    \hline
    \hline
  \end{tabular}
    \begin{tablenotes}
        \item \textbf{References:} [1] \citet{Wevers2017}; [2] \citet{Nicholl2020}; [3] \citet{Payne2021}; [4] \citet{Short2020}; [5] \citet{Charalampopoulos2024}; [6] \citet{Leloudas2019}; [7] \citet{Wevers2020}; [8] \citet{Yao2023TidalFunction}; [9] \citet{Kruhler2018}; [10] \citet{Nicholl2019}; [11] \citet{Hammerstein2023}; [12] \citet{French2020}.
    \end{tablenotes}
    \end{threeparttable}    
\label{tab:mbhs}
\end{table}

We investigate the stellar mass distributions (SMDs) in the near vicinity of the SMBHs by fitting the nuclear spectra (see Figure~\ref{fig:nuclear_spec}) with the \texttt{Starlight} spectral synthesis package \citep{CidFernandas2005, CidFernandes2008}. \texttt{Starlight} models the observed spectrum (after masking the emission-line-dominated regions and correcting for dust extinction) with a synthetic spectrum created by summing together contributions of simple stellar populations (SSPs). Accumulating the SSPs will provide more realistic SMDs than simply using an initial mass function (IMF) to represent the stellar distribution as it takes into account the evolution of the populations. The results between the two methods would be similar, but with our methodology fewer massive stars, arising from young stellar populations, are present as they explode quickly after formation. On the other hand, low-mass stars are more numerous as they are accumulated from multiple SSPs. In the \texttt{IFUanal} implementation of \texttt{Starlight}, the SSP base models are taken from \citet{Bruzual2003}\footnote{2016 update at \url{http://www.bruzual.org/bc03/.}} as created for the MILES spectral library \citep{Sanchez-Blazquez2006}, and use a \citet{Chabrier2003} IMF in the range of $0.1$\,--\,$100M_\odot$. The base models are split into 16 ages that range from 1\,Myr to 13\,Gyr at a roughly logarithmic spacing. We fit for three sets of metallicities ($Z=0.004$, $0.02$, $0.05$). However, as the metallicity in the regime from slighly sub-solar to slightly super-solar has only a small effect on the overall SSP age distribution when modelling optical spectra, we focus solely on the ages. While the wavelength range of the MUSE spectra is not blue enough to distinguish between very young populations ($\lesssim100$\,Myr), the nuclear spectra of the TDE galaxies are mostly visibly red (Figure \ref{fig:nuclear_spec}). This implies that the nuclear regions of the galaxies are dominated by old stellar populations, and in lack of young, hot stars, the derived SFHs should be reliable. An example \texttt{Starlight} model is provided in Figure \ref{fig:SL_AT2023clx}. For more details on the methodology, see \citet{Lyman2018}. 

\begin{figure*}
    \centering
    \includegraphics[width=\textwidth]{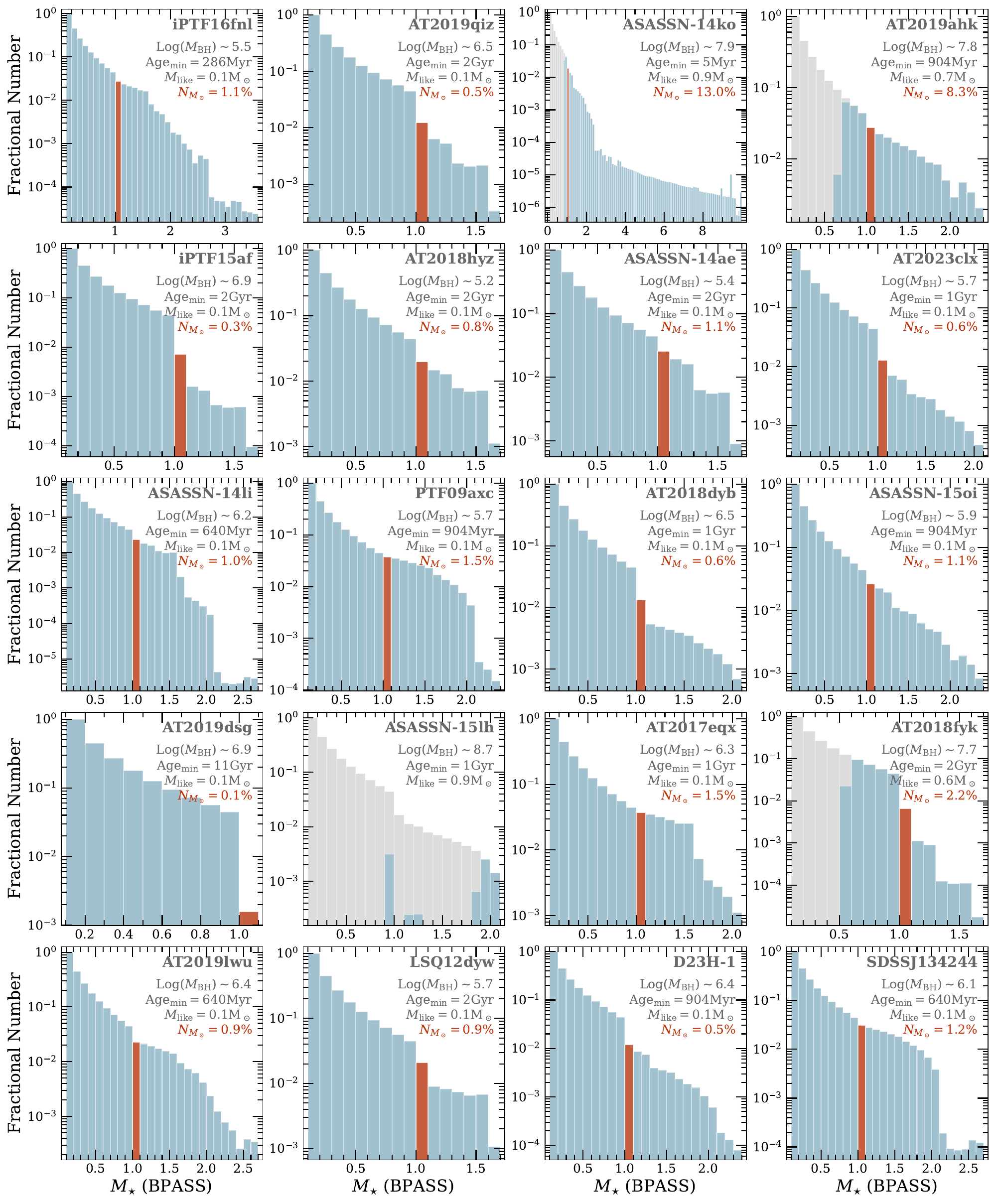}
    \caption{The total stellar mass distributions (SMDs) in the nuclear regions of the galaxies based on \texttt{Starlight} SFHs and BPASS SMDs of individual subpopulations (lighter shade). The population of stars that can generate visible TDEs are highlighted (darker shade). Stars in the $1.0$\,--\,$1.1M_\odot$ bin are shown (red). Key details such as BH mass, youngest stellar populations, likeliest mass to be disrupted as well as the number density of $1M_\odot$ stars that can create a visible TDE are shown. Apart from the few galaxies with very massive BHs ($M_\mathrm{BH}\gtrsim10^8M_\odot$), the stellar populations that can create a visible TDE  are dominated by subsolar-mass stars. The distribution of ASASSN-15lh is noticeably different due to the high SMBH mass prohibiting the production of visible TDEs for all stars except the small subpopulation that have reached the giant phase in the BPASS models, increasing the stellar radii sufficiently to allow an observable TDE.}
    \label{fig:smd_hists}
\end{figure*}

\begin{figure}
    \centering
    \includegraphics[width=0.46\textwidth]{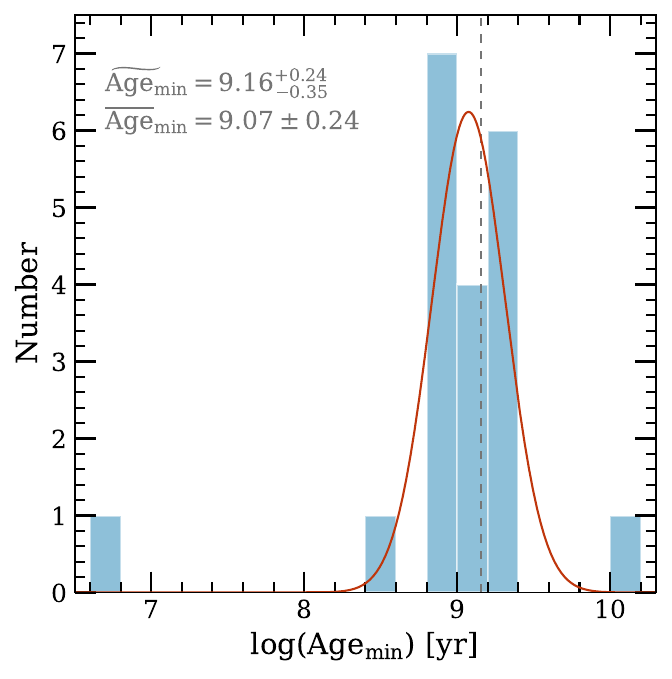}
    \caption{The logarithmic distibution of identified youngest stellar populations (Age$_\mathrm{min}$) in the nuclear regions of the TDE host galaxies. The distribution peaks sharply at $\sim1$\,Gyr. The median of the distibution ($\widetilde{\mathrm{Age}}_\mathrm{min}$) is shown with the $1\sigma$ percentiles, together with mean and standard deviation of a Gaussian fit ($\overline{\mathrm{Age}}_\mathrm{min}$).  }
    \label{fig:age_min_hist}
\end{figure}

To identify the stellar populations that can produce observable TDEs, we utilise the Hills mass \citep{Hills1975}:  the maximum mass of a non-rotating BH that can disrupt a star outside the event horizon, which is a function of the stellar mass and radius ($M_\mathrm{H} \propto R_\star^{1.5} \times m_\star^{-0.5}$). \texttt{Starlight} returns SFHs in the form of light ($x_\mathrm{j}$) and mass fractions ($M_\mathrm{j}$) of the SSPs of different ages, but it does not provide the mass and radius distributions of the actual stars we require. These are not quantities that are readily available from the \citet{Bruzual2003} models. The mass distribution could be approximated knowing the form of the IMF, and the mass of the stars on the Main Sequence turn-off at each age, but the stellar radii require an assumption in the form of a mass-radius relation. Instead, we reconstruct them with data from the Binary Population and Spectral Synthesis \citep[\texttt{BPASS};][]{Eldridge2008, Eldridge2017, Stanway2018} code to investigate the total stellar population seen around the nuclei. The \texttt{BPASS} stellar models are detailed 1-dimensional stellar evolution models calculated using a modified version of the Cambridge \texttt{STARS} code. Using the v2.2.1 data release, we extracted the mass (initial and present value), radius and \citet{Chabrier2003} IMF weighting for all stellar models at each of the time bins used in the \texttt{Starlight} fits. The \texttt{BPASS} models in the Chabrier IMF variant contain stars ranging from $0.1$--$100M_\odot$, roughly logarithmically spaced in mass. Dividing the IMF weighting assigned to each model by its initial mass returns the number of stars of that birth mass which exist within a stellar mass of $10^6M_\odot$. Since we only care about the distribution of stellar mass, this scaling value is arbitrary. Taking the radius and the remaining mass of each star at the specified age steps, the corresponding Hills mass can be determined. Note that we calculate the SMDs using only the \texttt{BPASS} single star evolution models, and do not include \texttt{BPASS} binary star models for this calculation. Binary evolution will broaden the mass distribution of living stars at a given age, but for old stellar populations comprising almost exclusively low-mass stars with a relatively low binary fraction \citep[$\sim25-30$\%;][]{Moe2017}, the impact should only have a small effect on the SMDs which do not change the overall conclusions. At $1.4$\,Gyr (approximate median age of the youngest stellar populations in the galaxy sample; see Figure \ref{fig:age_min_hist}) the number of stars with $M>0.3M_\odot$ is marginally higher in the model that includes stellar binaries. For example, the number ratio $N_{0.5}/N_{0.1}$ increases by $\sim11\%$, while $N_{1.5}/N_{0.1}$ by about $2\%$ if binaries are included, highlighting that the impact of binary star models is limited for this analysis.

The subpopulation of stars that can produce a visible TDE can then be determined by comparing the Hills masses to the SMBH masses mostly collected from literature, shown in Table \ref{tab:mbhs}. Most of these values are based on the $M-\sigma$ relation, where the BH mass is estimated from the velocity dispersion $\sigma$ of the spectral lines in high-resolution spectra. In the absence of more reliable values, the BH masses for AT\,2019ahk and LSQ12dyw were estimated based on the \texttt{Starlight} fits to the nuclear MUSE spectra using the $M-\sigma$ relation presented in \citet{Kormendy2013}. As shown in Table \ref{tab:mbhs}, the MUSE estimates are in general consistent with those from the literature, and the two values should be reliable. For AT\,2017eqx and AT\,2019lwu \texttt{Starlight} fits did not converge on a solution of $\sigma$ due to low S/N spectra, and we used literature values derived from light curve fits. In the case of ASASSN-14ko we adopted the value estimated by \citet{Payne2021} based on several scaling relations. 

The total stellar mass distributions together with the subpopulation that can generate an observable TDE are shown in Figure \ref{fig:smd_hists}. Given that most BHs are low mass in the context of SMBHs ($\lesssim10^7M_\odot$), the likeliest stars to create an observable TDE are 0.1$M_\odot$ for the majority of host galaxies. For these galaxies, all the stars in the BPASS models can create a visible TDE. Such low mass stars completely dominate the population by numbers and in comparison the $1M_\odot$ stars are only a few percent. The only clear exceptions are the hosts of ASSASSN-14ko, ASASSN-15lh, AT\,2018fyk and AT\,2019ahk where the BH masses are sufficient to exclude low-mass stars. Interestingly, three of these TDEs exhibit multi-peaked light curves. ASASSN-14ko is a known repeating source \citep[e.g.][]{Payne2021}, while ASASSN-15lh \citep[e.g.][]{Leloudas2017} and AT\,2018fyk \citep{Wevers2019, Wevers2022, Pasham2024, Wen2024} show pronounced secondary peaks in the UV. The relation possibly implies that multi-peaked light curves may be related to the high BH mass and/or the repeated partial disruptions of higher mass stars. However, even for them TDEs from low-mass stars are possible, as the Hills mass increases if one assumes significant BH spin.  This is perfectly exemplified by ASASSN-15lh. \citet{Leloudas2017} argues that the event is a TDE, but due to the extreme SMBH ($10^{8.7}M_\odot$), only a rapidly rotating BH would sufficiently increase the Hills mass to allow a TDE of a star likely present in the passive environment to occur. Our SMDs support a similar conclusion. The only stars that do not violate their Hills masses have reached the giant phase, and their stellar radii are sufficiently high to allow an observable TDE. However, as these stars are at a particular evolutionary stage, they are very rare in comparison to the whole SMD, and a TDE from a rotating BH is more likely.

The SMDs also allow to investigate the sample properties of the stellar populations. One of the key features, directly related to the massive stars, is the age of the youngest SSP in the fits. Based on the distribution shown in Figure \ref{fig:age_min_hist}, the Age$_\mathrm{min}$ clusters around $\sim~1$\,Gyr, implying absence of stars more massive than $\sim2-3M_\odot$ in the nuclear regions of the TDE hosts. However, we note that the fits may not capture the full stellar diversity, and some individual more massive stars are possible present in the near vicinity of the SMBHs. Regardless, their number should be very small, and TDEs from such progenitors should be rare unless there is a mechanism that would over-enhance the massive-star TDE rate. Possible mechanisms are discussed in Section \ref{subsec:SMD_caveats}. 

Finally, we note that we have decided to exclude the youngest 0.5\% stellar population in mass. \texttt{Starlight} has a tendency of finding a small amount of young stellar populations ($10^6-10^7$\,yr), that appear inconsistent with otherwise very old populations. This is likely caused by the fact that many of the cubes are taken soon after the TDE, and some residual blue emission from the TDE itself is likely present in the nuclear spectra (like AT\,2019ahk discussed in Section \ref{sec:EELRs}), mimicking the effect of young populations. Further, in some cases the S/N of the spectra are quite low, resulting in less reliable fits. While we cannot determine if some of these populations are not real, we consider them likely spurious. Were they included they would increase the maximum mass of stars in the SMDs, but these stars would be extremely rare.

\subsection{Comparison to masses of the disrupted stars derived from light curve models}

We can now compare the derived SMDs with the masses of the disrupted stars identified for the TDEs with light curve modelling. The commonly used light curve fitting routines are \texttt{MOSFiT} \citep[][]{Guillochon2018}, \texttt{TDEMass} \citep{Ryu2020} and \texttt{REDBACK} \citep{Sarin2024a}, all of which assume different mechanisms for powering the light curve. \texttt{MOSFiT} generates the light curves by assuming that the rate of energy production follows the fallback rate of the material on the BH at a constant conversion efficiency \citep{Mockler2019}, \texttt{TDEMass} adopts intersection of the debris streams \citep[e.g.][]{Piran2015, Jiang2016} as the powering mechanism, while \texttt{REDBACK} relies on a cooling envelope model, where the debris forms a pressure supported envelope that powers the UV/optical light curve \citep{Sarin2024}. 

\begin{figure}
    \centering
    \includegraphics[width=0.49\textwidth]{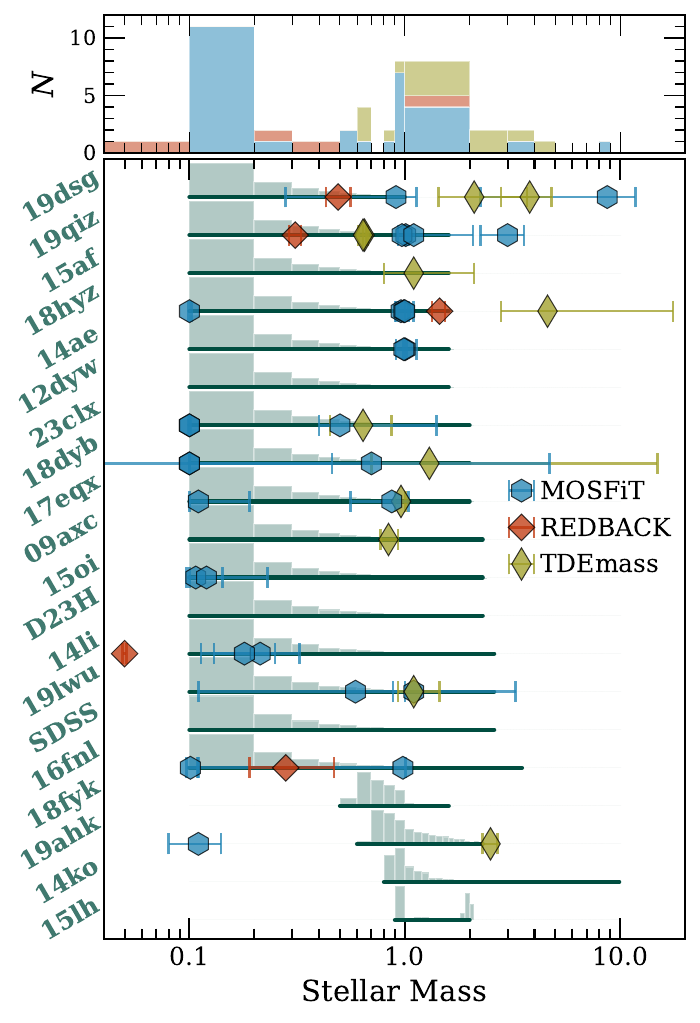}

    \caption{\textit{Bottom:} The allowed mass ranges of the disrupted stars for each TDE (solid line) based on the SMDs shown with histograms (as in Fig. \ref{fig:smd_hists}). Stellar mass estimates from light curve models collected from literature are shown. The values are broadly in agreement with the allowed masses, but they are concentrated at higher values than predicted by the SMDs. Stellar populations that violated the Hills mass are excluded. \textit{Top}:  The number of models per bin, where the bins have width $0.1M_\odot$ at $<1M_\odot$, and $1M_\odot$ at $>1M_\odot$. The stellar masses estimated with TDE light curves are collected from the following papers: \citet{Leloudas2019}; \citet{Mockler2019}; \citet{Nicholl2019};  \citet{Gomez2020}; \citet{Holoien2020}; \citet{Hung2020};   \citet{Nicholl2020}; \citet{Ryu2020}; \citet{Nicholl2022}; \citet{Charalampopoulos2024}; \citet{Hammerstein2023}; \citet{Hoogendam2024};  \citet{Sarin2024}; \citet{Zhong2024}. } 
    \label{fig:M_star_range}
\end{figure}

As shown in Figure \ref{fig:M_star_range}, the estimated values are broadly in agreement with the range of allowed masses. However, apart from some models at $\sim0.1M_\odot$, the fit values are mostly found at $\sim1M_\odot$ or above, where our SMDs predict only a fractional amount of stars if any. Further, clear model-specific tendencies can also be identified. \texttt{MOSFit} is the most used in the literature, and the large number of fits appear to cluster around $0.1M_\odot$, and  $1M_\odot$. The peak at $\sim0.1M_\odot$ is in line with the SMDs, but the feature is enhanced due to the \texttt{MOSFiT} model setup. The radii of stars below $0.1M_\odot$ are constant in the model, and as a consequence stellar masses of $\sim0.1M_\odot$ are favoured for fast-rising TDEs \citep[for details see][]{Mockler2019}. The latter cluster, however, is inconsistent with the SMDs. It is possibly an artefact of the fitting routine itself, as the polytropic index to define the stellar structure in the \texttt{MOSFiT} model changes at $1M_\odot$.  On the other hand, \texttt{TDEmass} finds only relatively high stellar masses ($0.7$\,--$5.0M_\odot$), clearly at odds with the predictions from the SMDs. Finally, for \texttt{REDBACK} there are fewer publicly available fit parameters, and comparison is difficult, but the few values are evenly distributed between $0.05$ and $2M_\odot$. 

\subsection{Causes for the apparent tension}
\label{subsec:SMD_caveats}
The apparent tension between the masses of the disrupted stars based on the light curve models and the derived SMDs can be a result of a number of different causes. First, it is possible that light curve models have difficulties in capturing the masses of the disrupted stars. This is especially likely as the stellar mass is considered a degenerate parameter at least in the \texttt{MOSFiT} model \citep{Mockler2019}, and deriving an exact solution might not be realistic. As such, these values should be treated with some caution. It is also possible that TDEs of very low-mass stars are in fact very prominent, but we have a detection bias against them \citep[e.g.][]{Nicholl2022}. The TDE luminosity scales according to the fallback rate as $\dot{M} \propto M_\star$ \citep[e.g.][]{Law-Smith2020StellarRates, Ryu2020TidalMass}, while the duration of TDEs is expected to scale as $t \propto M_\star^{-1}$. As such TDEs from low-mass stars are possibly faint and slow-evolving making them more difficult to identify than their brighter, faster evolving cousins arising from disruption of more massive stars. However, we note that disruption of low-mass stars have been suggested for fast-rising TDEs such as AT\,2020neh \citep{Angus2022} and AT\,2023clx \citep{Charalampopoulos2024}, and it is unclear how large a fraction of the population might be slow-evolving.

It is also possible that our SMDs do not represent the true stellar mass function near the SMBH. One possibly affecting assumption is the used IMF. We have assumed Chabrier IMF \citep{Chabrier2003}, as it predicts fewer low-mass stars than the Kroupa IMF \citep{Kroupa2001OnFunction}, and our choice should thus reduce the number of low-mass stars in comparison. It has also been suggested that the IMF in nuclear regions of galaxies is more top-heavy \citep[see e.g.][]{Lu2013StellarFunction}. However, subsolar mass stars will still dominate the distribution regardless, unless their production is very significantly stifled. It is unlikely the case here, as the nuclear regions of our TDE hosts exhibit mostly red spectra as is expected of old stellar populations that are dominated by very low-mass stars. Furthermore, the extracted nuclear spectra may also include a significant contribution from stellar populations in the Line-of-Sight (LoS). Although it is difficult to ascertain the amount, we have evaluated the significance qualitively by comparing the nuclear spectra, that should consists of both the nuclear and LoS populations, against spectra extracted with annuli just outside the nuclei that should probe only stellar populations similar to the LoS.  The annuli exhibit significantly lower flux per spaxel than the nuclei, implying that LoS population do not contribute significantly to the nuclear spectra. While their contribution is still present, we consider that the nuclear spectra are dominated by stellar populations near the central region. 

While the overall stellar populations in central regions are likely old, we may have missed small-scale nuclear stellar clusters that can contain younger populations of stars than in the larger-scale region probed by the MUSE cubes. Such star clusters are coincident with the nuclei and they are found mostly at sizes $r\lesssim10$\,pc  \citep[e.g.][]{Neumayer2011Two-dimensionalGalaxies}. In late-type galaxies the stellar clusters are dominated by old \citep[more than a few Gyr;][]{Neumayer2020NuclearClusters}, low-mass stars, but they also contain stars younger than a Gyr \citep[e.g.][]{Rossa2006HubbleType}, even down to a few 100\,Myr \citep[e.g.][]{Bender2005HSTHole, Walcher2005MassesGalaxies, Kacharov2018StellarGalaxies}. In the early-type galaxies the clusters typically consists of only old \citep[$\gtrsim1$\,Gyr; e.g.][]{Monaco2009The205, Kacharov2018StellarGalaxies} populations, but some have shown substantial amount of younger stars $\lesssim1$\,Gyr \citep[e.g.][]{Spengler2017VirgoSpectroscopy}. As the measured FWHM of the MUSE cubes are at the smallest $\sim200$\,pc (see Table \ref{tab:fwhms}), nuclear clusters that concentrate on scales smaller than $20$\,pc are beyond our resolving power. If such clusters existed in the host galaxies, it is possible they house a substantial population of younger, more massive stars than probed by the derived SMDs. In fact, using HST images, \citet{Newsome2025} identified $\sim340$\,Myr old starburst within $44$\,pc of the ASASSN-14li host nucleus, that is younger than the $\sim550$\,Myr stellar population found at an offset radius of $88$\,pc. The minimum age of the stellar population in our \texttt{Starlight} fit is $\sim640$\,Myr, and our nuclear spectrum is clearly dominated by the more distant older stellar population, and the younger population is diluted by the physical resolution of the MUSE cube. However, as discussed above, it is unlikely that IMF of the clusters would be top-heavy enough for very low-mass stars not to dominate the population.

While low-mass stars should be present in the nuclear star clusters, mass segregation has been hypothesised to enhancence the rate of TDEs from of supersolar-mass stars. \citet{Rom2025SegregationEMRIs} estimated that TDEs from stars in $1$\,--\,$3M_\odot$ range are enhanced by a factor $\sim M_\star$, and at $\gtrsim3M_\odot$ by a factor $\sim9$ independent of the stellar mass. While significant mass segregation can thus be expected, the study concludes that the most likely disrupted stars are still the very low-mass stars that dominate the population. Based on our SMDs, $1M_\odot$ stars would have to be enhanced by a factor $\sim100$ (depending on the host) over the $0.1M_\odot$ stars for them to be equally likely. As this is more than predicted by \citet{Rom2025SegregationEMRIs}, it is unclear if it can be the only process affecting which stars produce TDEs. Similarly, \cite{Kochanek2016} investigated the rates of TDEs as function of mass of the disrupted star and the SMBH. The study concludes that at $M_\mathrm{BH}<10^{6.5}M_\odot$ likeliest TDEs arise from $M_\star\sim0.1M_\odot$, and at $M_\mathrm{BH}$ in range of $10^{6.5}$\,--\,$10^{7.5}M_\odot$ the typical TDE is caused by an $M_\star\sim0.3M_\odot$ M-dwarf, although the mass function is reasonably flat below $1M_\odot$. Given most of our SMBHs are $<10^{7.5}M_\odot$, the results are directly comparable.  While the study predicts fewer TDEs from low-mass stars than our SMDs due to the relatively flat TDE rate below $\sim0.5M_\odot$, the TDEs from such stars are still the likeliest and thus should still dominate the observed distribution.

Observational evidence for the disruption of moderately massive stars ($1$\,--\,$2M_\odot$) have also been presented in the literature. \citet{Mockler2022EvidenceHoles} investigated the stellar masses of three TDEs in the MUSE sample (ASASSN-14li, iPTF15af and iPTF16fnl) using N/C ratio with \ion{N}{III} $\lambda1750$ and \ion{C}{III} $\lambda 1980$ emission lines present in the HST spectra. Given the strict abundance ratio of $\mathrm{N/C}\geq10$ \citep[see][]{Yang2017TheEvents}, they concluded that the TDEs were a result of disruption of stars with $M_\star\gtrsim1.3M_\odot$. Using the Kroupa IMF, the study determined that disruptions of moderately massive stars are over-represented over the galactic stellar population by a factor of $\gtrsim100$. With our SMDs, we can estimate comparable enchancement factors. Using only stars that can generate visible TDEs, stars of $\geq1.3M_\odot$ require constant factors of a minimum few tens to be equally likely as less massive stars. However, note that this estimate excludes hosts with massive, $\log(M_\mathrm{BH}>7.5$) SMBHs where the factors are much smaller due to the absence of low-mass stars. For the three systems analysed by \citet{Mockler2022EvidenceHoles}, our SMDs imply enhancement factors of $\sim70$ (ASASSN-14li), $\sim1200$ (iPTF15af) and $\sim30$ (iPTF16fnl). As such the suggested enhancement seems broadly comparable to our SMDs as it would make TDE from moderately massive stars the likeliest option for most of the host systems. 

In the absence of fully reliable distribution of the disrupted star masses, it is difficult to conclude how much of an effect the discussed mechanisms have.  Given the theoretical and observational support, it is likely that supersolar-mass stars contribute significantly to the TDE population through some mechanism of mass segregation. However, our results directly imply that TDEs from low-mass stars should be abundant and even dominate the TDE distribution, unless there is a mechanism that stifles production of low-mass TDEs or hampers their detection significantly.
 
\section{Conclusions}
\label{sec:conclusions}
We have presented a detailed investigation of MUSE data of 20 TDE host galaxies, focusing on two key aspects: the extended emission line regions surrounding the galaxies and stellar mass distributions near the SMBHs. Although collected from multiple programmes, the sample appears to be representative of TDEs in terms of the TDE spectral types, and the subset of 16 TDEs with literature classifications are divided into  6.3\% He-TDEs, 62.5\% of H+He-TDEs, 31.3\% H-TDEs. The sample is also statistically consistent with the largest TDE host galaxy compilation \citep{French2020} comprising of 35\% \say{star-forming} (i.e. strong H$\alpha$ emission), 45\% of QBS/PSB implying a significant, recent ($<1$\,Gyr) starburst and 15\%  of Quiescent. However, it is possible that the sample is slightly biased towards QBS/PSB galaxies.   

In the first part of the analysis we characterised the population of extended emission line regions seen in the TDEs. We identified EELRs in 5/20 of the host galaxies including two previously unreported systems. The systems are found below $z=0.045$, indicating a possible distance bias, and low-luminosity EELRs cannot be excluded above this redshift. The EELRs are only identified in systems that have likely undergone a galaxy merger, and all post-merger galaxies below $z=0.045$ show EELRs (5/11 of the sample). Given the probable distance bias, we conclude that a minimum 45\% of post-merger TDEs exhibit EELRs in the sample. If the sample is representative in terms of the TDE host galaxies, the value reflects the occurance rate for all post-merger TDEs. 

In all five hosts, at least some EELRs require a hard AGN-like ionising continuum to explain the diagnostic line strengths. Assuming the projected offsets are the actual light-travel distances, the ionising continuum had to be emitted 2000 to 75000\,yr before the TDE, clearly indicating that significant nuclear activity had to occur in the galaxies for the EELRs to be ionised. However, in three of the systems the current nuclear luminosity is not high enough to explain the EELRs, and in the remaining two the current nuclear luminosity is likely over-estimated due to TDE emission present in the observations. Following \citet{Wevers2024ExtendedEvents}, the significant over-enhancement of EELRs in post-merger TDE hosts implies that either the turn-off of the AGN phase has to be causally related to the TDEs (or efficient detection of TDEs), or the TDE rate in these galaxies is high enough to provide the ionising continuum.

In the second part of the analysis we focused on the stellar mass distributions in the near-vicinity of the SMBH, to understand the stellar populations that can produce observable TDEs. Based on \texttt{Starlight} fits of the nuclear spectra, we derive the SFHs as well as the light and mass fractions of the simple stellar populations (SSPs) contributing to the integrated nuclear spectrum. 
We then used  \texttt{BPASS} stellar models, weighted with a Chabrier IMF, to obtain the mass and radius distributions of the stellar populations near the black hole. The youngest stellar populations in the host nuclei have typical ages of 1 Gyr ($\log(\mathrm{Age}_\mathrm{min}/\mathrm{yr}) = 9.16^{+0.24}_{-0.35}$) and the maximum stellar masses present are typically below $2.0 - 2.5M_\odot$. By computing the Hills masses for the stellar populations, we determined that the population that can be disrupted outside the event horizon of a non-rotating SMBH is dominated by very low-mass stars down to $0.1M_\odot$. Only for four TDEs, the host SMBH is massive enough to exclude such stars, but even for them low-mass TDEs are possible if one assumes significant BH spin. 
Interestingly, three of these TDEs exhibit peculiar multi-peaked light curves, possibly due to partial repeating disruptions of more massive stars.  

Our SMDs are in tension with stellar masses of the disrupted stars derived using light curve models \citep[][]{Guillochon2018,Ryu2020,Sarin2024a}, especially \texttt{TDEMass}, which assumes that the optical light curve is produced by shock collisions. If the observed TDEs come predominantly from approximately solar-mass stars, then the TDE rate from more massive stars has to be significantly over-enhanced (by a factor of $\sim 100$), as supported by theoretical predictions for mass stratification \citep[e.g.][]{Rom2025SegregationEMRIs}, and by observational constraints \citep{Mockler2022EvidenceHoles}. Alternatively, TDE models have to be revised, at least concerning the determination of the disrupted star masses. Regardless, our SMDs directly imply that TDEs from low-mass stars should be abundant and contribute to a significant fraction of all TDEs. This study represents the most spatially resolved (mostly sub-kpc scales) IFU analysis of TDE host stellar populations to date, highlighting the importance of post-merger, old stellar populations in producing TDEs.

\section*{Acknowledgements}
We thank the anoynmous referee for helpful comments.
Based on observations collected at the European Southern Observatory under ESO programmes 096.D-0296, 097.D-1054, 099.D-0115, 0103.D-0440, 0105.20GS and 106.2155. 
MP and GL acknowledge support from a VILLUM FONDEN research grant (19054).
GL acknowledges support from a VILLUM FONDEN research grant (VIL60862).
MP and JL acknowledge support from a UK Research and Innovation Fellowship (MR/T020784/1).
CMB acknowledges funding from the UK Science and Technology Facilities Council (STFC) through Consolidated Grant Number ST/X001121/.
PC acknowledges support via the Research Council of Finland (grant 340613).
PR acknowledges support from STFC grant 2742655.
This work was funded by ANID, Millennium Science Initiative, ICN12\_009. 
FEB acknowledges support from ANID-Chile BASAL CATA FB210003, FONDECYT Regular 1241005, and Millennium Science Initiative, AIM23-0001.
MN is supported by the European Research Council (ERC) under the European Union’s Horizon 2020 research and innovation programme (grant agreement No.~948381).
HK was funded by the Research Council of Finland projects 324504, 328898, and 353019.
JLP acknowledges support from ANID, Millennium Science Initiative, AIM23-0001.
\section*{Data Availability}
The MUSE data analysed in this manuscript are publicly available in the ESO archive. 
 



\bibliographystyle{mnras}
\bibliography{references} 




\appendix

\section{Figures}
\label{appsec:figs}

\begin{figure*}
    \centering
    \begin{subfigure}[b]{0.246\textwidth}
         \centering
        \includegraphics[width=\textwidth]{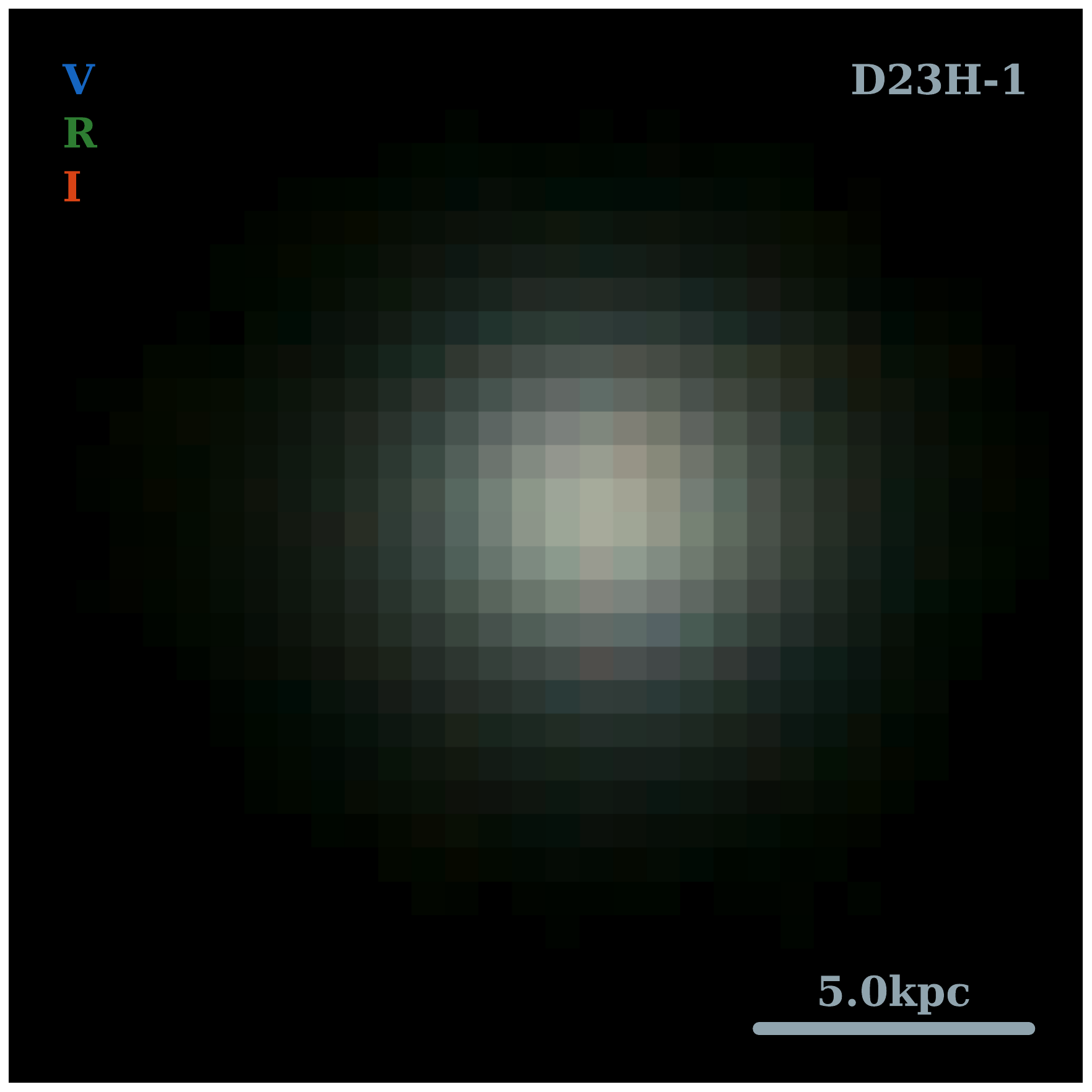}
    \end{subfigure} %
    \begin{subfigure}[b]{0.246\textwidth}
         \centering
        \includegraphics[width=\textwidth]{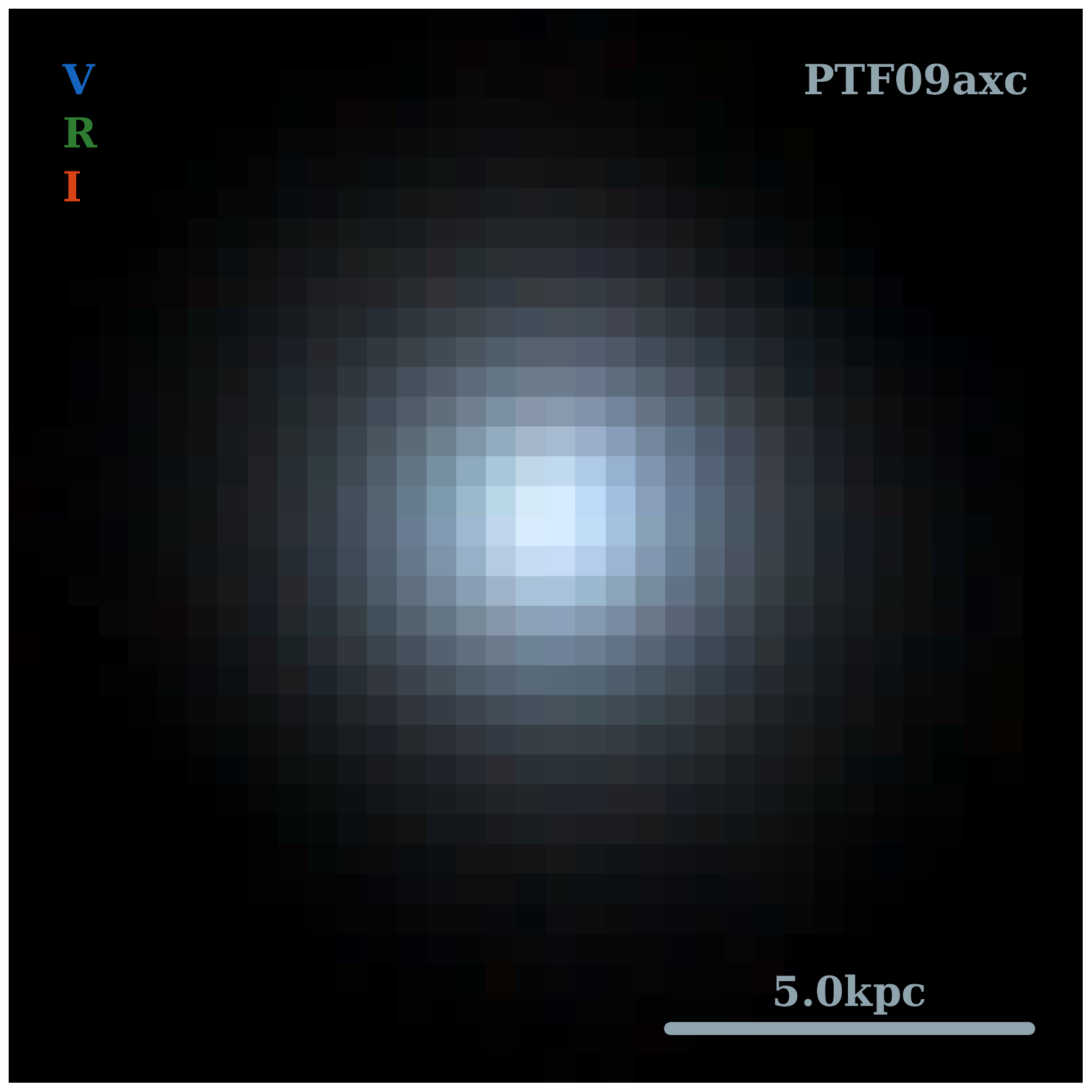}
    \end{subfigure} %
    \begin{subfigure}[b]{0.246\textwidth}
         \centering
        \includegraphics[width=\textwidth]{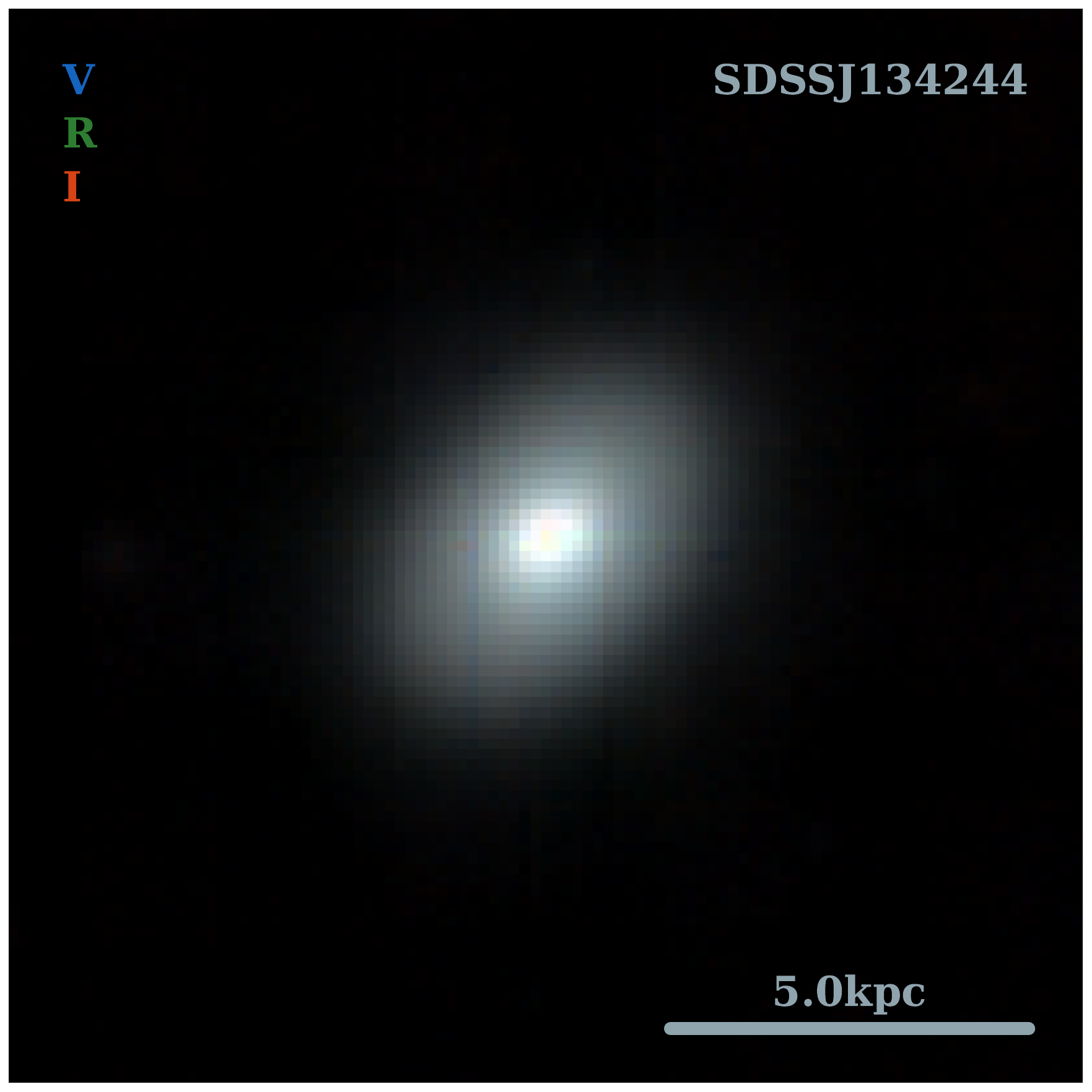}
    \end{subfigure} %
    \begin{subfigure}[b]{0.246\textwidth}
         \centering
        \includegraphics[width=\textwidth]{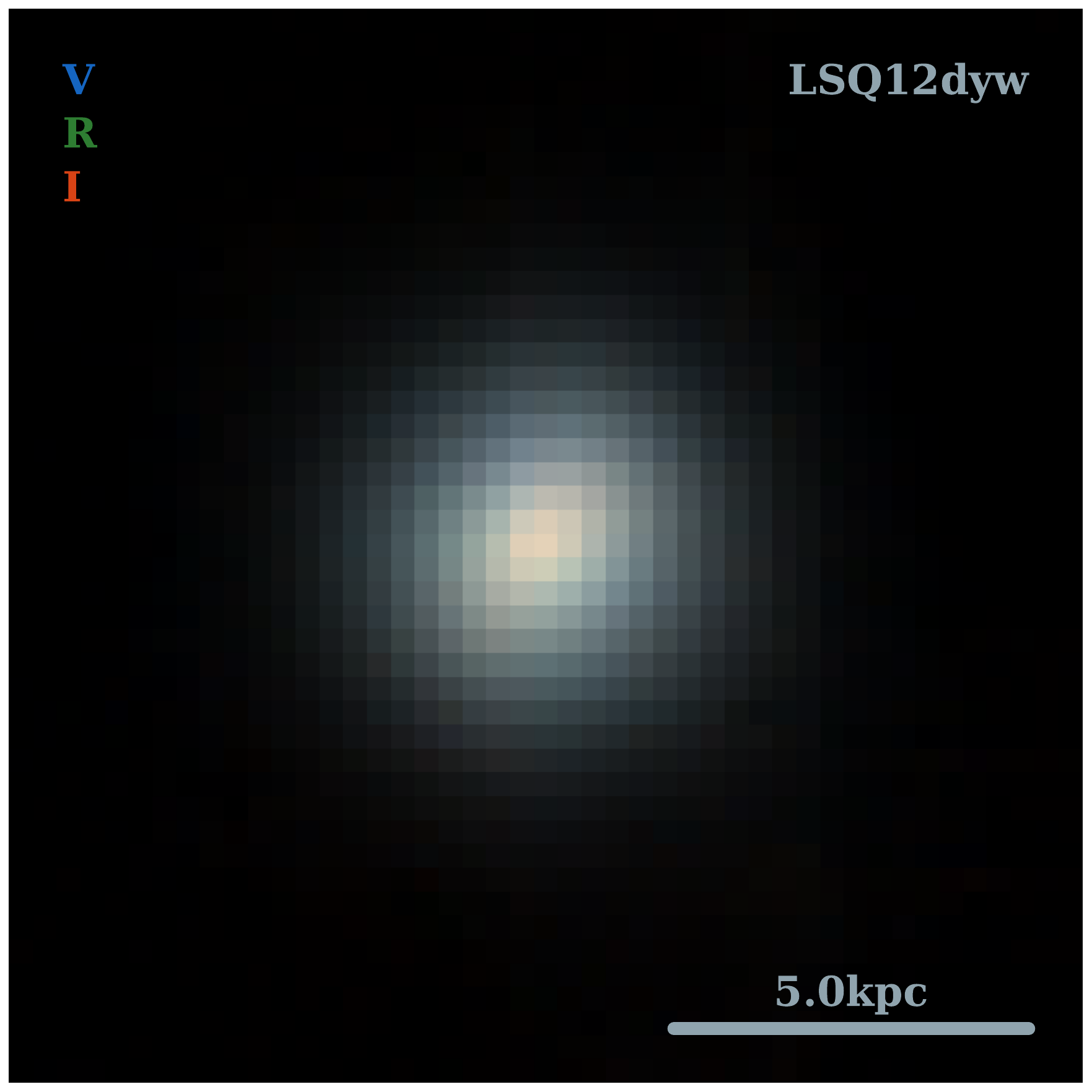}
    \end{subfigure} %
    
    \begin{subfigure}[b]{0.246\textwidth}
         \centering
        \includegraphics[width=\textwidth]{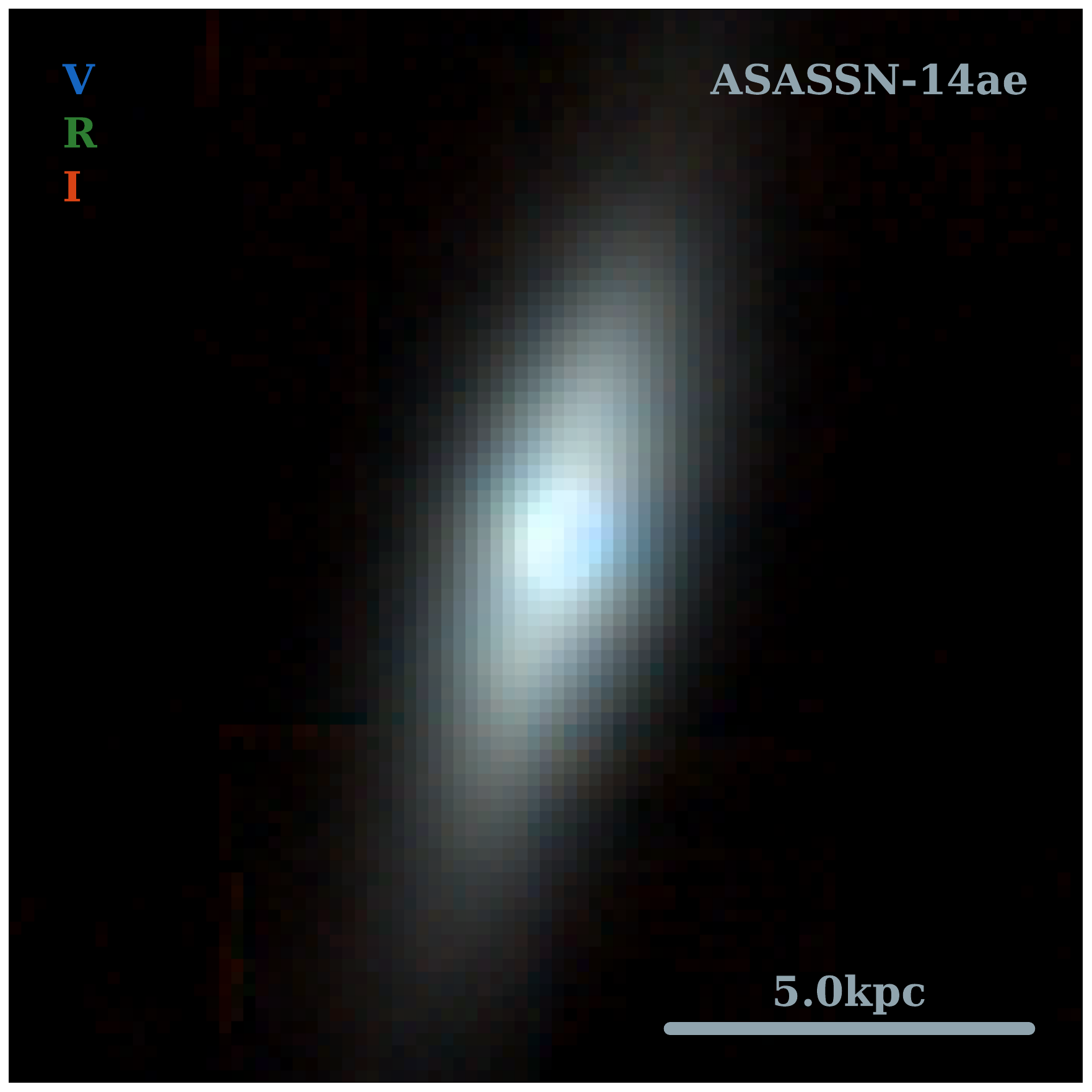}
    \end{subfigure} %
    \begin{subfigure}[b]{0.246\textwidth}
         \centering
        \includegraphics[width=\textwidth]{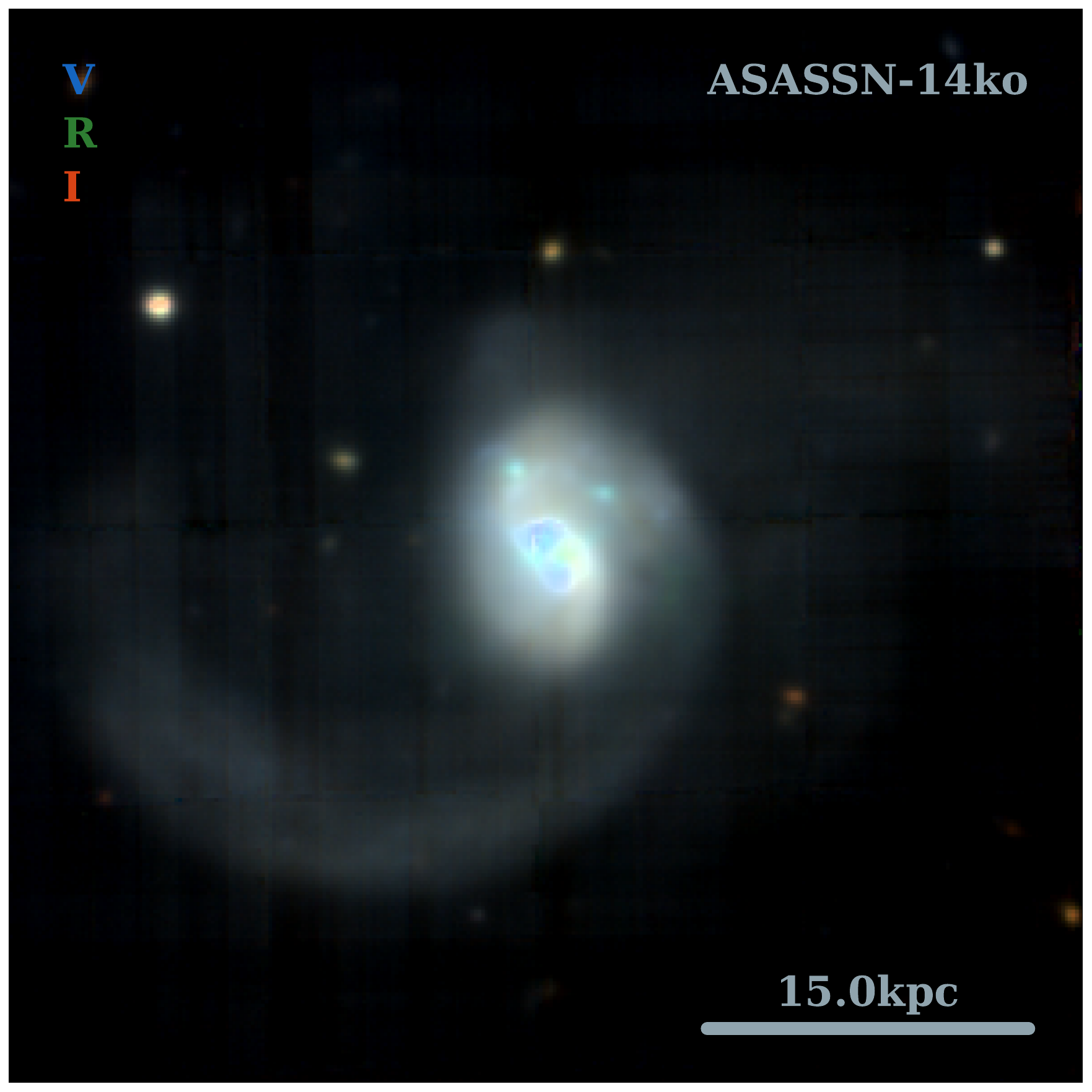}
    \end{subfigure} %
    \begin{subfigure}[b]{0.246\textwidth}
         \centering
        \includegraphics[width=\textwidth]{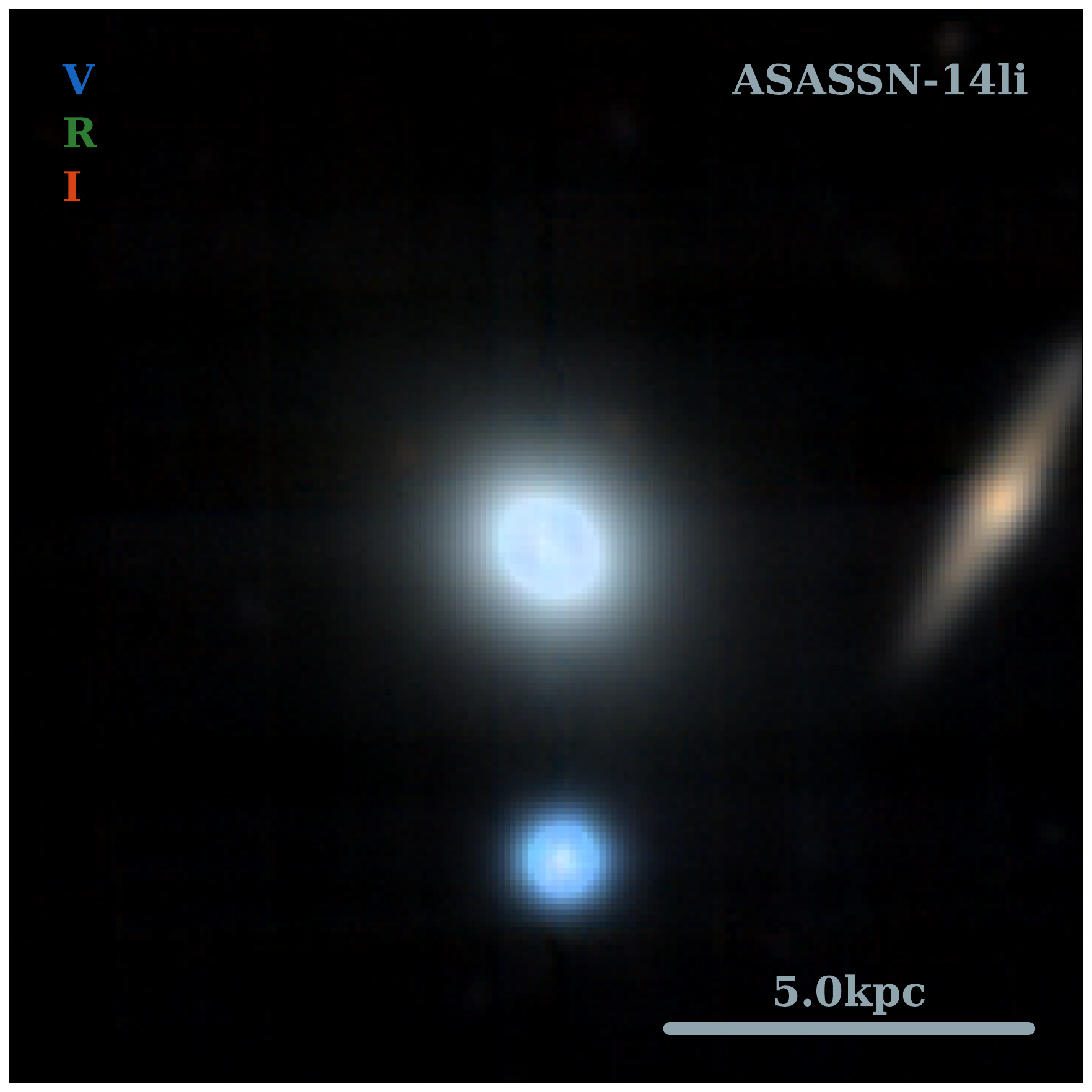}
    \end{subfigure} %
    \begin{subfigure}[b]{0.246\textwidth}
         \centering
        \includegraphics[width=\textwidth]{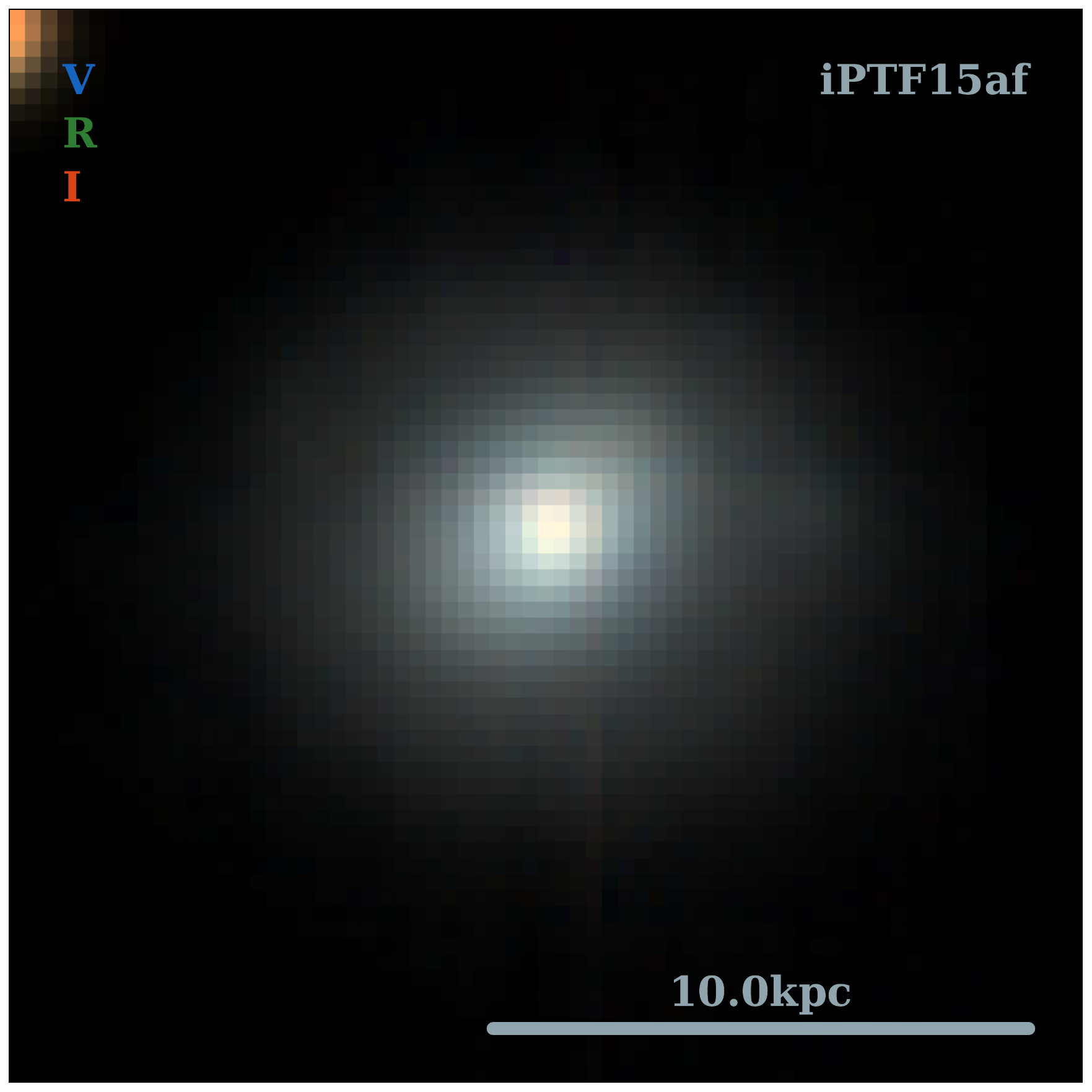}
    \end{subfigure} %

    \begin{subfigure}[b]{0.246\textwidth}
         \centering
        \includegraphics[width=\textwidth]{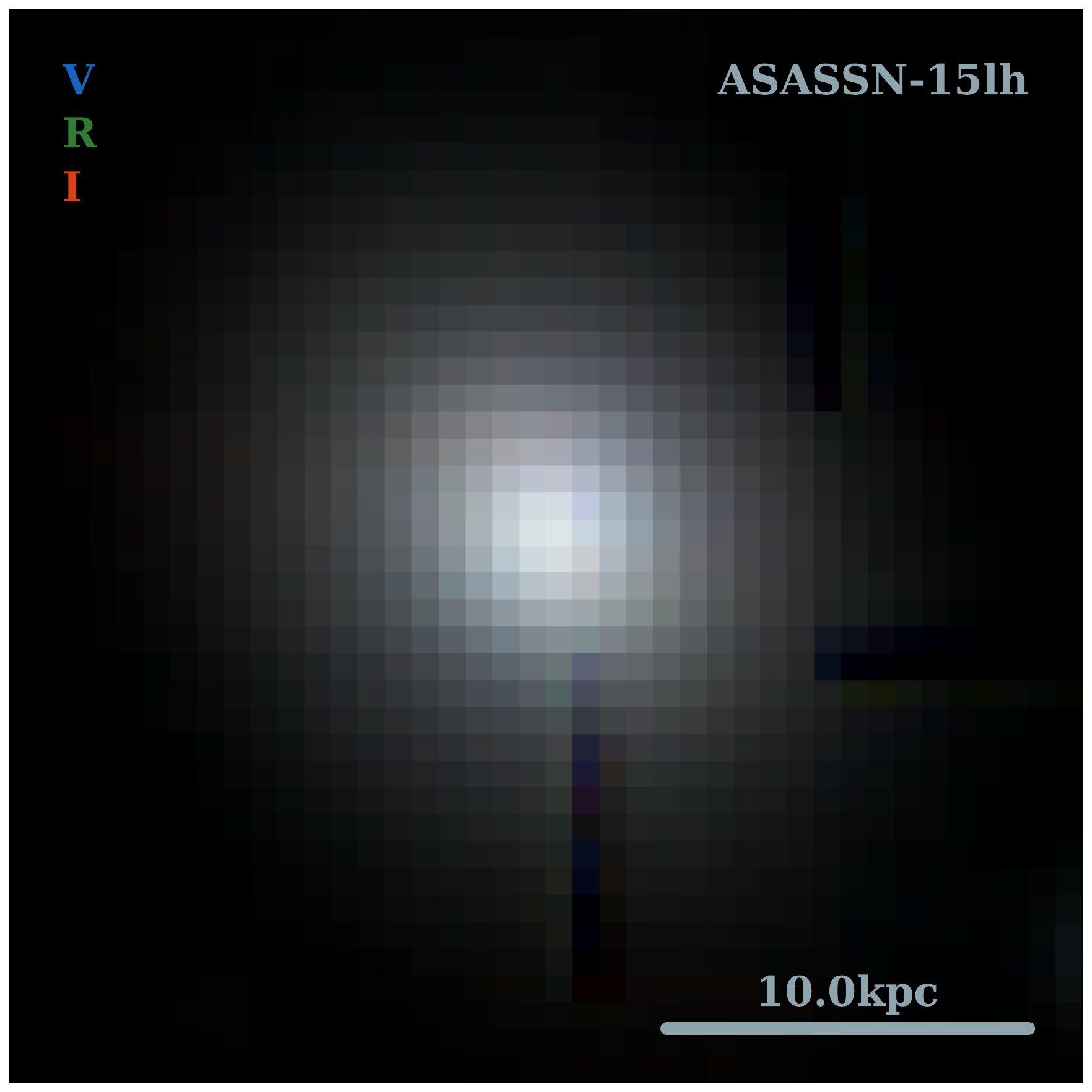}
    \end{subfigure} %
    \begin{subfigure}[b]{0.246\textwidth}
         \centering
        \includegraphics[width=\textwidth]{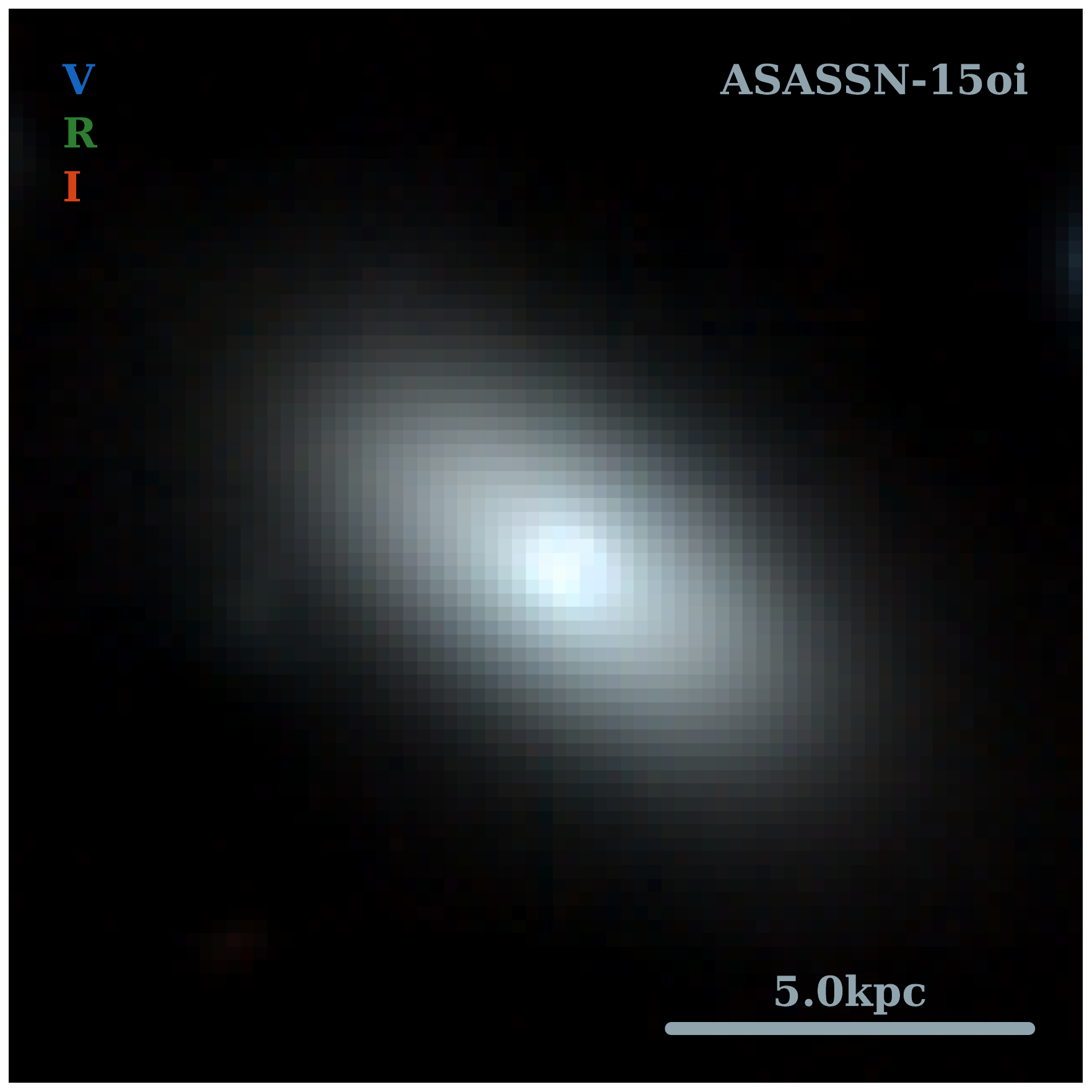}
    \end{subfigure} %
    \begin{subfigure}[b]{0.246\textwidth}
         \centering
        \includegraphics[width=\textwidth]{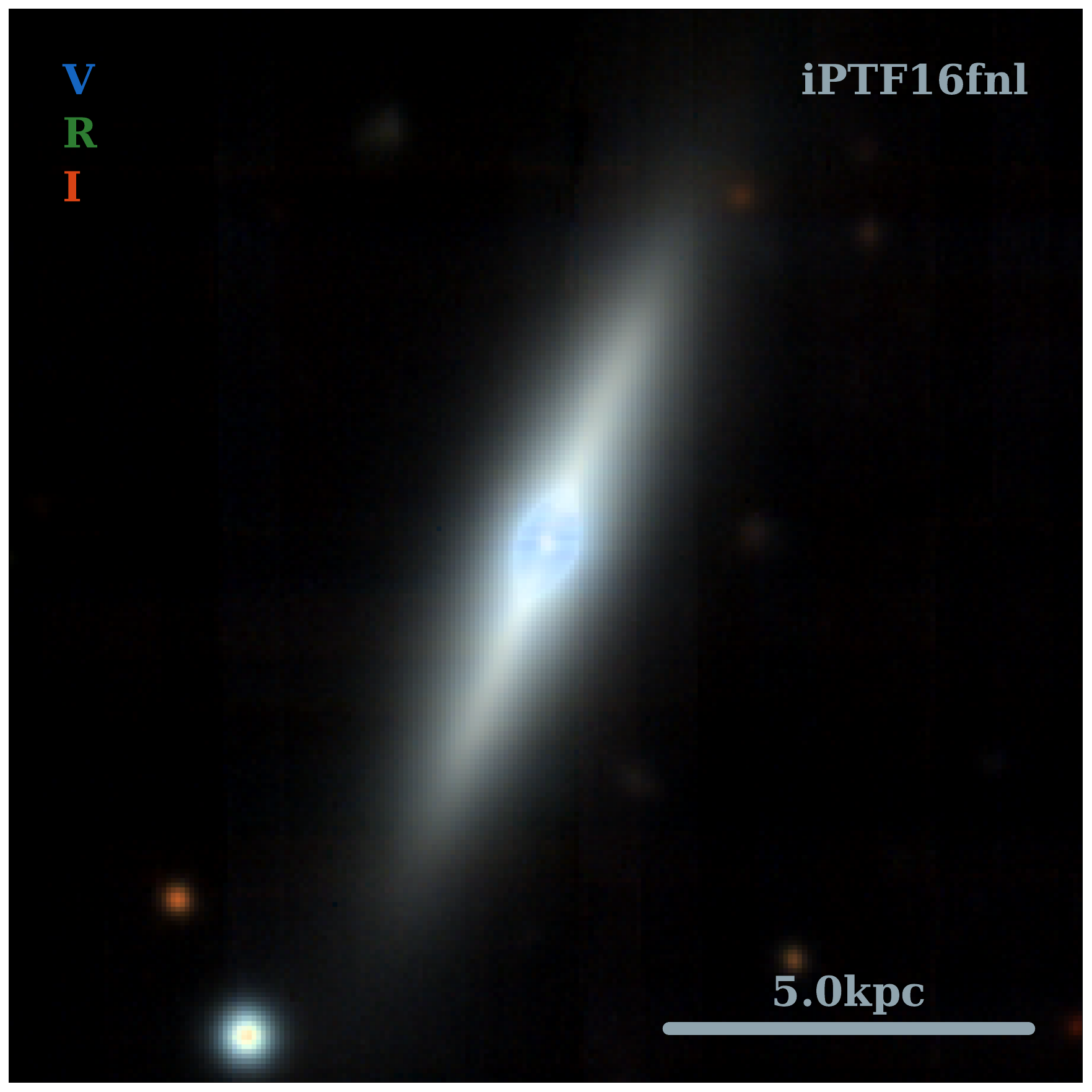}
    \end{subfigure} %
    \begin{subfigure}[b]{0.246\textwidth}
         \centering
        \includegraphics[width=\textwidth]{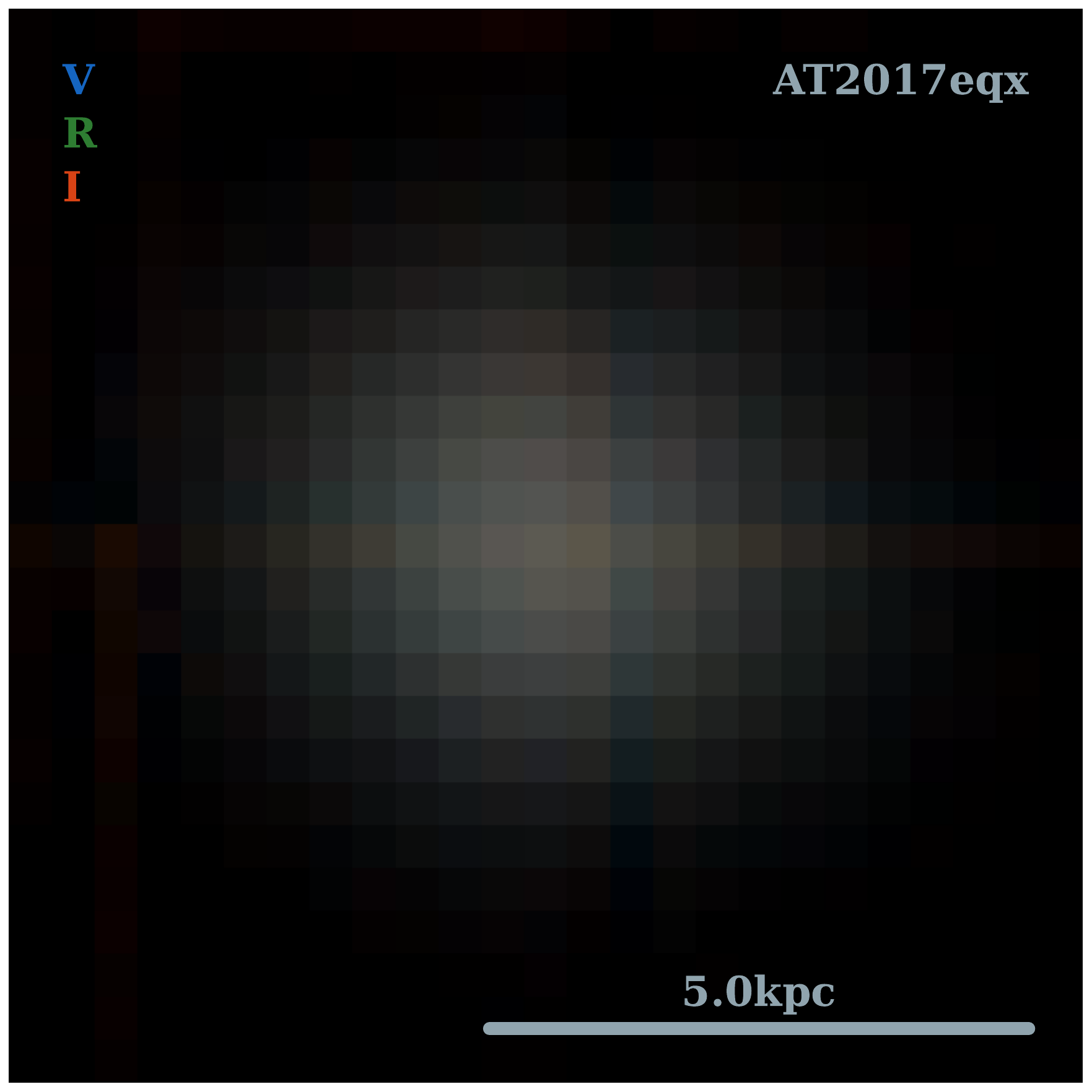}
    \end{subfigure} %

    \begin{subfigure}[b]{0.246\textwidth}
         \centering
        \includegraphics[width=\textwidth]{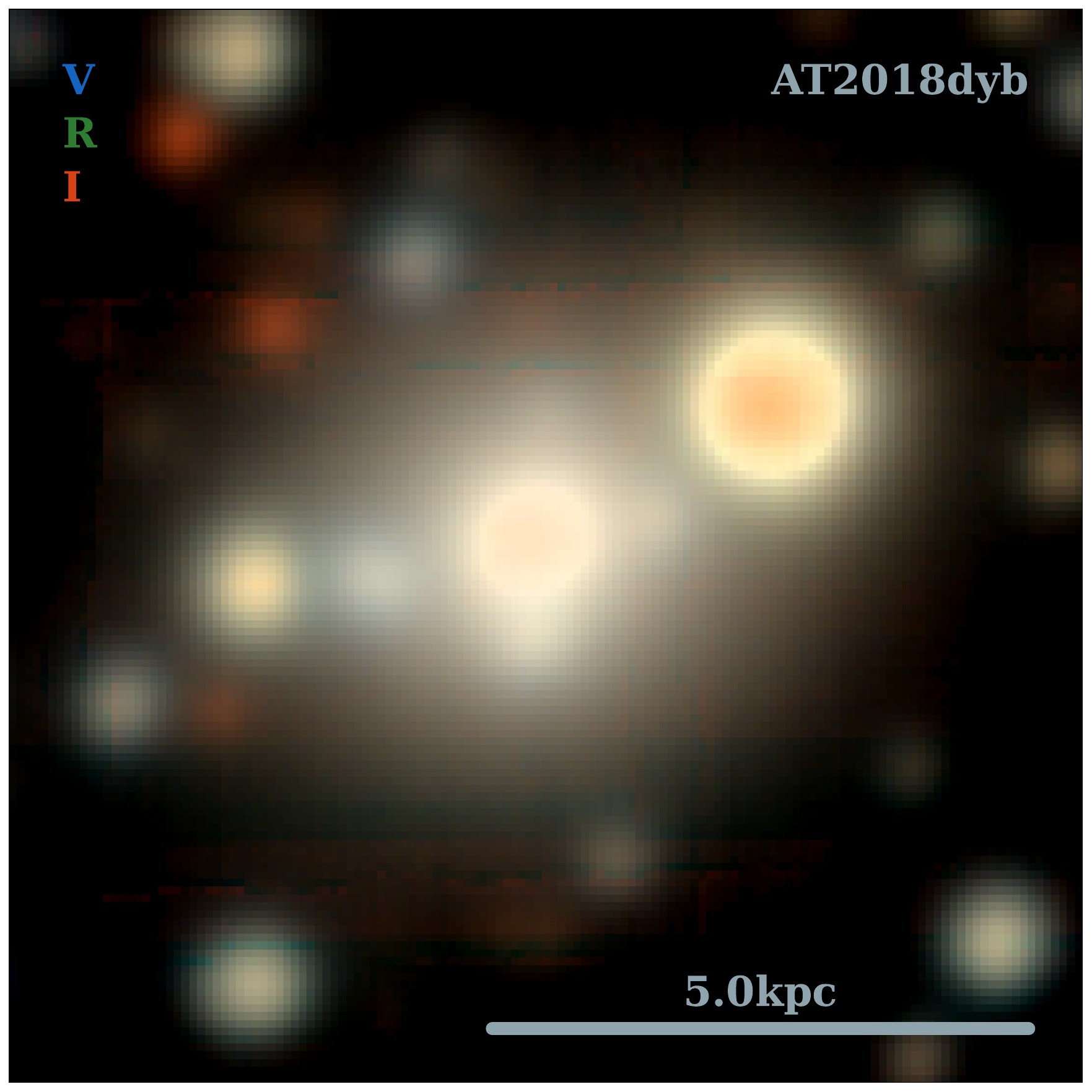}
    \end{subfigure} %
    \begin{subfigure}[b]{0.246\textwidth}
         \centering
        \includegraphics[width=\textwidth]{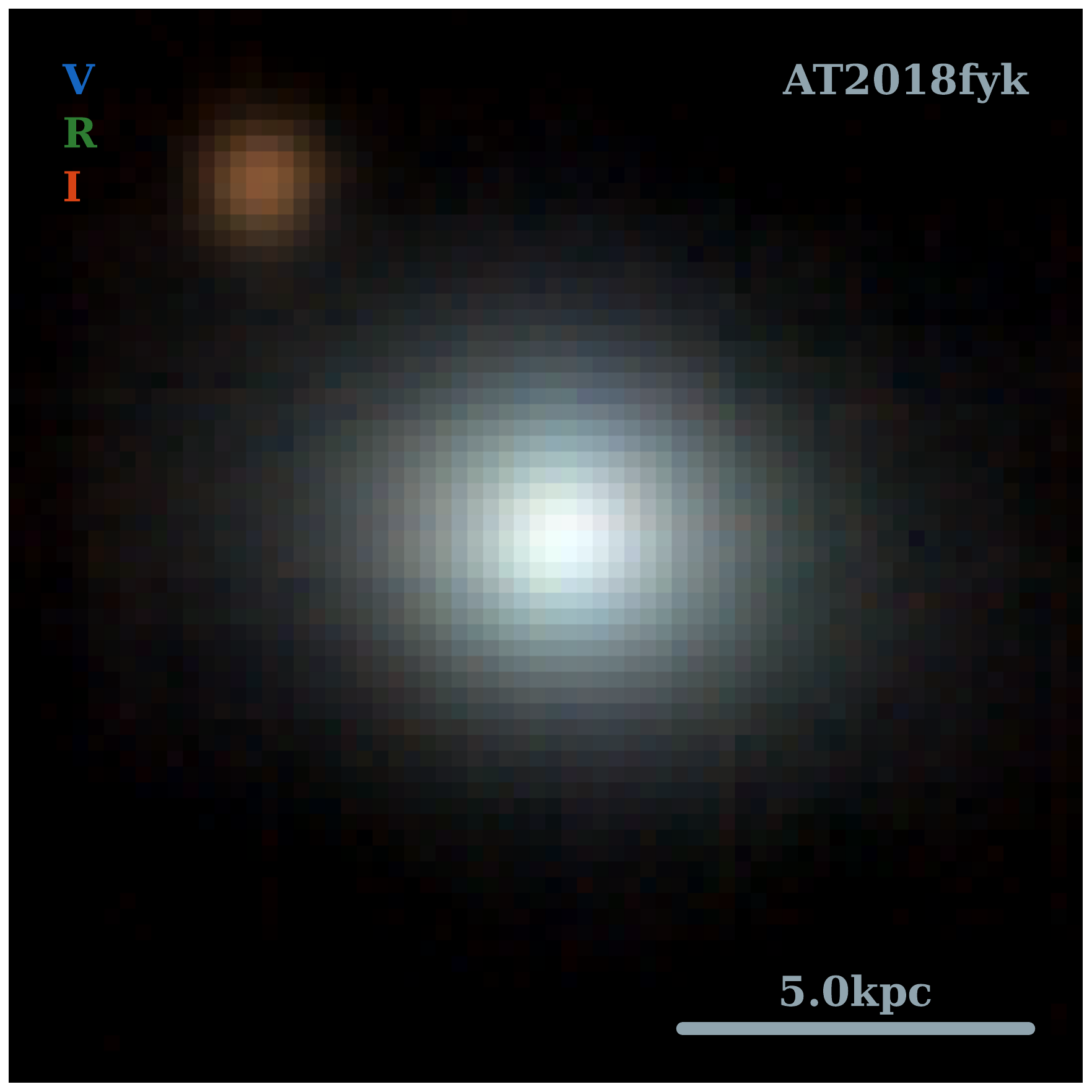}
    \end{subfigure} %
    \begin{subfigure}[b]{0.246\textwidth}
         \centering
        \includegraphics[width=\textwidth]{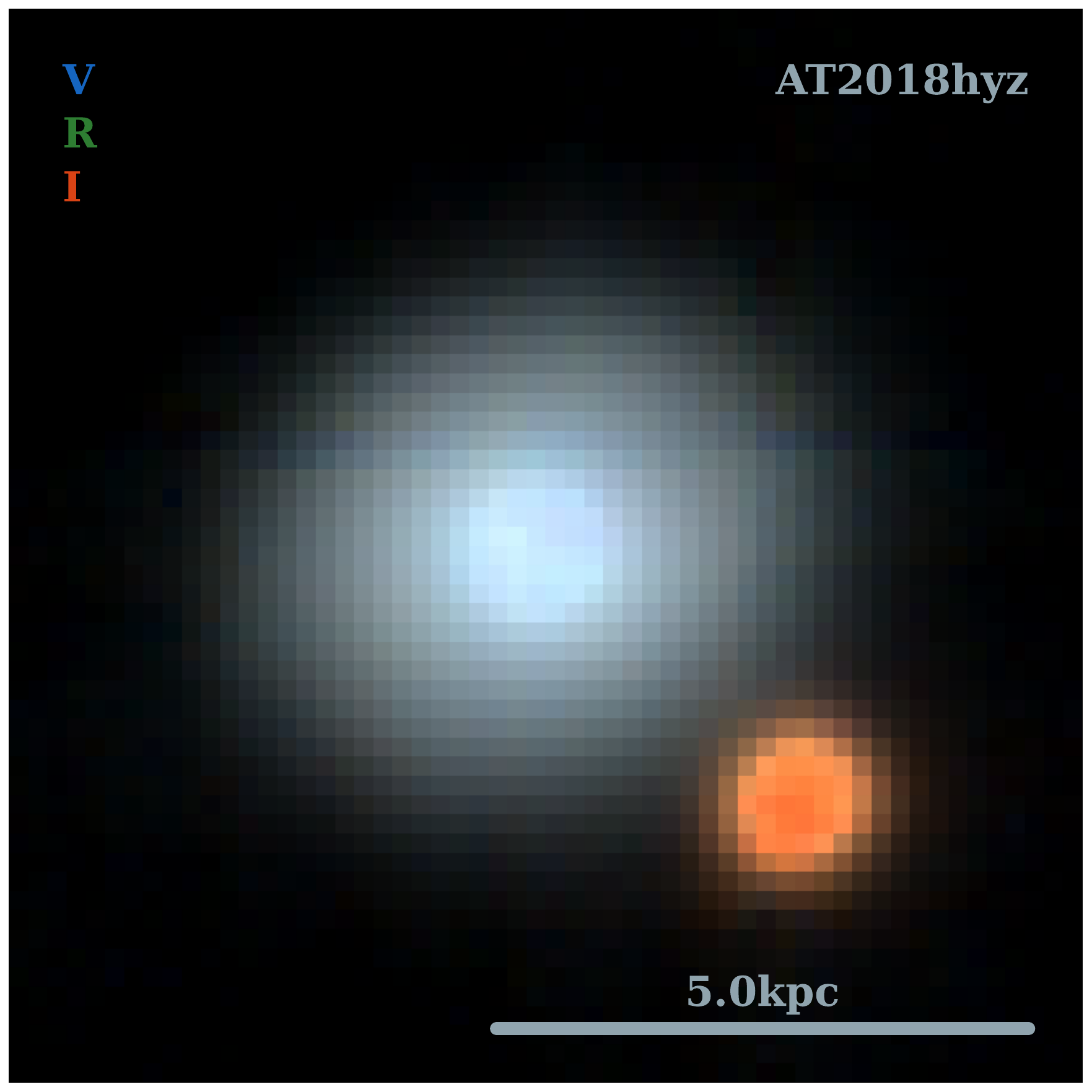}
    \end{subfigure} %
    \begin{subfigure}[b]{0.246\textwidth}
         \centering
        \includegraphics[width=\textwidth]{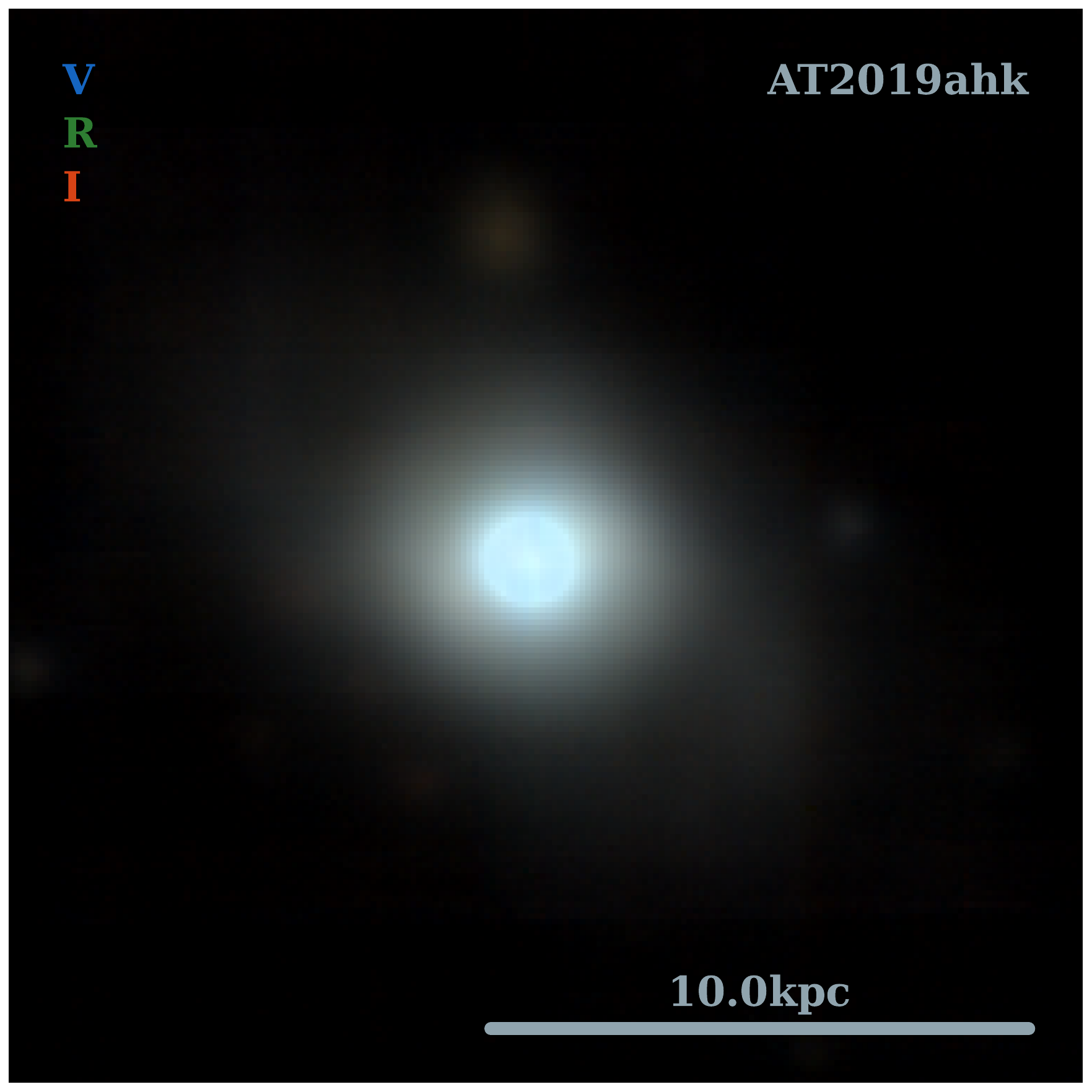}
    \end{subfigure} %
    
    \begin{subfigure}[b]{0.246\textwidth}
         \centering
        \includegraphics[width=\textwidth]{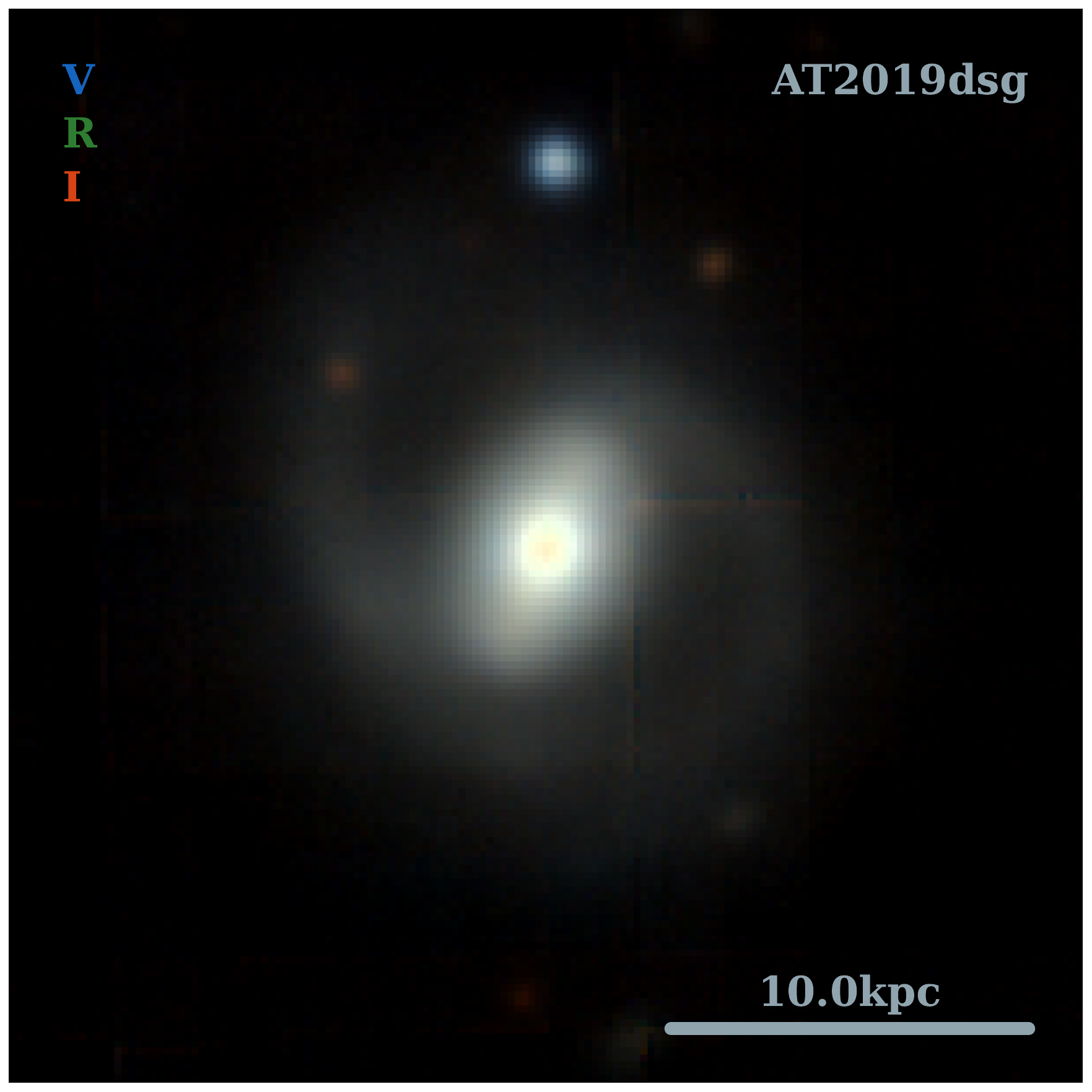}
    \end{subfigure} %
    \begin{subfigure}[b]{0.246\textwidth}
         \centering
        \includegraphics[width=\textwidth]{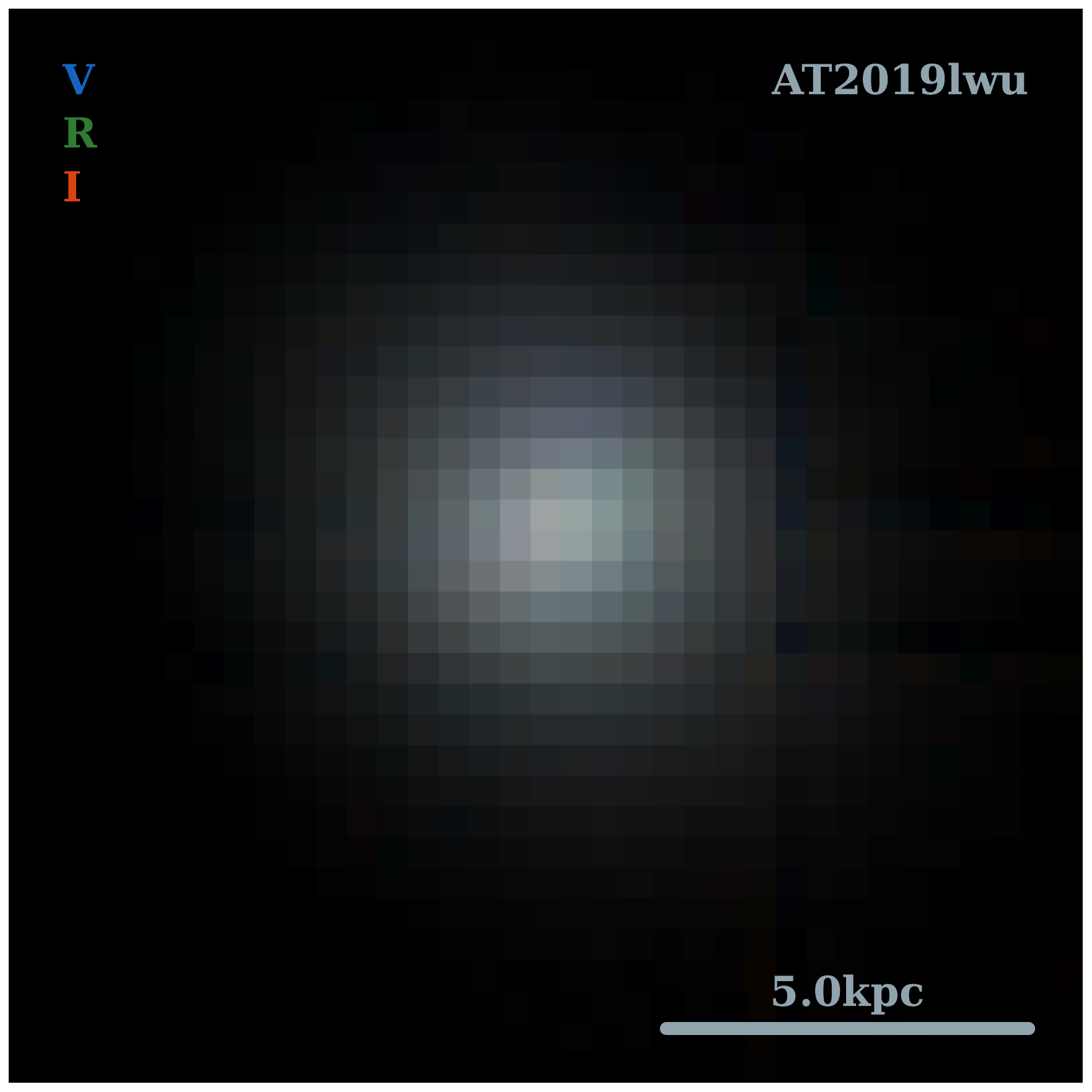}
    \end{subfigure} %
    \begin{subfigure}[b]{0.246\textwidth}
         \centering
        \includegraphics[width=\textwidth]{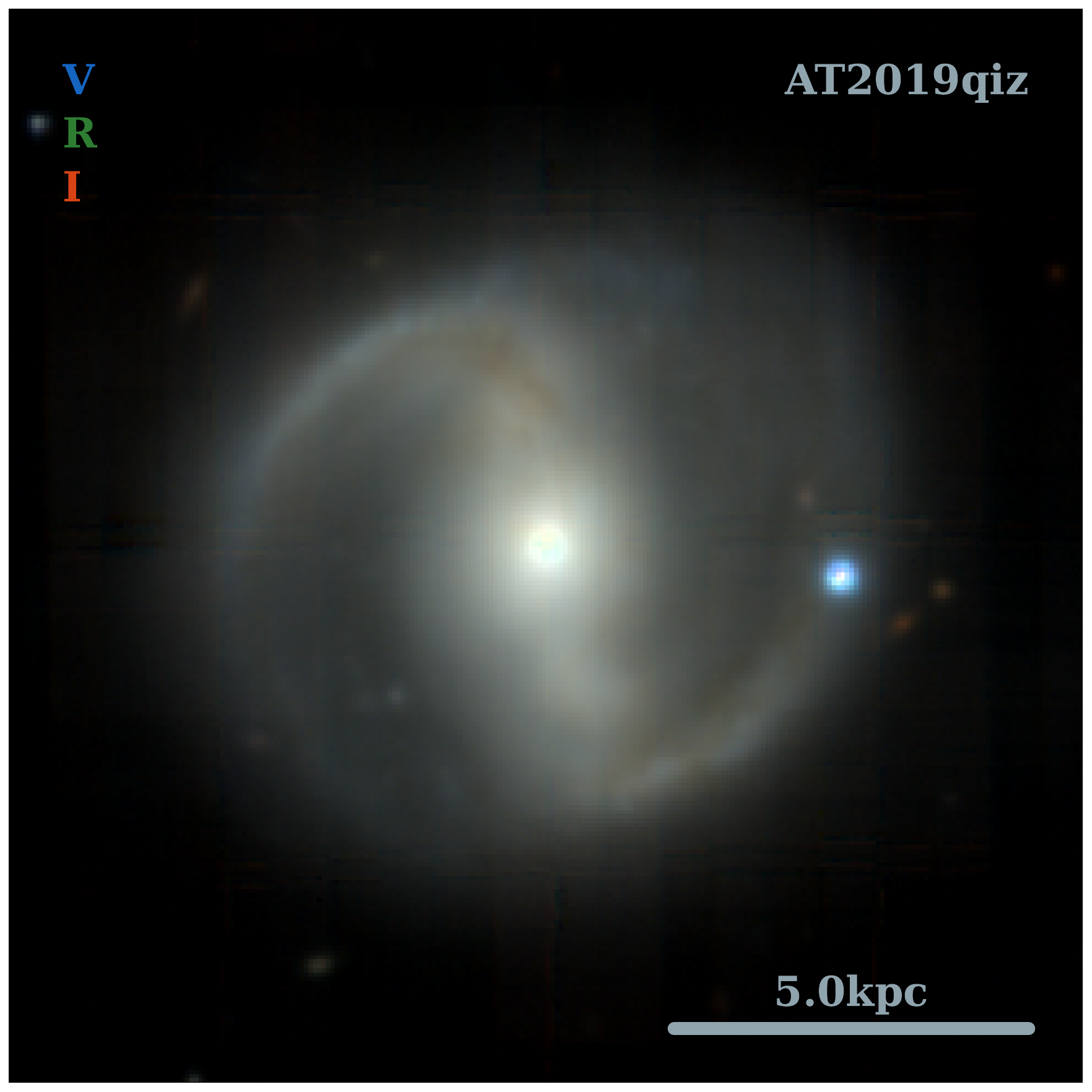}
    \end{subfigure} %
    \begin{subfigure}[b]{0.246\textwidth}
         \centering
        \includegraphics[width=\textwidth]{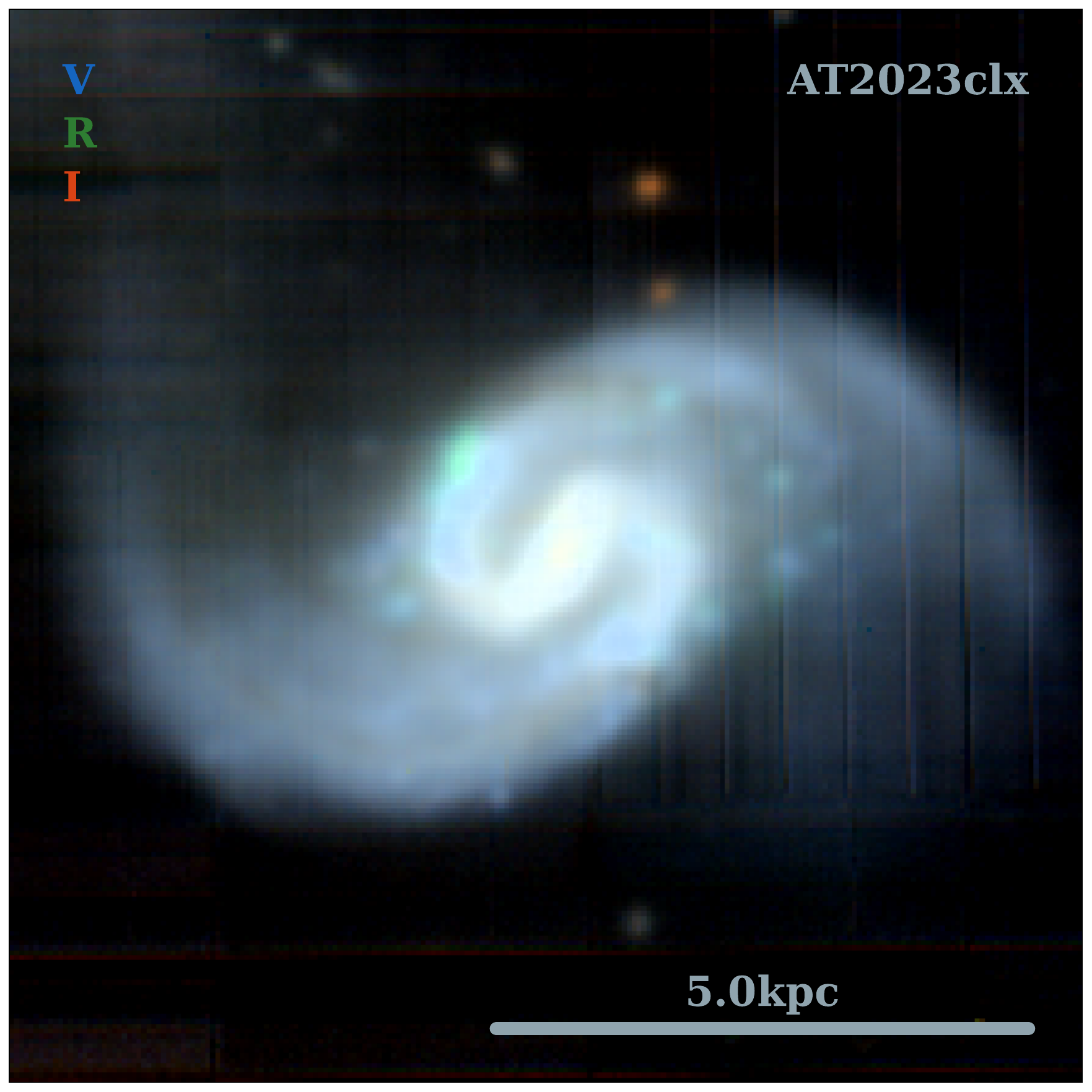}
    \end{subfigure} %

    \caption{$VRI$ false colour images of the TDE host galaxies generated from the MUSE cubes. The images are cropped to the near vicinity of the galaxies.}
    \label{fig:VRI_crop}
\end{figure*}

\begin{figure*}
    \centering
    \begin{subfigure}[b]{0.425\textwidth}
         \centering
        \includegraphics[width=\textwidth]{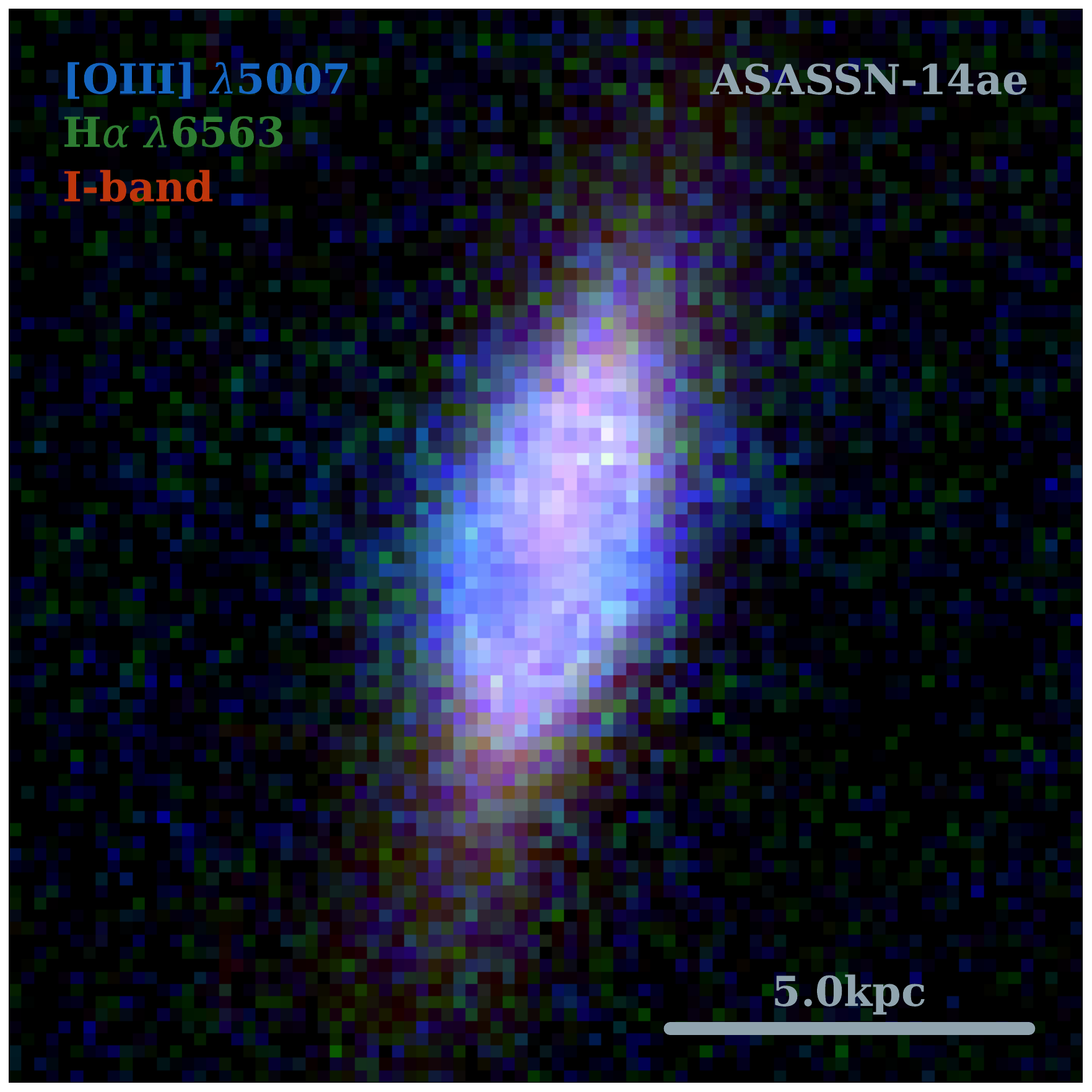}
    \end{subfigure} %
    \begin{subfigure}[b]{0.425\textwidth}
         \centering
        \includegraphics[width=\textwidth]{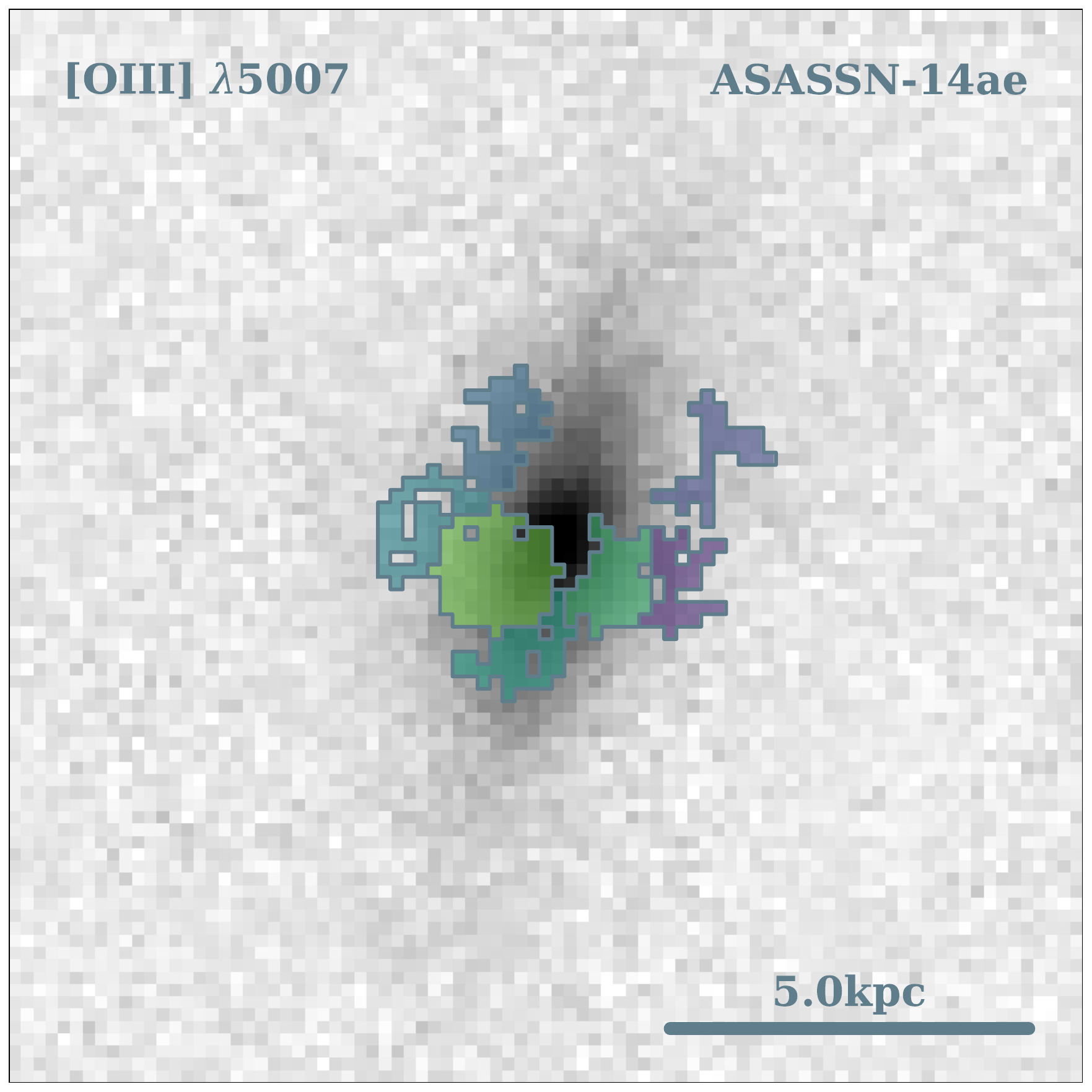}
    \end{subfigure} %

    \begin{subfigure}[b]{0.425\textwidth}
         \centering
        \includegraphics[width=\textwidth]{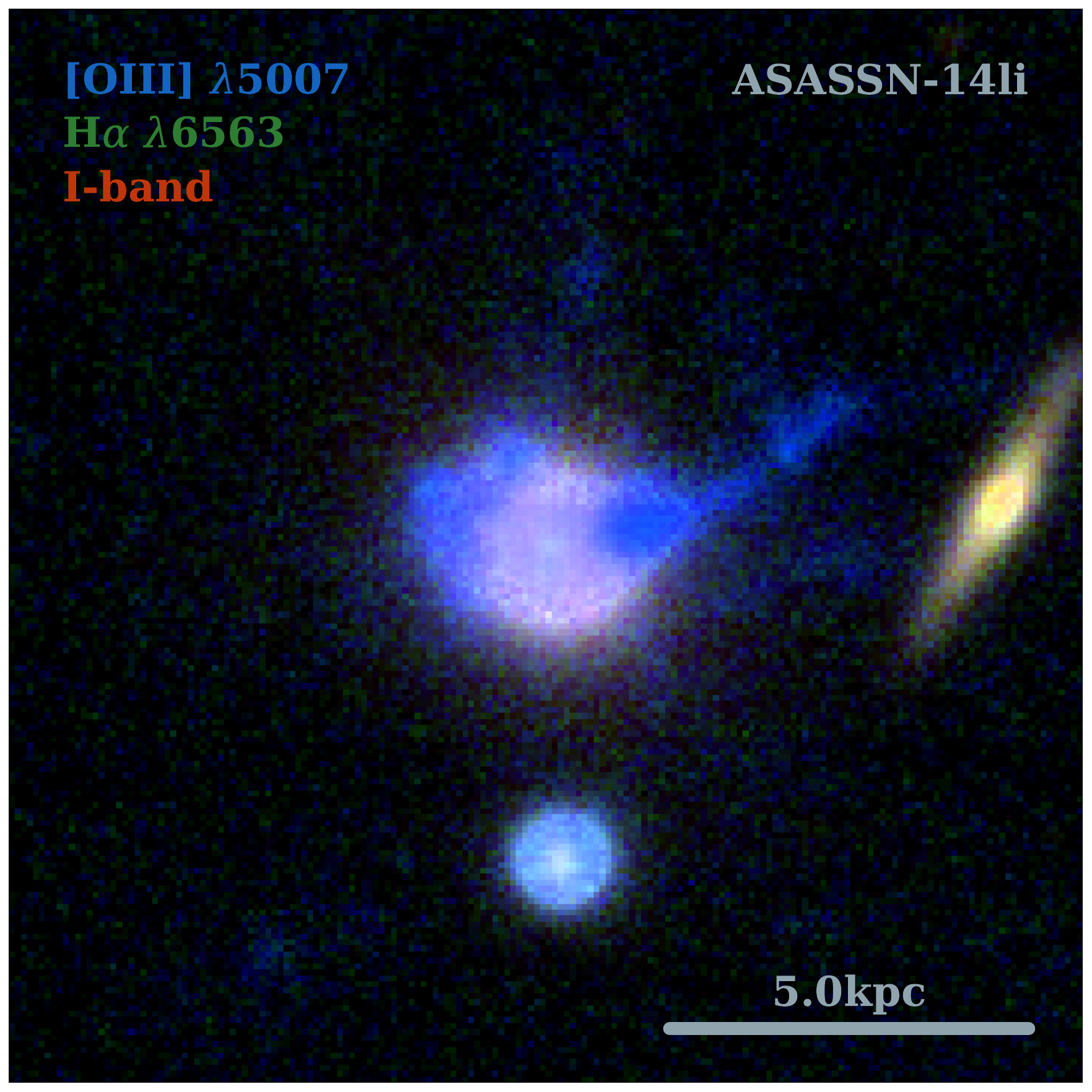}
    \end{subfigure} %
    \begin{subfigure}[b]{0.425\textwidth}
         \centering
        \includegraphics[width=\textwidth]{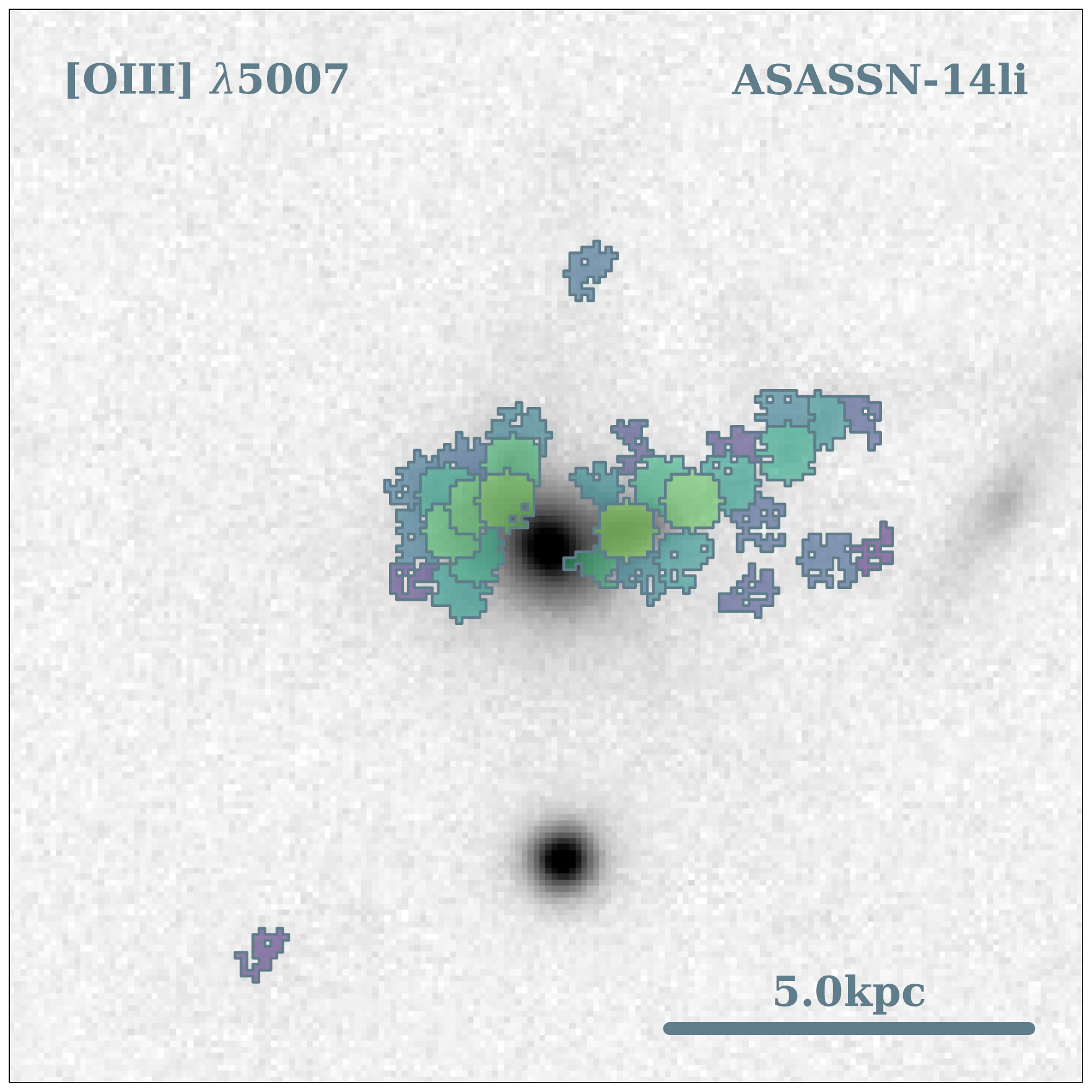}
    \end{subfigure} %

    \caption{The identified\ion{[O}{III]} $\lambda5007$ emission line regions for ASASSN-14ae and ASASSN-14li \textit{Left:} False colour image highlighting the \ion{[O}{III]} $\lambda5007$ (blue) and H$\alpha$ emission (green). \textit{Right}: Identified emission line bins over a greyscale image of \ion{[O}{III]} $\lambda5007$ emission. The colours refer to the spectra shown in Figure \ref{fig:emission_line_spectra_ASASSN-14ae_14li_14ko}.}
    \label{fig:EELR_figs1}
\end{figure*}

\begin{figure*}
    \centering

        \begin{subfigure}[b]{0.425\textwidth}
         \centering
        \includegraphics[width=\textwidth]{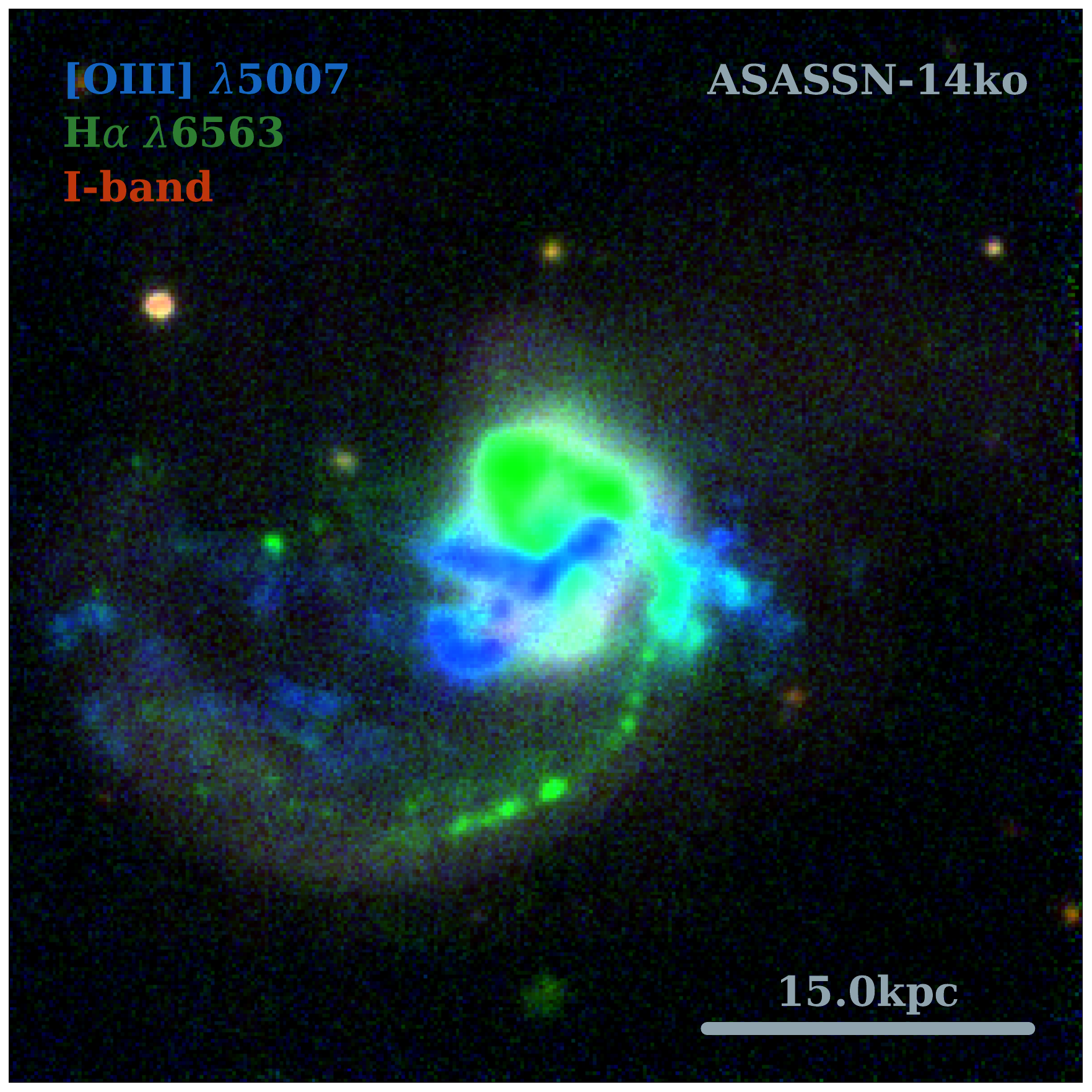}
    \end{subfigure} %
    \begin{subfigure}[b]{0.425\textwidth}
         \centering
        \includegraphics[width=\textwidth]{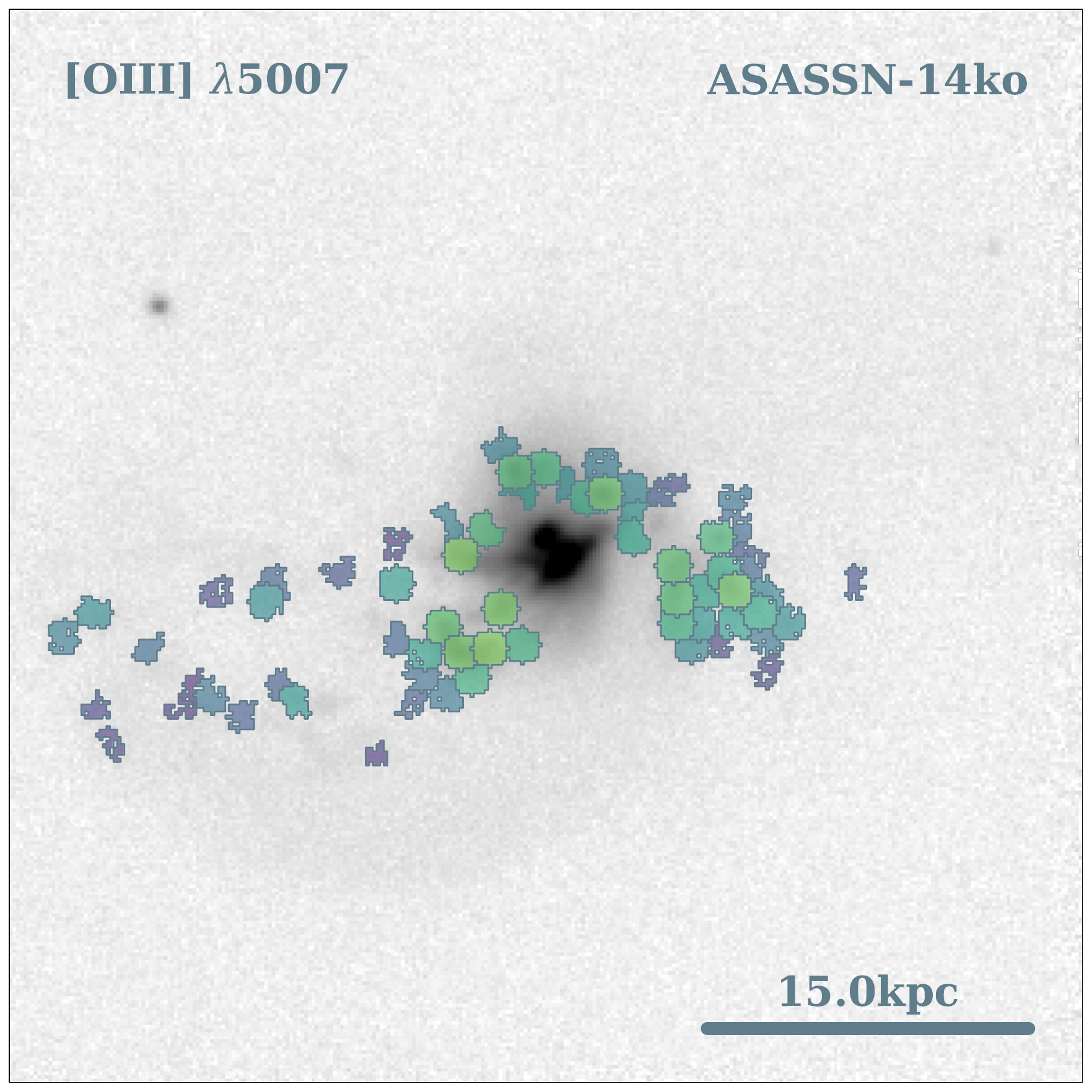}
    \end{subfigure} %

    \begin{subfigure}[b]{0.425\textwidth}
         \centering
        \includegraphics[width=\textwidth]{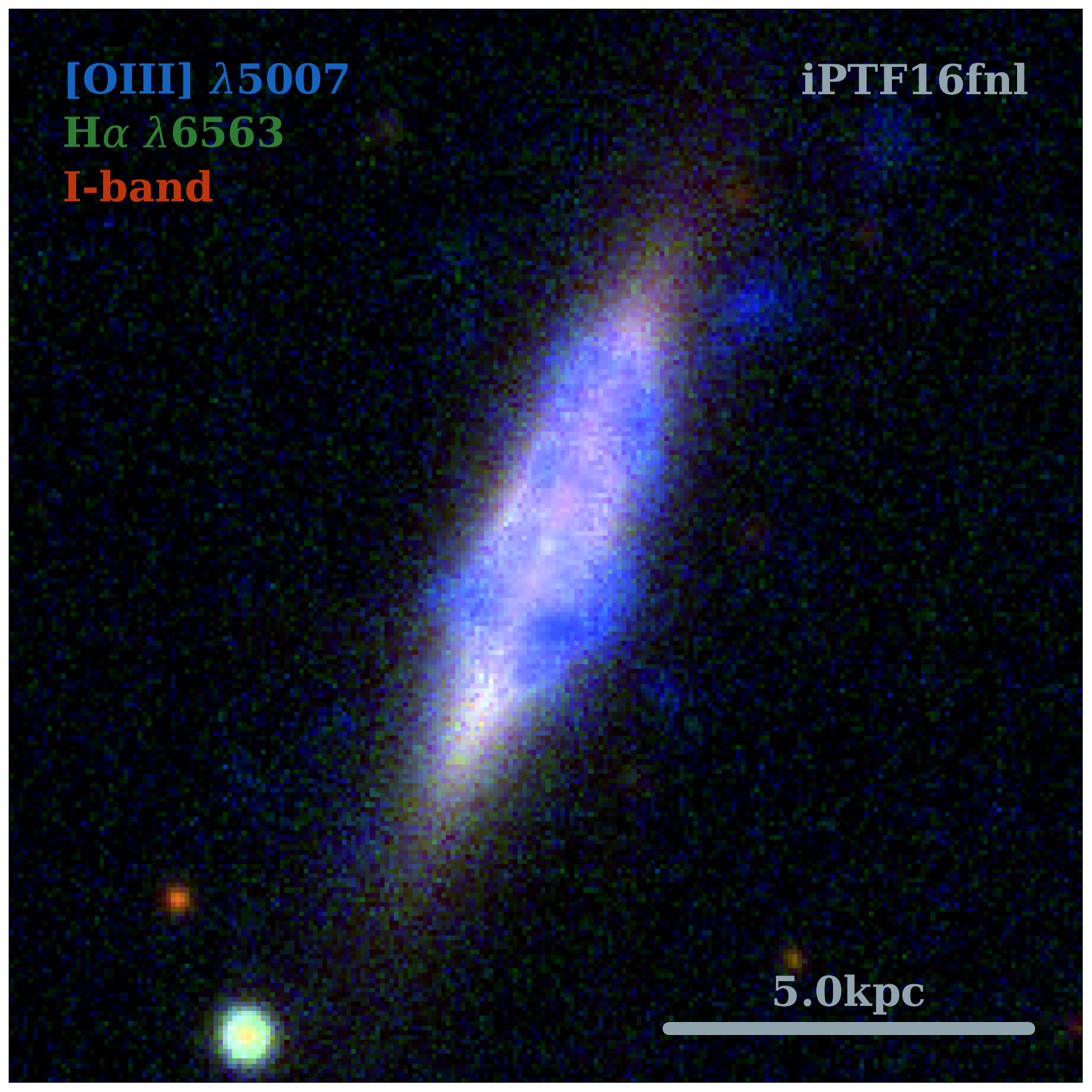}
    \end{subfigure} %
    \begin{subfigure}[b]{0.425\textwidth}
         \centering
        \includegraphics[width=\textwidth]{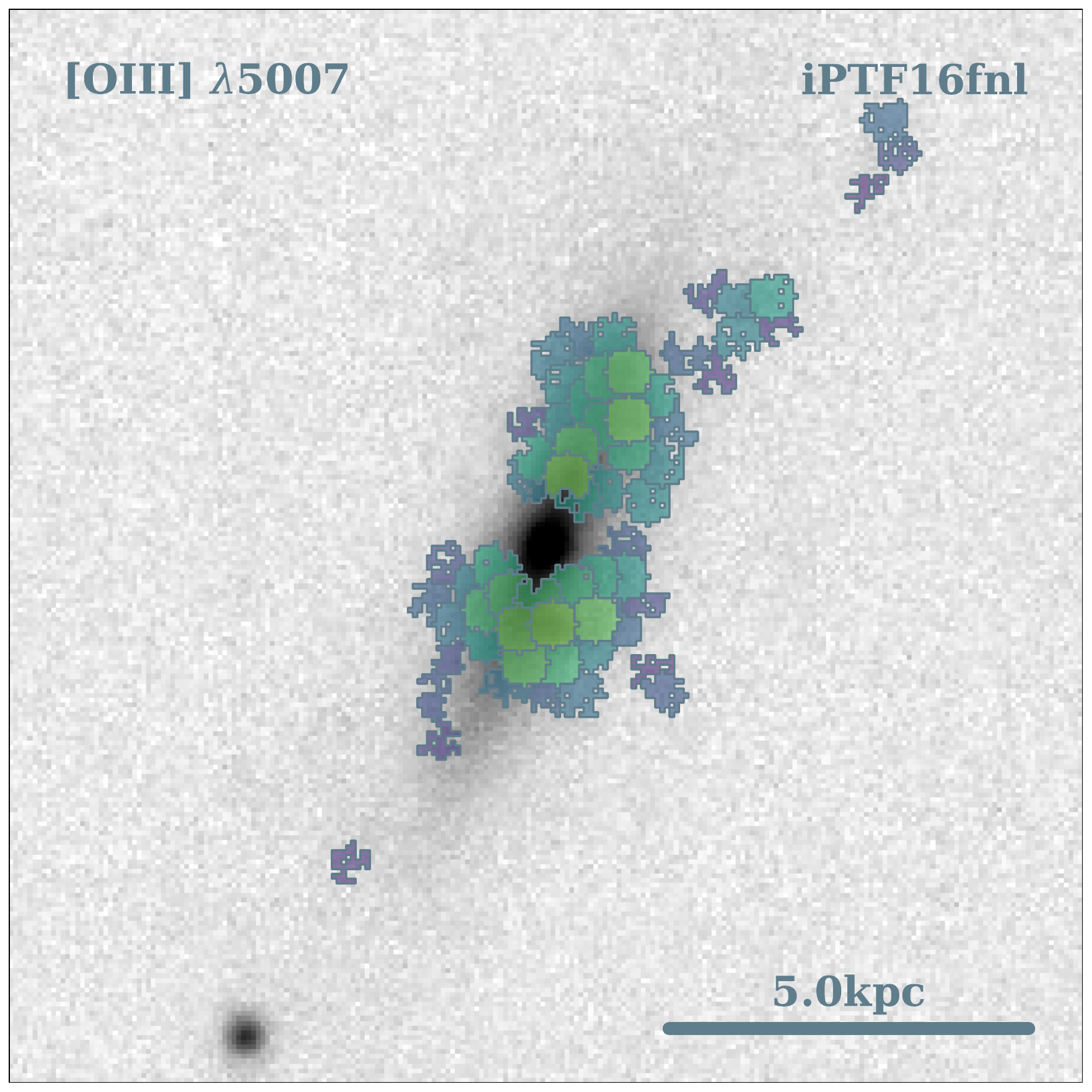}
    \end{subfigure} %

    \begin{subfigure}[b]{0.425\textwidth}
         \centering
        \includegraphics[width=\textwidth]{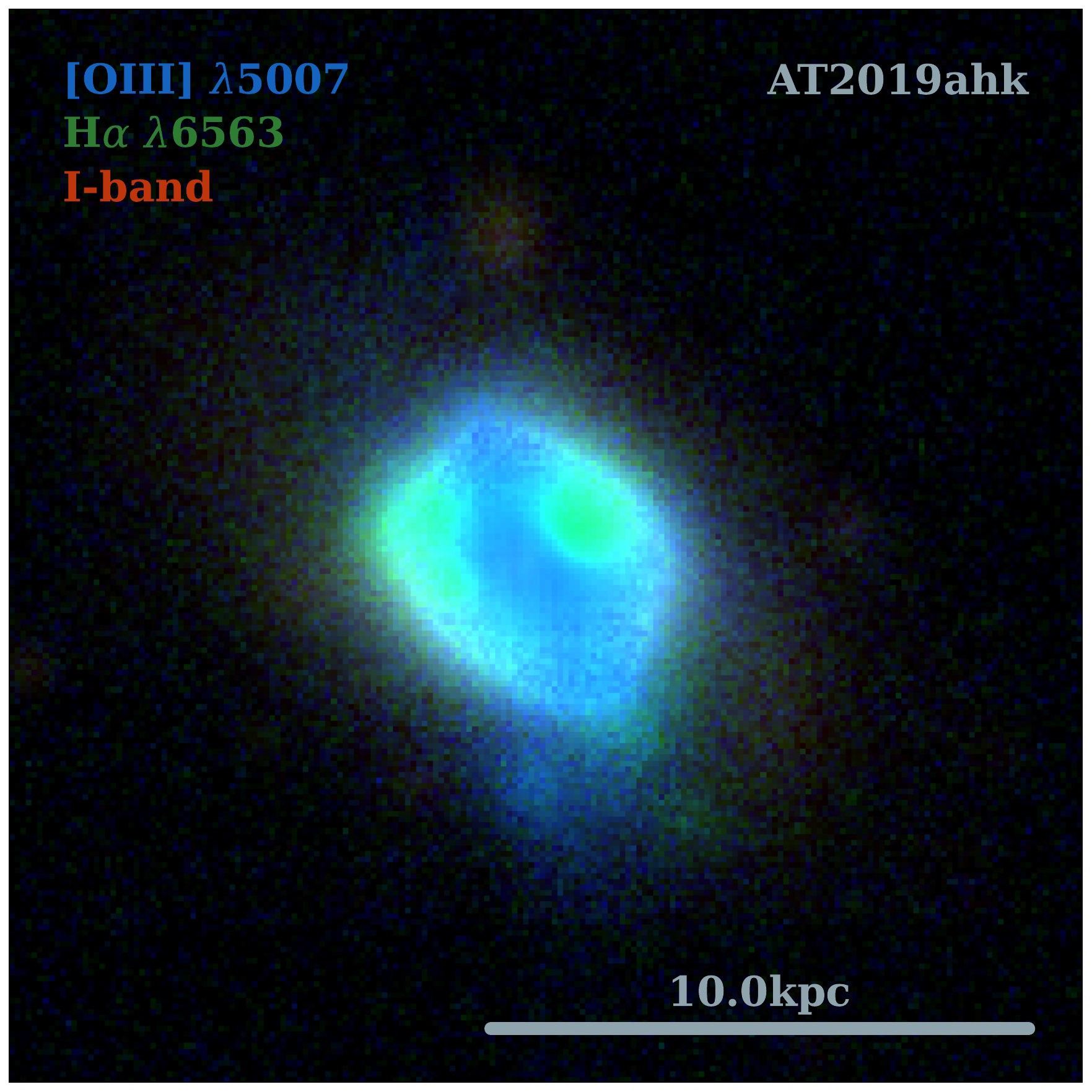}
    \end{subfigure} %
    \begin{subfigure}[b]{0.425\textwidth}
         \centering
        \includegraphics[width=\textwidth]{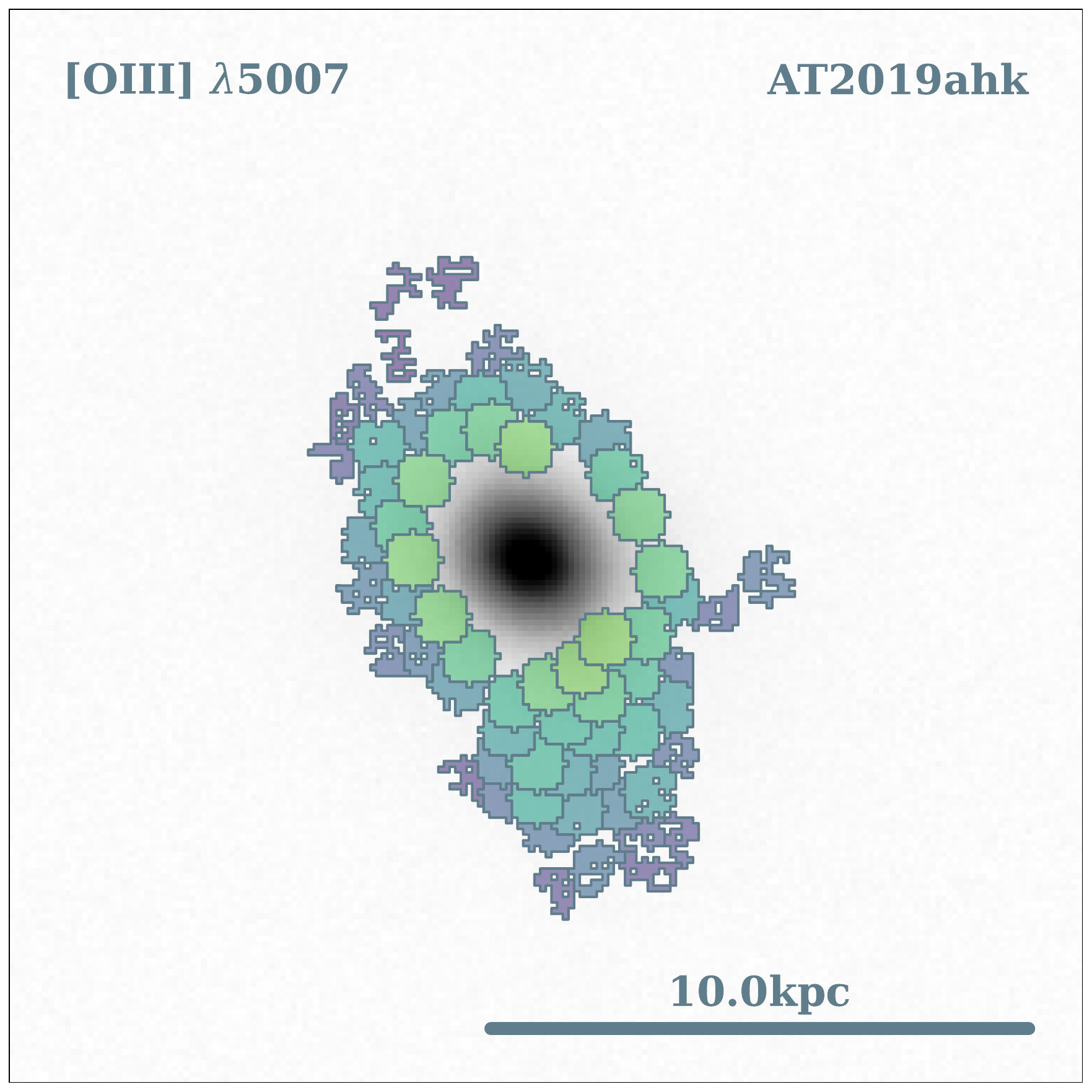}
    \end{subfigure} %
      
    \caption{Same as \ref{fig:EELR_figs1} for ASSASN-14ko, iPTF16fnl and AT\,2019ahk. The corresponding spectra are shown in Figures \ref{fig:emission_line_spectra_ASASSN-14ae_14li_14ko} and \ref{fig:emission_line_spectra_iPTF16fnl_AT2019ahk}}
    \label{fig:EELR_figs2}
\end{figure*}

\begin{figure*}
    \centering
    \begin{subfigure}[b]{0.49\textwidth}
        \centering
        \includegraphics[width=\textwidth]{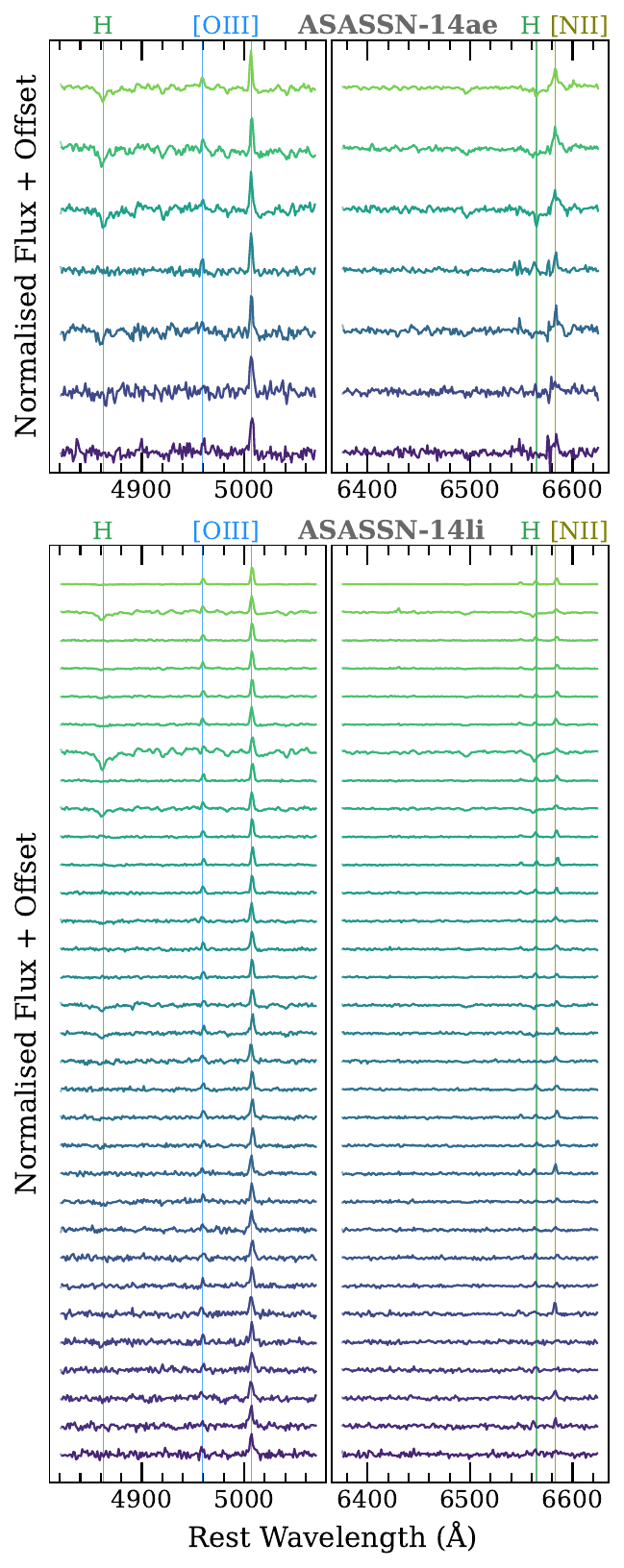}
    \end{subfigure} %
    \begin{subfigure}[b]{0.49\textwidth}
        \centering
        \includegraphics[width=\textwidth]{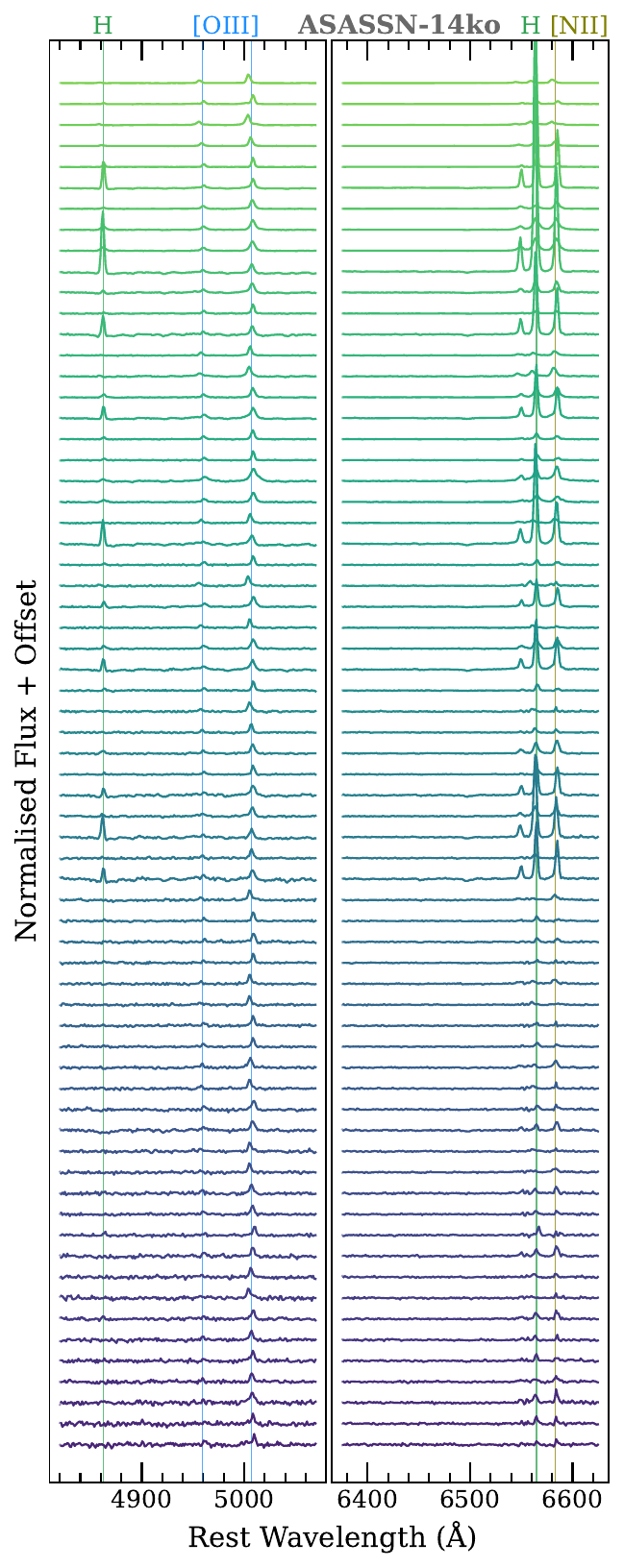}
    \end{subfigure} %
    \caption{Spectra of the identified emission line bins in ASASSN-14ae, ASASSN-14li and ASASSN-14ko. The spectra are shown in the order of highest \ion{[O}{III]} $\lambda5007$ luminosity from top to bottom, and they are normalised by the measured line flux. \ion{[O}{III]} $\lambda5007$ emission is apparent in all the bins, even if other lines are not identified.}
    \label{fig:emission_line_spectra_ASASSN-14ae_14li_14ko}
\end{figure*}

\begin{figure*}
    \centering
    \begin{subfigure}[b]{0.49\textwidth}
        \centering
        \includegraphics[width=\textwidth]{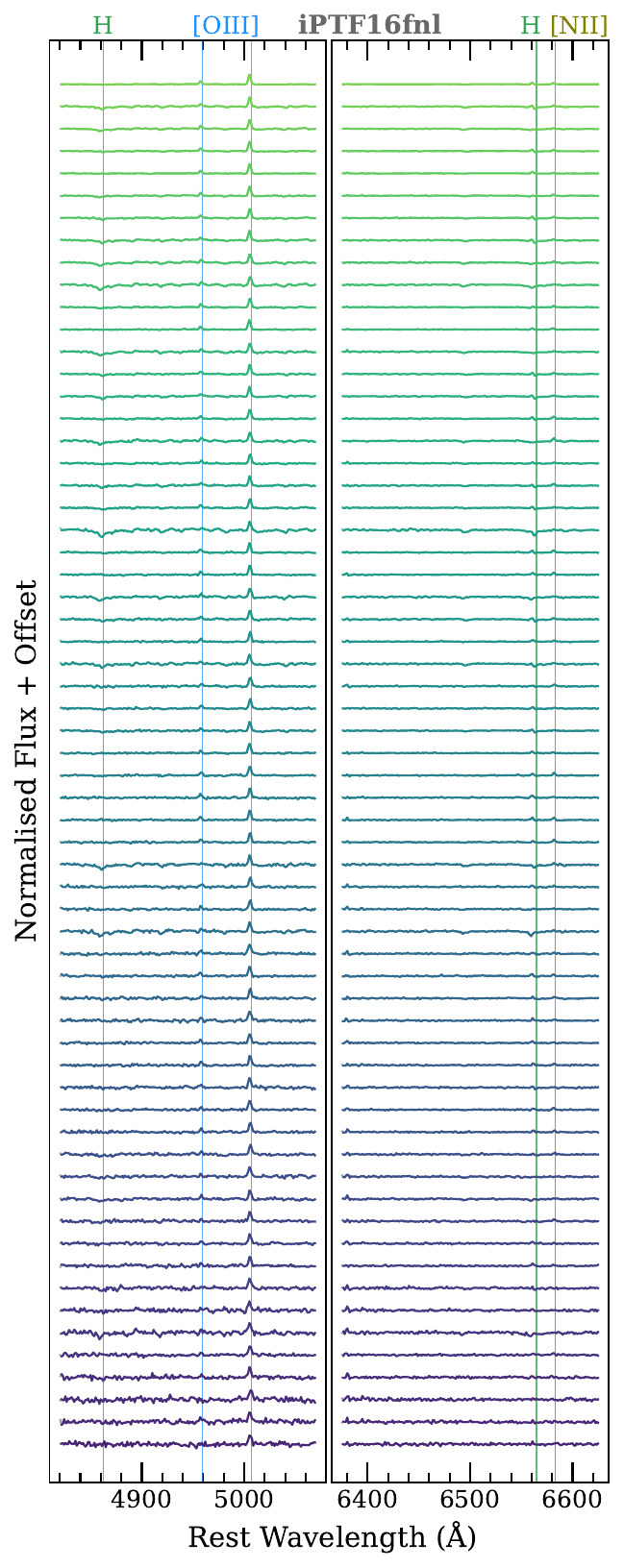}
    \end{subfigure}
    \begin{subfigure}[b]{0.49\textwidth}
        \centering
        \includegraphics[width=\textwidth]{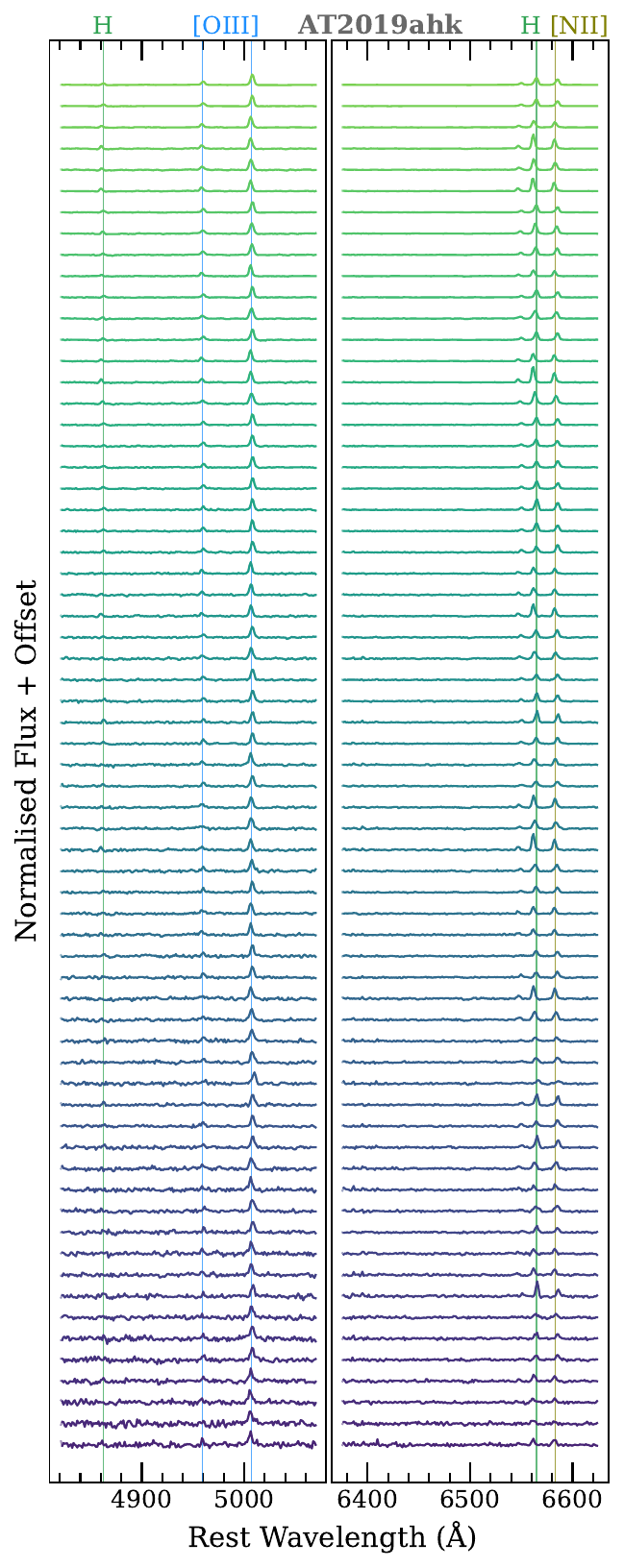}
    \end{subfigure}
    \caption{Same as Figure \ref{fig:emission_line_spectra_ASASSN-14ae_14li_14ko}, but for iPTF16fnl and AT2019ahk}
    \label{fig:emission_line_spectra_iPTF16fnl_AT2019ahk}
\end{figure*}

\begin{figure*}
    \centering
    \includegraphics[width=1\textwidth]{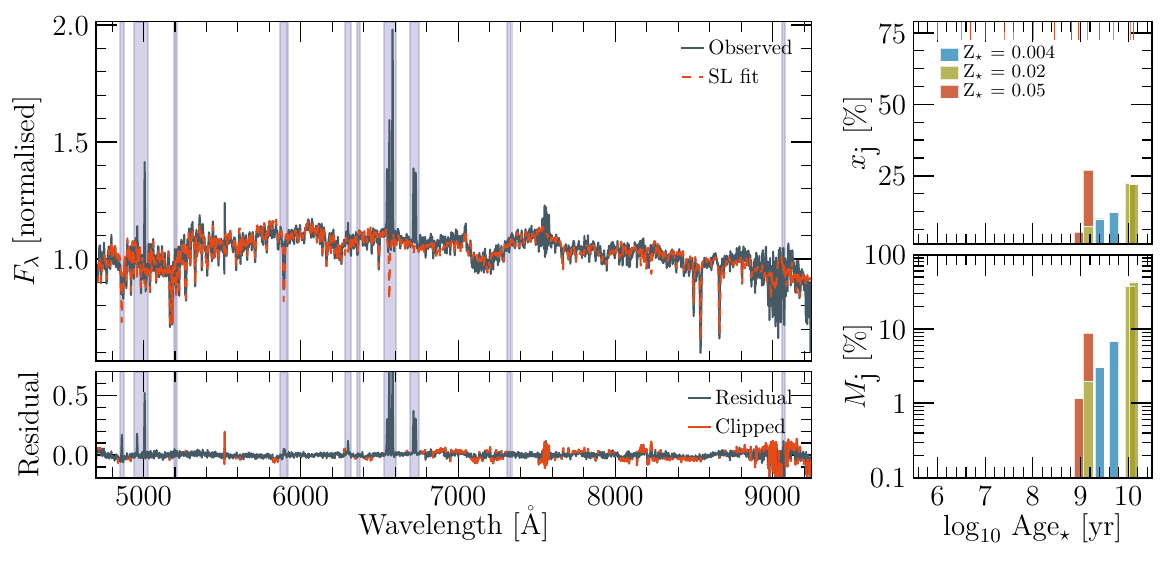}
   
    \caption{The best-fitting \texttt{Starlight} model of the nuclear spectrum of the AT\,2023clx host galaxy. \textit{Left:} The spectrum, the fit and the resulting residuals. \textit{Right:} The light ($x_\mathrm{j}$) and mass ($M_\mathrm{j}$) fractions of the different SSPs. Three metallicities were used, but populations were merged into one in the analysis. The regions of strong galaxy lines (shaded) are masked in the fitting routine.}
    \label{fig:SL_AT2023clx}
\end{figure*}


\bsp	
\label{lastpage}
\end{document}